\documentclass[acmlarge,10pt]{acmart} 

\usepackage{xcolor}
\usepackage{xurl}
\usepackage{tikz}
\usepackage{array} 
\usetikzlibrary{patterns}
\usepackage{booktabs}
\usepackage[ruled]{algorithm2e}
\usepackage{algpseudocode}
\usepackage{enumerate}
\usepackage[english]{babel}
\usepackage{blindtext}
\usepackage{bm}
\usepackage{amsmath}

\usepackage{amssymb}
\usepackage{xspace}
\usepackage{multirow}
\usepackage{threeparttable}
\usepackage{float}
\usepackage{flafter}
\usepackage{comment}
\usepackage[skip=12pt]{caption}
\usepackage{subcaption}
\usepackage{graphicx}
\usepackage{ulem}
\usepackage{verbatim}
\usepackage[export]{adjustbox}
\usepackage{enumitem}

\usepackage{lipsum}

\def\fig{Fig.\xspace}
\def\eqn{Eq.\xspace}
\def\sec{Sec.\xspace}
\def\tab{Tab.\xspace}

\def\ie{{\textit{i.e.}\xspace}} 
\def\eg{{\textit{e.g.}\xspace}}

\def\etc{{\textit{etc}\xspace}}

\newcommand{\head}[1]{{\noindent \textbf{#1:}}}

\graphicspath{{./figures/}}
\graphicspath{{./pdf/}}
\DeclareGraphicsExtensions{.pdf,.jpeg,.png}

\ifodd 0
\newcommand{\rev}[1]{#1}
\newcommand{\com}[1]{\textbf{\color{red}(COMMENT: #1)}} 
\newcommand{\todo}[1]{\textbf{{\color{orange}(TODO: #1)}}}
\newcommand{\unused}[1]{{\color{gray}#1}}
\newcommand{\sheng}[1]{\textbf{\color{olive}(Sheng: #1)}} 
\else
\newcommand{\rev}[1]{#1}
\newcommand{\com}[1]{}
\newcommand{\todo}[1]{}
\newcommand{\unused}[1]{}
\newcommand{\sheng}[1]{}

\fi

\ifodd 0
\newcommand{\imwut}[1]{{\color{blue}#1}} 
\else
\newcommand{\imwut}[1]{#1}
\fi

\newcolumntype{B}[1]{>{\centering\arraybackslash}p{#1}}

\newcommand{\drawfilledbar}[2][3]{
    \begin{tikzpicture}[baseline=-0.5ex]
        \fill[lightgray!30] (0,0) rectangle (#1,0.4); 
        
        \fill[black] 
            (0,0) rectangle (#1*#2+0.2,0.4); 
        
        \draw[black!30] (0,0) rectangle (#1,0.4);
        
        \node at (-0.3,0.2) {0}; 
        \node at (#1+0.3,0.2) {1}; 
        \node[white] at (#1*#2/2,0.2) {\small #2};
    \end{tikzpicture}%
}

\setcopyright{acmlicensed}

\acmDOI{10.1145/3749475}

\acmISBN{2474-9567/2025/9-ART115}

\acmYear{2025}
\copyrightyear{2025}

\acmJournal{IMWUT}
\acmVolume{9}
\acmNumber{3}
\acmArticle{115}
\acmMonth{9}

\acmPrice{}

\def\sysname{\texttt{ASE}\xspace}
\begin{document}

    \title[\sysname: Practical Acoustic Speed Estimation Beyond Doppler \rev{via Sound Diffusion Field}]{\sysname: Practical Acoustic Speed Estimation Beyond Doppler \rev{via Sound Diffusion Field}}

\begin{anonsuppress}
	\author{Sheng Lyu}
        \affiliation{%
		\institution{The University of Hong Kong}
		\country{Hong Kong SAR, China}
	}
	\email{shenglyu@connect.hku.hk}
	
        \orcid{0000-0002-9559-1987}

	\author{Chenshu Wu}
        \affiliation{%
		\institution{The University of Hong Kong}
		\country{Hong Kong SAR, China}
	}
	\email{chenshu@cs.hku.hk}
        \orcid{0000-0002-9700-4627}
\end{anonsuppress}

\renewcommand{\shortauthors}{Sheng and Chenshu}

\begin{abstract}

\rev{
Passive human speed estimation plays a critical role in acoustic sensing. 
Despite extensive study, existing systems, however, suffer from various limitations:  
First, the channel measurement rate proves inadequate to estimate high moving speeds. 
Second, previous acoustic speed estimation exploits Doppler Frequency Shifts (DFS) created by moving targets and relies on microphone arrays, making them only capable of sensing the radial speed within a constrained distance. 
To overcome these issues, we present \sysname, an accurate and robust Acoustic Speed Estimation system on a single commodity microphone. We propose a novel Orthogonal Time-Delayed Multiplexing (OTDM) scheme for acoustic channel estimation at a high rate that was previously infeasible, making it possible to estimate high speeds. We then model the sound propagation from a unique perspective of the acoustic diffusion field, and infer the speed from the acoustic spatial distribution, a completely different way of thinking about speed estimation beyond prior DFS-based approaches. We further develop novel techniques for motion detection and signal enhancement to deliver a robust and practical system. We implement and evaluate \sysname through extensive real-world experiments. 
Our results show that \sysname  
reliably tracks walking speed, independently of target location and direction, with a mean error of 0.13 m/s, a reduction of 2.5x from DFS, and a detection rate of 97.4\% for large coverage, \eg, free walking in a 4m $\times$ 4m room. 
We believe \sysname pushes acoustic speed estimation beyond the conventional DFS-based paradigm and inspires exciting research in acoustic sensing. Code is available at \url{https://github.com/aiot-lab/ASE}.

}

\end{abstract}


\maketitle

\begin{figure}[t]
    \centering
    \includegraphics[width=0.8\linewidth]{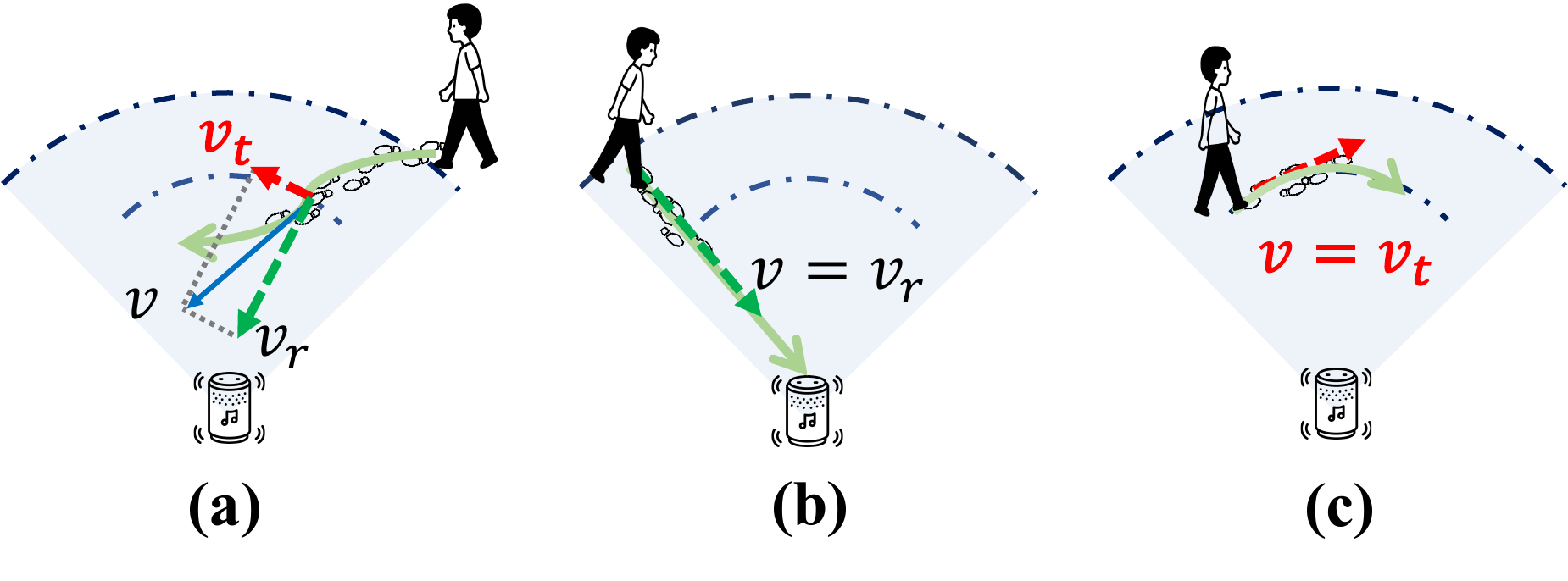}
    \caption{\sysname vs. DFS. {\rm Walking speed $\vec{v}$ can be decomposed to radial speed $v_r$ and tangent speed $v_t$. DFS can only capture radial speed $v_r$, but fails to capture $v_t$. Conversely, \sysname can capture both $v_r$ and $v_t$, a complete estimation of $v$.
    }}
    \label{fig: dfs_acf}
\end{figure}
\section{Introduction}
\label{sec:intro}

Capturing the portraits of indoor human activities is an enduring task in the wide sensing community \cite{liLASensePushingLimits2022, xu2023practically, wu2022WifiCanDoMore, zhang2023addressing, qian2018widar2}. 
Frequently, human subjects are in motion, rendering \textbf{passive speed estimation} one of the most fundamental components in human sensing. 
At the heart of understanding the physical state of moving subjects, speed provides valuable insights into human behaviors and health. 
With speed profiles, a broad range of applications can be accommodated, such as gait recognition \cite{xu2019acousticid,chiang20203d}, fall detection \cite{lianFallDetectionInaudible2021, hu2021defall}, human activity recognition \cite{zheng2019zero,ghosh2019ultrasense, ouyang2022cosmo}, tracking \cite{qian2018widar2, mao2017indoor,chenEchoTrackAcousticDevicefree2017} and
fitness tracking \cite{xuAcousticIDGaitbasedHuman2019,wuGaitWayMonitoringRecognizing2021a,umairbinaltafAcousticGaitsGait2015}, \etc. 

Particularly, walking speed estimation plays an important role in well-being monitoring. 
It is increasingly perceived as the sixth vital sign \cite{middleton2015walking, fritz2009white} 
which is closely associated with and predictive of one's health conditions \cite{wuGaitWayMonitoringRecognizing2021a}. 
Slowing walking speed suggests increased frailty, leading to potential physical and cognitive decline \cite{rosso2017slowing,rasmussen2019association}. 
Moreover, walking speed acts as a biomarker for gait recognition \cite{wu2020gaitway} and an effective indicator of risky falls \cite{hu2021defall}. 
Enabling these applications calls for accurate and robust speed estimation, preferably in a contactless and passive manner.

Passive speed estimation, however, is a long-standing open challenge. 
Various approaches have been proposed for indoor speed estimation using vision systems \cite{ViconAwardWinning, MotionCaptureSystems}, wireless signals \cite{zhangWiSpeedStatisticalElectromagnetic2018,wangWiFallDeviceFreeFall2017,wuGaitWayMonitoringRecognizing2021a}, and acoustic signals \cite{yunTurningMobileDevice2015,chenEchoTrackAcousticDevicefree2017,maoCATHighprecisionAcoustic2016}. 
Camera-based approaches, such as the VICON motion capture system \cite{ViconAwardWinning}, usually provide the most accurate speed but require complex and expensive hardware setups and are limited within a functional area. 
Wireless sensing has recently been extensively studied, yet mostly relies on specialized radars \cite{xie2023boosting, IWR1843DataSheet2023, X4M03LaonuriCom2021, wang2020uwhear, kuang2024brush, kuang2025air},  or certain WiFi chipsets.

Acoustic devices, especially smart speakers and IoT audio, are now widely and interactively available in our everyday lives, often as plug-and-play devices with co-located microphones and speakers, making acoustic sensing an increasingly hot topic in recent years \cite{zhang2023addressing, fan2023towards, ferlini2021eargate,shenVoiceLocalizationUsing2020, wangContactlessInfantMonitoring2019a, wangMilliSonicPushingLimits2019}. 
The widespread usage
of such audio devices hold substantial potential for enabling significant applications, provided that accurate and robust speed estimation can be achieved through acoustic sensing.

Existing acoustic speed estimation methods mostly rely on the Doppler Frequency Shifts (DFS) caused by target movements to derive the speed \cite{yunTurningMobileDevice2015,lianFallDetectionInaudible2021}. 
These approaches, however, suffer from three fundamental limitations in acoustic speed estimation: 
\begin{itemize}
    \item The low sound speed imposes an innate limit on the maximum Channel State Information (CSI) rate achievable on an acoustic channel \cite{zhang2023addressing}, which unfortunately, is insufficient for estimating high speeds (\eg, typical walking speed of around 1 m/s).

    \item Depending on the specific moving direction and location, Doppler-based approaches can only capture the partial speed projected in a radial direction that creates reflection path length changes and thus frequency shifts \cite{qian2018widar2,wang2015understanding}, as shown in \fig\ref{fig: dfs_acf}.

    \item DFS often only utilizes one or a few reflection paths off the target for sensing, leading to performance degradation with the increase of distance, 
\eg, over 1 m \rev{\cite{li2022experience, zhang2023addressing,liLASensePushingLimits2022, wangDevicefreeGestureTracking2016, maoCATHighprecisionAcoustic2016}}, and thereby confining them for short-range sensing. 
\end{itemize}

To overcome these limitations, we pose a crucial research question: \textit{Can we achieve \rev{robust and practical} acoustic speed estimation beyond DFS-based approaches?} We present \sysname, a completely novel pipeline that achieves practical acoustic speed estimation using only a \textit{single} microphone channel. At a high level, it features three distinct components: 1) an innovative modulation scheme that breaks the upper limit of acoustic CSI, 2) a novel theoretical model that can derive the whole speed, rather than solely radial speed, while being independent of target/device locations and moving orientations, and 3) a set of innovative techniques that make \sysname more robust in practice.

Regarding the first challenge, we propose a novel Orthogonal Time Delayed Multiplexing (OTDM) scheme. Based on in-depth analysis, we reveal that the CSI rate is inherently limited by the travel speed of sound and the propagation distance. This limitation restricts acoustic devices from estimating high speeds, presenting a fundamental challenge in acoustic speed estimation.  Inspired by OFDM, we develop a specialized modulation framework to mix the signals effectively without extra cost. We leverage the separation ability of orthogonal signals to send two signals concurrently, while delaying one sequence for half of the frame time. Through this approach, we can effectively atomize the original frame to its half and boost 2x of the CSI rate.

To surpass Doppler-based speed estimation, we build a comprehensive model that can capture the entire speed. Our model builds upon the acoustic diffusion model, as illustrated in \fig\ref{fig: sound_diffusion_model}. Specifically, we investigate the fundamental properties of sound diffusion indoors
and introduce a concise approach for speed estimation. 
We draw aspiration from previous physical studies of room acoustics \cite{kuttruffRoomAcoustics2000}, which model the diffusive sound propagation.
Conceptually, \sysname employs the distinct spatial distribution of the sound pressure field. The analysis of spatial-temporal properties of the sound field shows that the correlation of the sound energy embodies the speed of a moving target. Although the Autocorrelation Function (ACF) has been employed for extracting periodic vital signs \cite{gongBreathMentorAcousticbasedDiaphragmatic2022,wang2022loear}, building the theoretical and practical bridge between the ACF of sound and speed is new and non-trivial.
In \sysname, we establish the theoretical relationship in the context of the acoustic diffusion field and model the ACF of acoustic CSI as a function of the moving speed. 
Different from DFS, the proposed model statistically leverages all the reflection signals, and integrates over all possible directions, thus making speed estimation less dependent on the moving location and direction and 
building a novel foundation for speed estimation with commodity acoustic devices.

We also incorporate several effective designs to transform \sysname into a practical system. We employ pseudo-noise code in \sysname with nice orthogonality and tolerance to interference for CSI estimation from inaudible sound signals. By careful modulation and filtering, the sensing signals are made hardly audible compared to previous designs like the widely used chirps \cite{wangContactlessInfantMonitoring2019a}, which are considerably intrusive to human ears in practice. Moreover, we identify robust motion indicators and devise an effective approach that embraces frequency diversity to significantly boost the weak signals for speed estimation, which largely extends the sensing coverage and improves the speed estimation accuracy.

We prototype \sysname on commodity audio devices. 
We conduct comprehensive experiments to evaluate \sysname for diverse walking behaviors in real-world indoor scenarios. 
The results show that \sysname achieves a mean speed accuracy of 0.13 m/s with a 90\%-tile error of $<0.2$ m/s and an overall detection rate of 97.4\% in a 4m $\times$ 4m room, significantly outperforming prior DFS-based approach, which yields a 2.5$\times$ higher mean error 
under the same conditions.
We further conduct case studies on human activity recognition, fall detection, and gait analysis
as potential applications by profiling the speeds of different human activities. 
The superior performance validates \sysname and its underlying model as a new paradigm of acoustic speed estimation for many applications. 

\head{Contributions} We believe \sysname lays a completely new foundation for acoustic speed estimation and offers new insights into the field by making the following core contributions: 
\begin{itemize}
    \item We develop OTDM, the first-of-its-kind modulation scheme that allows acoustic CSI estimation at a high rate exceeding the previous maximum possible rate.

    \item To the best of our knowledge, we are the first to employ the sound diffusion model for speed estimation and integrate it with the acoustic channel, which fundamentally differs from DFS-based approaches.  

    \item We design and implement \sysname system using a single commodity microphone. We incorporate a pipeline of distinct techniques and conduct experiments to validate its superior performance over prior approaches. 
\end{itemize}

In the rest of the paper, we first introduce a universal dilemma in acoustic speed estimation and our OTDM design in \S\ref{sec:otdm}. Then we present the theoretical model in \S\ref{sec:model} and the design of \sysname in \S\ref{sec:design}. Implementation details and evaluations are presented in\S\ref{sec:exp}, respectively, followed by discussions in \S\ref{sec:limitation} and related works in \S\ref{sec:related_works}. We conclude in \S\ref{sec:conclusion}.

\section{OTDM Design}
\label{sec:otdm}
We first discuss the contradiction between high speed estimation and the acoustic CSI rate. After that, we will present our novel solution of Orthogonal Time Delayed Multiplexing (OTDM) scheme to boost the CSI rate.

\subsection{High Speed \& Low CSI Rate Dilemma}
\label{subsec: max_speed}
Initially, we will discuss a universal problem of acoustic speed estimation. To profile speed information, a common practice is to estimate Channel State Information (CSI), which characterizes how sound propagates in a Tx/Rx system. The CSI rate $F_s$ refers to how many CSIs we can acquire per second in the system. However, the inadequacy of the acoustic CSI rate imposes a fundamental challenge for speed estimation. Conceptually, the lower CSI sampling rates we acquire, the smaller the maximum frequency shift can be observed. Given a CSI rate of $F_s$, the maximum frequency shift detectable on the Doppler spectrum is $F_s/2$. Assuming a carrier frequency of $f$, the maximum measurable speed is 
\begin{equation}
    v_{\max} = \frac{F_s}{2 \cdot f}\cdot c,
\end{equation}
which is only 0.4 m/s considering $f=20$ kHz and $F_s=$ 49 Hz. \imwut{Here $c$ denotes the speed of the sound.}When factoring in noise, this obviously falls short of measuring typical indoor walking speeds.

Then another question arises: can we increase the acoustic CSI rate? Unfortunately, slow sound speed imposes an inherent limitation on the maximum achievable CSI rate on an acoustic channel. Assuming the sampling rate of acoustic signal is $f_s$ (note that $f_s$ is very different from the CSI rate $F_s$), plus $x$-m propagation path, then the CSI rate is given by
\begin{equation}
    \label{eqn:csi_rate}
    F_s = \frac{f_s}{x / c \cdot f_s} = \frac{c}{x}.
\end{equation}
As we can see, if we want to increase the CSI rate, the only way is to limit the sensing distance. For example, considering the in-air sound speed of 343 m/s and the longest propagation path of 7 meters, the minimum channel measurement interval should be larger than 20 ms \imwut{, resulting in a CSI rate of 49 Hz. }
However, even if we narrow this range to 3.5 meters, the maximum possible CSI rate without causing signal mixture only increases to 100 Hz, still insufficient for estimating human walk speed around 1 m/s. 

This observation reminds us that the current CSI rate cannot support high speed estimation for daily activities (\eg, walking), and it is infeasible to increase the CSI rate given the sensing distance for the current channel design. To tackle this problem, we present a novel OTDM design in the next section.

\subsection{OTDM Transmission Scheme}

Based on the above discussion, we find a challenging contradiction for acoustic speed estimation: Given the current channel design, it is impractical to enhance the CSI rate, while not reducing the sensing range or causing signal mixture. The core of this dilemma is the failure to fully utilize the channel capacity.
\rev{This situation reminds us of concurrent transmission schemes in network data communication, which are achieved through multiplexing modulations to allow for greater resource utilization. These multiplexing schemes can be roughly divided into Time Division Multiplexing (TDM), Frequency Division Multiplexing (FDM), Code Division Multiplexing (CDM), \etc, depending on which dimension is being reused. Notably, Orthogonal Frequency Division Multiplexing (OFDM) is an extension of FDM that divides the frequency band into closely spaced, orthogonal subcarriers. Each subcarrier can be modulated independently, thereby providing a larger data rate within the same frequency bandwidth. 

\begin{figure*}[t]
    \centering
    \includegraphics[width=0.8\linewidth]{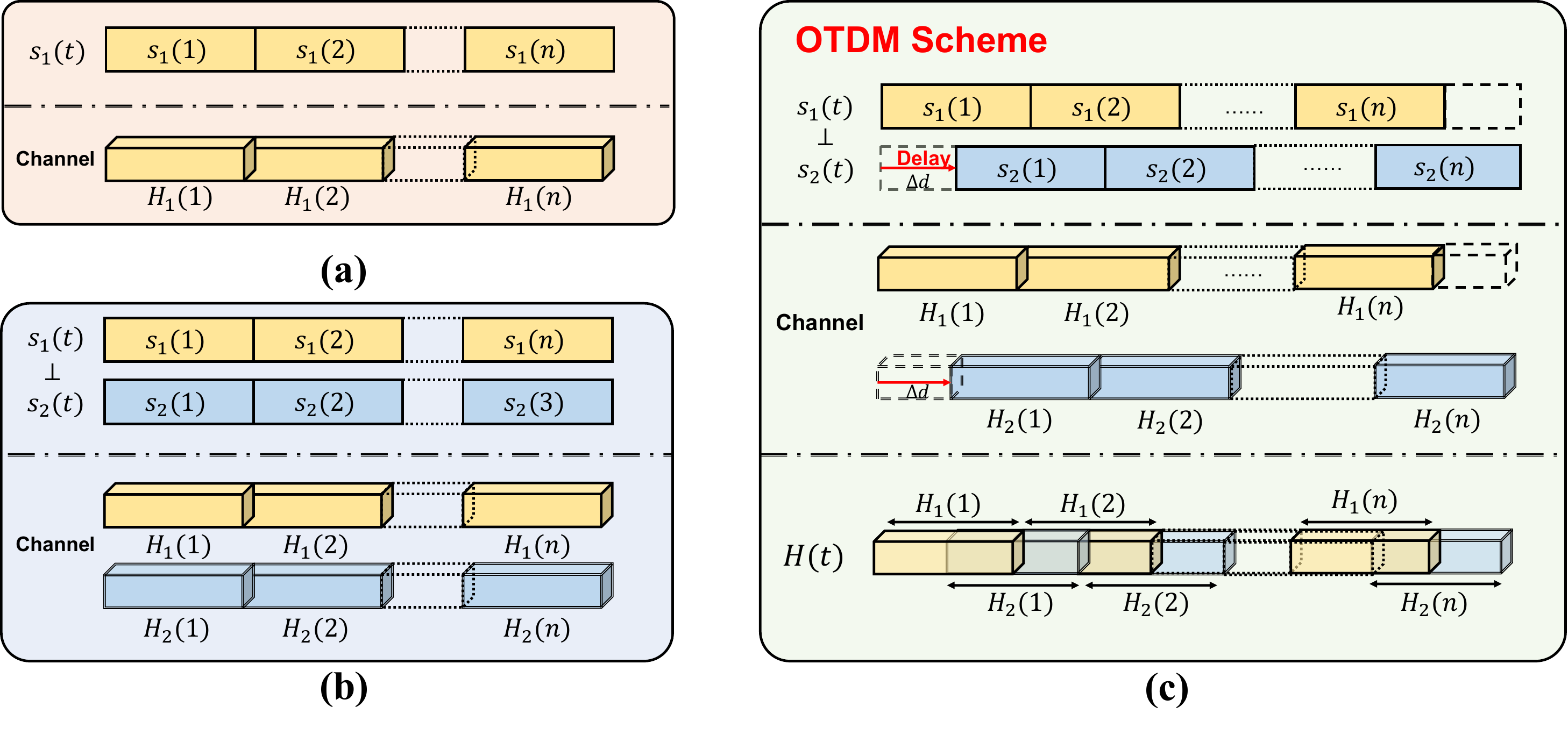}
    \caption{OTDM Scheme. \rm{(a) Given $s_1(t)$, the channel is estimated by frame. (b) Given orthogonal sequences $s_1(t)$ and $s_2(t)$, we can get two channel estimations at the same time. (c) Orthogonal sequences $s_1(t)$ and $s_2(t)$, with $s_2(t)$ delayed by $\Delta d$, are modulated into one sequence $s(t)$. $H_1(t)$ and $H_2(t)$ estimated from the two sequences are concatenated into $H(t)$ alternately.}}
    \label{fig:otdm_scheme}
\end{figure*}

In our context, we care about fully exploiting the \textit{temporal channel capacity} to achieve a higher CSI rate within the same time period. As shown in \fig\ref{fig:otdm_scheme}(a), transmitting a single probing signal yields one channel estimation $H_1$, which has proven insufficient. The question is, if we can estimate the channel simultaneously, can we get more channel estimations and increase the CSI rate? Thanks to the separability of orthogonal signals, with two orthogonal sequences  $s_1(t)$ and $s_2(t)$, we can indeed obtain two channel estimations simultaneously, as shown in \fig\ref{fig:otdm_scheme}(b).
However, this alone does not increase the channel rate because the measurements are aligned, resulting in redundant channel information. Inspired by OFDM, if we introduce spacing between the temporal estimations, we can observe the channel at different times and split the channel into more granular frame clips, potentially increasing the CSI rate. To achieve this, we are inspired to design a novel modulation scheme named Orthogonal Time-Delayed Multiplexing (OTDM). }

The key idea is that, by using orthogonal signals, we can transmit multiple sensing signals concurrently over the same physical channel (\ie, speaker-microphone pair) and perform channel estimation separately, which brings more samples of the channel measurements. 
If we further \textit{shift the multiple orthogonal signals by a certain amount of time}, we effectively obtain finer-grained
CSI measurements in the time domain, \ie, a higher CSI rate. 
By OTDM, the duration and interval of each separate sensing signal will not be shortened.

\begin{figure}[t]
    \centering
    \includegraphics[width=0.6\linewidth]{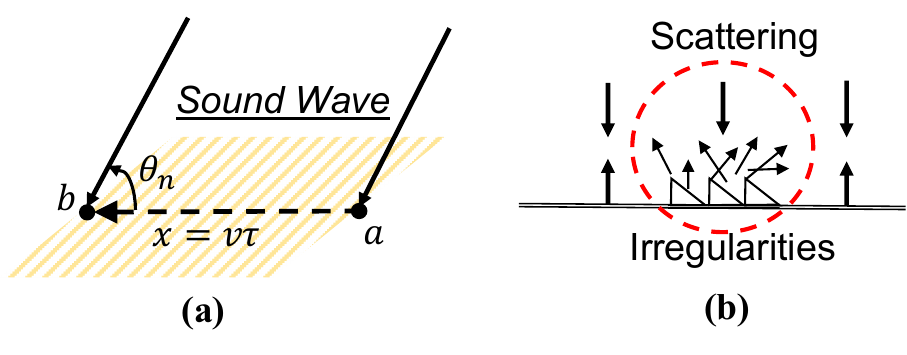}
    \caption{Sound Diffusion Model. \rm{(a) The sound wave traverses while an object is moving $a \rightarrow b$. (b) The sound wave is diffusive in all directions by irregular reflectors. }}
    \label{fig: sound_diffusion_model}
\end{figure}

Suppose we have two sequences $s_1(t)$ and $s_2(t)$ with decent orthogonality over time.
Then we can transmit them concurrently, while still being able to separate them on the receiver side by using correlation. 
Previous orthogonal transmission schemes typically transmit synchronized signals. 
Differently, as shown in \fig\ref{fig:otdm_scheme}, to enhance the CSI rate, we need to delay one sequence by $\Delta d = N_s /2$, where $N_s$ is the default sensing interval as described in \S\ref{sec:kasami}. Then we can acquire two sets of channel estimation: $H_1 \in \mathbb{R}^{n}$ and $H_2 \in \mathbb{R}^{n}$. 
The two CSI measurements, estimated from the two orthogonal sequences respectively, characterize the physical channel independently, yet at slightly shifted times. 
Therefore, by stacking the two series of CSI measurements, \ie, $[H_1(1),H_2(1),\cdots, H_1(n), H_2(n)]$, we can double the CSI rate from $1/N_s$ to $2/N_s$ without shortening the sequence length (\ie, signal duration). 
Our extensive experiment will show the effectiveness of OTDM design. 
We will detail the design for the most common case (\ie, one speaker) in \S\ref{subsec: tx_design}, while our scheme can generalize to more. 

\section{Acoustic Diffusion Speed Model}
\label{sec:model}
Conventional approaches for speed estimation relying on DFS can only measure the \imwut{radial} speed, making it a location- and direction-dependent solution, as shown in \fig\ref{fig: dfs_acf}. In this section, we will introduce a novel theoretical model for capturing the complete speed.
Our model is primarily inspired by 
room acoustics \cite{kuttruffRoomAcoustics2000}. 
Below, we first briefly review statistical room acoustics and then establish the model that works with channel measurements on commodity audio devices. 

\subsection{Limitations of DFS}
\begin{figure}[t]
    \centering
    \begin{subfigure}{.5\linewidth}
        \includegraphics[width=\linewidth]{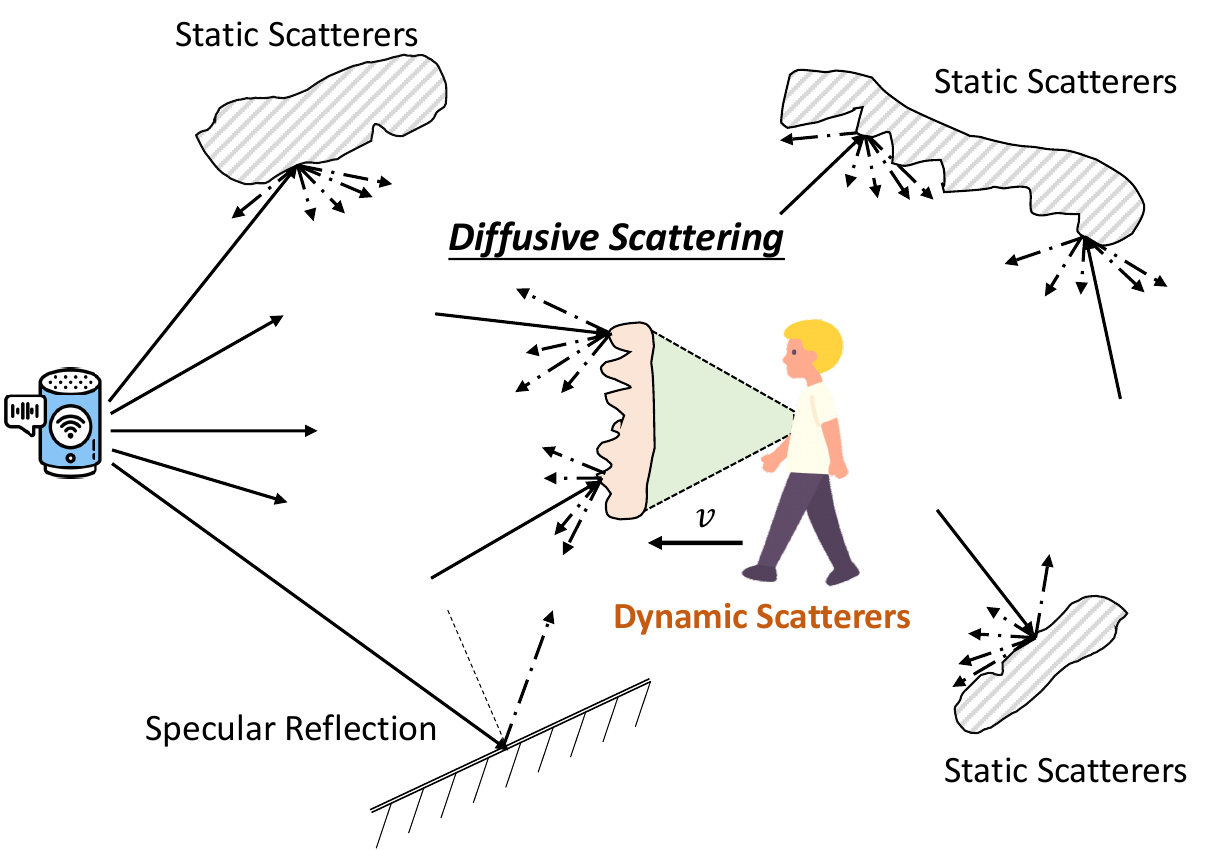}
        \caption{}
        \label{subfig:4a}
    \end{subfigure}\hfill
    \begin{subfigure}{.35\linewidth}
        \includegraphics[width=\linewidth]{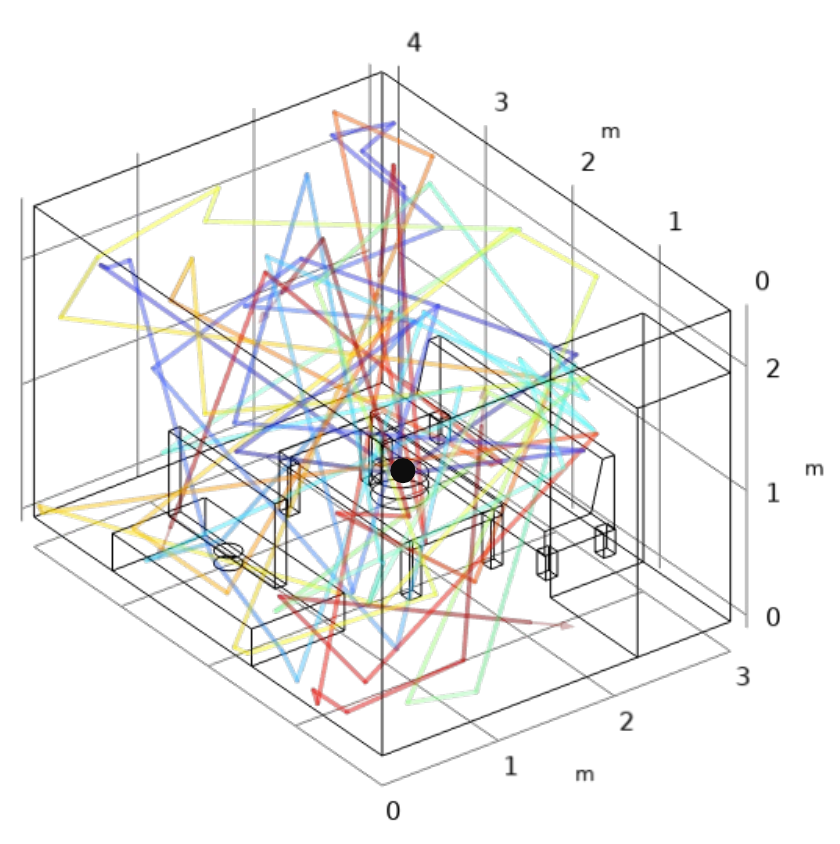}
        \caption{}
        \label{subfig:4b}
    \end{subfigure}\hfill

    \caption{Diffusive Scattering in the room. \rm{\textbf{(a)} Illustration of diffusion scattering and specular reflections. The acoustic waves not only experience specular reflections, but also diffusive scatterings on any non-flat planes. \textbf{(b)} A simulation result of how an acoustic wave is reflected and scattered in a common room. The black point emits the acoustic rays with a frequency range of 17 kHz- 20 kHz. By employing the rich multi-path semantics, these scatterers are creating "virtual speakers", thus creating different views of speed estimation.}}
    \label{fig: diffusion_scatter}
\end{figure}

Existing DFS-based acoustic speed estimation implicitly or explicitly relies on a simplistic propagation model. These methods typically involve identifying the primary reflection path between the human body and audio devices and deducing the DFS from the corresponding bins. Consequently, these techniques yield only partial information regarding speed estimation \cite{zhangWiSpeedStatisticalElectromagnetic2018, niu2022rethinking}, specifically the radial speed $v_r$ between the human and acoustic device pair, neglecting the tangent component $v_t$, as illustrated in \fig\ref{fig: dfs_acf}. 
Essentially, if we only consider one or a few reflection paths, the tangent component will inevitably be lost. Fortunately, we notice that the acoustic signals will be scattered by the numerous reflectors to different directions in the environment, as can be seen from \fig\ref{fig: sound_diffusion_model}(b) and \fig\ref{fig: diffusion_scatter}. If we can efficiently leverage the multi-path reflections, it equivalently creates various estimations of speed along different directions. Conceptually, the reflectors can be regarded as many "virtual speakers", offering numerous speed observations of different tangent components. From a statistical perspective, we can imagine these speed components are synthesized into a complete estimation of speed $v$, which contains both racial speed $v_r$ and tangent speed $v_t$. To this end, we explore the propagation properties of sound and derive the speed information from a statistical approach. We will first introduce the acoustic diffusion model in \S\ref{subsec: SAD} and generalize it to CSI in \S\ref{subsec: SAS}. 

\subsection{Sound Diffusion Field}
\label{subsec: SAD}

Given a bounded sound propagation scenario (\eg, indoor space), the sound energy is partly reflected/scattered by the obstacles, as shown in \fig\ref{fig: sound_diffusion_model}. 
Considering a moving target in space that continuously distorts the sound propagation, a diffuse sound field will be created due to the continuous redistribution of sound energy, especially in an environment with rich diffuseness such as a room. 
Thus, according to the acoustic wave equation  \cite{pierceAcousticsIntroductionIts2019}, we can express the sound pressure as 
$\mathbf{p}=p(x, t)=A e^{j(\omega t-\vec{k} \cdot \vec{r})}$
, where $A$ denotes the amplitude, $\omega$ is the frequency of the wave, $\vec{k}$ indicates the propagation vector, and $c$ is the propagation speed of sound. $r$ is the position vector, \ie, $\vec{r} = x\vec{i} + y\vec{j} + z\vec{k}$.
Assume the space is excited by a limited-band signal, and consider two adjacent observation points $a$ and $b$ separated by a distance of $x$, as shown in \fig\ref{fig: sound_diffusion_model}. 
The sound pressure is expressed as $p_a(t) = A \cos (\omega t - \phi)$ and $p_b(t) = A \cos\left(\omega t - \phi - k x \cos(\theta)\right)$, respectively, 
where $\theta$ denotes the direction of the incident sound wave, $A$ and $\phi$ are random amplitudes and angles, and $k$ and $\omega$ represent the wave number and center frequency of the sound signals. 
The correlation coefficient of the sound energy over space provides insight into the degree of diffuseness. 
Specifically, the spatial correlation over $p_a(t)$ and $p_b(t)$ is expressed as
\begin{equation}
    \label{eq:acf_pressure}
        \psi_p =\dfrac{\overline{p_a(t) p_b(t)}}{\sqrt{\overline{p_a(t)^2} \cdot \overline{p_b(t)^2}}},
\end{equation}
where $\psi_p$ represents the correlation coefficient, and $\overline{\left.\cdot\right.}$ denotes operation of time average. 
Since sound is a plane wave, we can derive the time average of $p_a(t)^2$ and $p_b(t)^2$ both as $\frac{A^2}{2}$, and meanwhile, the average of the product of sound pressure is computed as $\overline{p_a(t) p_b(t)} = \frac{1}{\Delta t}\int_{t}^{t+\Delta t}p_a(t) p_b(t) dt =  \frac{A^2}{2}\cos(kx\cos\theta)$. Therefore, by substituting them into \eqn\eqref{eq:acf_pressure}, we derive the correlation for the direction of incidence $\theta$ as
\begin{equation}
    \label{eq:acf_sound_theta}
    \psi_p\left(x, \theta\right)=\cos \left(k x \cos \theta\right).
\end{equation}
Then by integrating \eqn\eqref{eq:acf_sound_theta} over all directions, we can obtain the spatial \rev{correlation} of the total sound pressure. If we consider that the incident sound waves are distributed in a plane, namely 2D model, we can get
\begin{equation}
    \label{eq:acf_sound_2d}
    \psi_p\left(x\right)= \frac{1}{2\pi}\int_{0}^{2\pi} \psi\left(x, \theta\right) d\theta = J_0(kx),
\end{equation}
where $J_0(x) = \frac{1}{2\pi}\int_{0}^{2\pi} \cos(x\cos\theta)d\theta$ is the $0^\mathrm{th}$-order Bessel function of its first kind. 
If we consider 3D scattering model, \ie, the sound waves are distributed in a sphere, we can acquire the correlation coefficient as,
\begin{equation}
    \label{eq:acf_sound_3d}
    \psi_p\left(x\right)= \frac{1}{4\pi}\int_{0}^{\pi}\int_{0}^{2\pi} \psi\left(x, \theta\right) d \phi d\theta = \frac{\sin(kx)}{kx},
\end{equation}
where $\phi$ is the azimuth angle, varying from 0 to 2$\pi$. 
Further considering a target moving from point $a$ to $b$ at a speed of $v$, we have $x=v\tau$, where $\tau$ is the traveling time. 
Thus, the above equation can be written as
\begin{equation}
    \label{eq:acf_sound}
    \psi_p(\rev{v; }\tau) = \frac{\sin(kv\tau)}{kv\tau}.
\end{equation}
It bridges the \rev{spatial correlation} of the sound pressure $\psi_p$ with a target's moving speed $v$, independent of the moving direction and location. Note that in both scenarios, we average diffuseness across \textit{all directions}, in contrast to DFS, that only takes the radial speed into consideration, promising a potential approach for speed estimation beyond the Doppler-based approach. 

\begin{table}[t]
\caption{\imwut{Examples of different scattering coefficients indoor \cite{faiz2012measurement, ProSoundTraining2015}}}
\label{tab:indoor_scatter_coeff}
\resizebox{0.5\textwidth}{!}{%
\begin{tabular}{c|B{6cm}}
\hline
\textbf{Room Components} & \textbf{Scattering Coefficient} \\ \hline
Cabinet                  & \drawfilledbar{0.2} \\ \hline
4 Chairs                 & \drawfilledbar{0.45} \\ \hline
1 table + 3 chairs + PC  & \drawfilledbar{0.45} \\ \hline
Audience                 & \drawfilledbar{0.7} \\ \hline
Carpets                  & \drawfilledbar{0.3} \\ \hline
Irregular Books          & \drawfilledbar{0.5} \\ \hline
\end{tabular}%
}
\end{table}

\imwut{
\head{Understanding Sound Diffuseness in Room Environment} Indoor environments inherently exhibit scattering effects. As established by modern room acoustic studies \cite{faiz2012measurement, ProSoundTraining2015, kuttruffRoomAcoustics2000}, the level of scattering increases with increasing sound frequency and the roughness of surfaces. 
In common room settings, these scattering effects are ubiquitous and pervasive. Curved walls, textured surfaces, or objects with uneven shapes do not reflect sound in a single, coherent direction. Instead, they disperse sound energy across a wide range of angles, much like how a crumpled piece of paper scatters light compared to a smooth mirror. This significantly increases the scattering coefficient, meaning more sound energy is diffused rather than specularly reflected. As summarized in \tab\ref{tab:indoor_scatter_coeff}, typical scattering coefficients for indoor materials range from 0 (no scattering) to 1 (full scattering). Ordinary objects such as chairs and tables act as scatterers, diffusing incident sound energy throughout the space (see \fig\ref{fig: sound_diffusion_model}(b)). The scattering becomes more pervasive in complex room settings and with irregular reflectors. This implies that modeling sound diffusion in a room offers a new perspective on its acoustic behavior. By adopting a statistical perspective on the sound field, we can analyze the acoustic environment without considering the specific material types of objects, whether they are on Line of Sight (LoS) or Non-Line of Sight (NLoS) paths. In other words, this statistical approach enables us to estimate overall acoustic properties in a more comprehensive way than the DFS method and simplifies the processing by eliminating the need to separate specific paths from the received signal.
} 

To achieve \sysname, however, we need to investigate how to obtain $\psi_p$ on commodity audio devices.

\subsection{Acoustic Speed Estimation}
\label{subsec: SAS}
In this section, we investigate CSI measured on commodity audio devices and extend \eqn\eqref{eq:acf_sound} to the ACF of acoustic CSI. 

\begin{figure}[t]
    \centering
    \includegraphics[width=0.6\linewidth]{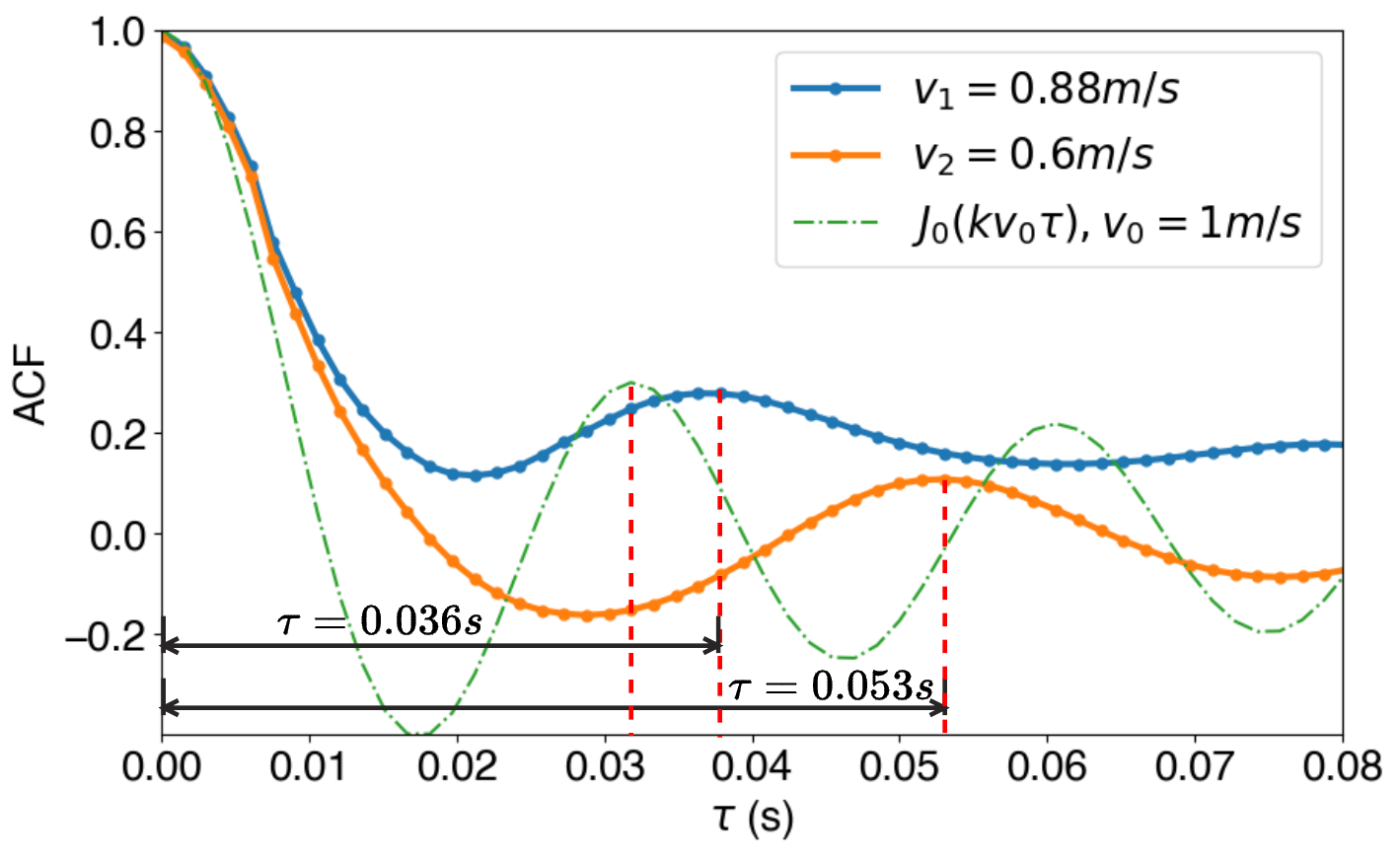}
    \caption{ACF curves for two different speeds and the theoretical function for $v_0 = 1m/s$.}
    \label{fig:acf_curve_speed}
\end{figure}

\head{Channel \rev{Power}}
Indoor space proves to be an environment with rich diffuseness \cite{kuttruffRoomAcoustics2000}, where the above acoustic diffusion model applies. 
However, it is infeasible to directly measure the sound pressure without professional equipments. It is crucial to efficiently estimate sound pressure on common devices.
In \sysname, we approximate the sound pressure as the power of the sound diffusion field. 
Sound pressure can be considered as the equivalent of the power of CSI \cite{leightonFeynmanLecturesPhysics1965}. 
CSI is the linear aggregation of multi-path components, which can then be decomposed into two parts, \ie, the static part $\Lambda_S$ attributed to static scatterers and the dynamic part $\Lambda_D$ contributed by dynamic scatterers, typically from human subjects, as shown in \fig\ref{fig: diffusion_scatter}. To sum up, we can model 
the relationship between channel power and sound pressure as 
\begin{equation}
    \label{eq:CFR and sound pressure}
    \begin{aligned}
        G(f,t) &= |H(f,t)|^2 + \epsilon(f,t) \approx |p(f,t)|^2 + n(f, t)\\
               &= \left|\sum_{i \in \Lambda_D} p_i(f,t) +  \sum_{i \in \Lambda_S} p_i(f,t)\right|^2 + n(f, t),
    \end{aligned}
\end{equation}
where $H(f,t)$ represents CSI at time $t$ and subcarrier $f$. Practically, the static components, including the direct path between the speaker and microphone, are cancelled by removing the time average from $H(f,t)$.  $\epsilon(t,f)$ and $n(t,f)$ are noise terms, with a variance of $\Sigma_N(f)$. $p(f,t)$ is the sound pressure, which is further decomposed as the dynamic and static components. $p_i(f,t)$ is the sound pressure contributed by the $i^\mathrm{th}$ scatterer, which can also be deemed as mutually independent of $p_j(f,t)$, $\forall i\neq j$, in sound diffusion field. 
Next, we model the spatial properties of $G(f,t)$ for speed estimation.

\head{Speed Estimation Model}
To learn the spatial distribution of channel power $G(f,t)$, we compute the ACF function, \ie,
\begin{equation}
    \label{eq:ACF_G}
    \psi_{G}(f,t,\tau) = \frac{\mathbb{E}\left[G(f,t) G^H(f,t+\tau) \right]}{\mathbb{E}\left[G(f,t) G^H(f,t) \right]} \triangleq \frac{\mathcal{R}_1}{\mathcal{R}_2}.
\end{equation}
 Given that $p_i(f,t)$ for any $ i \in \Lambda_D$ and $p_j(f,t)$ for any $ j \in \Lambda_S$ are mutually independent, and considering that sound pressures induced by static scatterers are statistically characterized by a zero mean, \ie, $\mathbb{E}[p_j(f,t)] = 0, \forall j \in \Lambda_S$, it follows that any term involving a static component from $\Lambda_S$ in a product will be canceled in expectation. Specifically, $\forall i \in \Lambda_D$ and $\forall j \in \Lambda_S$, we have $\mathbb{E}[p_i(f,t)p_j(f,t)] = 0.$
 Let $\bm{P_D}(t) = \sum_{i \in \Lambda_D} p_i(f,t)$, we have
\begin{equation}
\begin{aligned}
        \mathcal{R}_1 &\approx \mathbb{E}\left[\bm{P_D}(t)\bm{P_D}^H(t+\tau)\right] + \mathbb{E}\left[\bm{N}(t)\bm{N}^H(t+\tau)\right]\\
        &= \bm{\Theta_D^H}\cdot \bm{\Psi_{D}} + \Sigma_N(f)\cdot \mathbf{I_n},
\end{aligned}
\end{equation}
where $\bm{N}(t) = \left[n(f_1,t), n(f_2,t), \cdots, n(f_{N_f},t)\right]$, $N_f$ is the number of subcarriers. $\mathbf{I_n} \in \mathbf{R}^{N_{\tau}}$ is identity matrix, $N_\tau$ is the number of time lag for ACF. $\bm{\Psi_D} \in \mathbf{R}^{N_{\tau}\times N_D}$ is the ACF matrix of sound pressure incurred by dynamic scatterers, where $N_D$ is the number of dynamic scatterers, \ie, $\bm{\Psi_D} = \bm{\Psi_D}(\bm{\tau}; \bm{v}), \bm{\tau} \in \mathbf{R}^{N_\tau}, \bm{v} \in \mathbf{R}^{N_D}$. $\bm{\Theta_D^H} \in \mathbf{R}^{N_D \times N_f}$ is the frequency gain to normalize ACF matrix $\bm{\Psi_D}$. Similarly, we can compute $\mathcal{R}_2$ as 
\begin{equation}
    \mathcal{R}_2 = \bm{\Theta_D^H}\cdot \mathbf{I_{N_D}} + \Sigma_N(f).
\end{equation}
$\psi_{G}(f,t,\tau)$ is irrelevant with $t$ and can be seen as linear combination of $\bm{\Psi_D}$. 
At $\tau_j$, $\bm{\Psi_D}$ is composed of $\psi_p \in \mathbf{R}^{N_D}$, \ie,
\begin{equation}
\label{eqn: multi_speed}
\begin{aligned}
    \bm{\Psi_D}|_{\tau=\tau_j} &= \bm{\Psi_D}(\tau=\tau_j;\bm{v}) \\
                            &= \psi_p(v_i; \tau_j), \: i = 1,\cdots, N_D.
\end{aligned}
\end{equation}
Considering a single moving target, we can assume that all the dynamic scatterers contributed by the target share approximately the same speed $v_i$ as the walking speed $v$. The assumption holds in practice as for human targets, the major reflection energy from the torso dominates that from limbs. Notably, prior DFS-based work implicitly adopted similar assumptions, \eg, hand gesture works usually neglect body motions \cite{zhang2023addressing, yunStrataFineGrainedAcousticbased2017, li2022room}.  To this end, we get $\bm{\Psi_D}|_{\tau=\tau_j} = \psi_p(v; \tau_j)$. With this, we can compute \eqn\eqref{eq:ACF_G} as 

\begin{equation}
    \label{eq:acf_csi_velocity}
    \psi_{G}(f,\tau) = \bm{\widehat{\Theta_D^H}} \cdot \left.\bm{\Psi_{D}}\right|_{\tau} = \bm{\widehat{\Theta_D^H}} \cdot \psi_p(\rev{v; }\tau),
\end{equation}
where $\bm{\widehat{\Theta_D^H}}$ is the broadcast aggregation of $\bm{\Theta_D^H}$ and $\Sigma_N(f)$. 

With the above \eqn\eqref{eq:acf_csi_velocity}, we bridge the ACF of the CSI and that of sound pressure as a function of speed, for the first time, offering a completely different approach to acoustic speed estimation by calculating the ACF of the CSI measured on commodity audio devices. 
Practically, with the CSI time series as input, we can use the sample ACF $\tilde{{\psi}_{G}}(f,\tau)$, where a noise term $\mu(f,\tau)$ will be added to ${\psi}_{G}(f,\tau)$.

\fig\ref{fig:acf_curve_speed} shows the ACF of CSI under different speeds. 
As can be seen, the shape of $\psi_G(f,\tau)$ well resembles \rev{$\psi_p(\rev{v; }\tau) $} as expected. 
Thus, to derive the speed $v$, we can find a reference point to align the calculated $\tilde{{\psi}_{G}}(f,\tau)$ to the theoretical ${\psi}_{p}(v,\tau)$. 
In our work, we use the first peak of ${\psi}_{G}(f,\tau)$, yet the first valley or their combination will also work. Let $x_0$ denote the reference point on ${\psi}_{p}(v,\tau)$, then the speed can be acquired to solve the optimization problem:
\begin{equation}
    \label{eq:speed_tau}
    \begin{gathered}
    \min_{v} \left| x_0 - \frac{\lambda(f)}{2 \pi} v \tau_s \right|, \\ 
        \begin{aligned}
            &\text{s.t.} \quad x_0  = \min_x \left\{x: \frac{d\psi_p(x)}{dx} = 0 \land \frac{d^2\psi_p(x)}{dx^2} < 0 \right\}, \\
            &\phantom{\text{s.t.} \quad} \tau_s  = \min_\tau \left\{\tau : \frac{\partial\psi_G(f, \tau)}{\partial\tau} = 0 \land \frac{\partial^2\psi_G(f, \tau)}{\partial\tau^2} < 0 \right\},
        \end{aligned}
    \end{gathered}
\end{equation}
where $\lambda(f)$ represents the wavelength of sound related to its subcarrier frequency $f$ and $\tau_s$ is the counterpart of $x_0$ on $\psi_G(f,\tau)$, \ie, the delay corresponding to the first peak. 
This is \rev{independent from either 2D or 3D model we choose} and allows an efficient approach for speed estimation, as we only need to calculate the ACF of the CSI and localize $\tau_s$. 
\begin{figure}[t]
    \centering
    \includegraphics[width=1\linewidth]{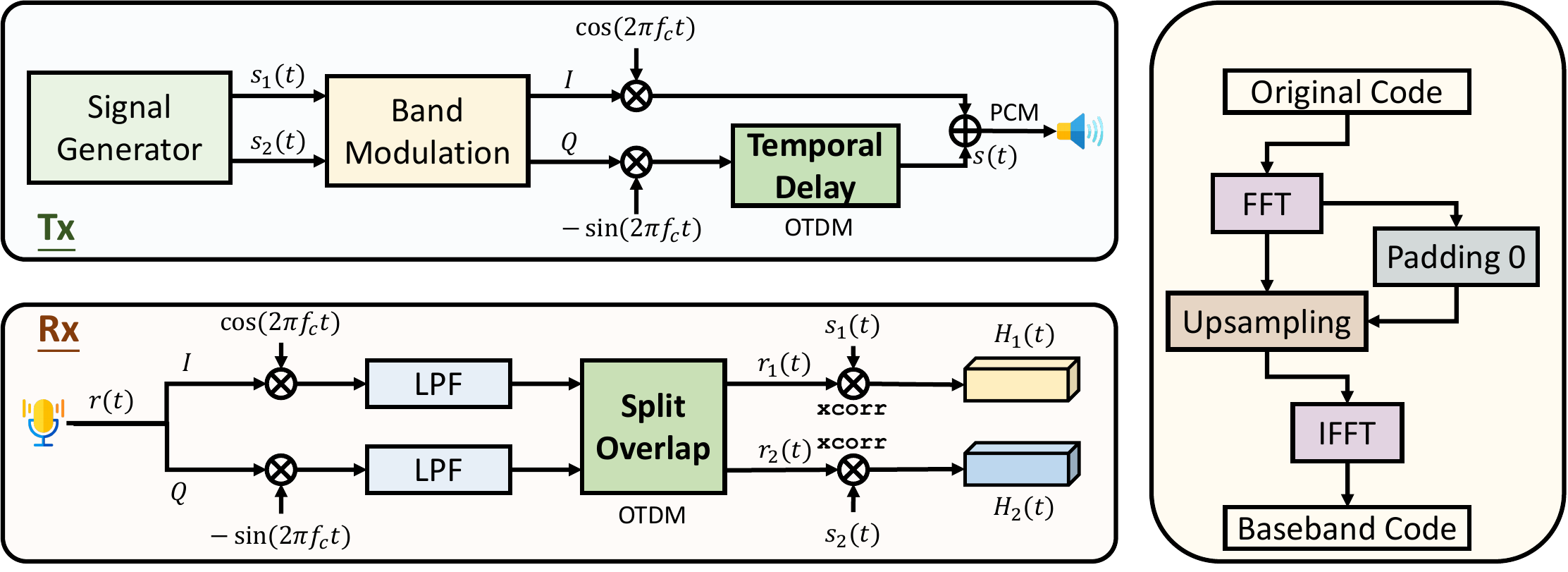}
    \caption{\imwut{Transmitter and Receiver Design of \sysname. \rm{ Tx: Band modulation (\S\ref{sec:kasami}) and I/Q modulation (\S\ref{subsec: tx_design}); Rx: Receiver Demodulation (\S\ref{subsec: tx_design}) and Channel Estimation (\S\ref{sec:kasami}).}} }
    \label{fig:modulation_scheme}
\end{figure}

\imwut{
\head{Multipath Effect} The previous practice separates different paths from the CSI for modeling the multipath propagation. The multipath effect can be harmful for those approaches, as the multipath components will be hard to extract in complicated scenarios. However, our sound diffusion model does not involve decomposing each multipath, instead aggregates the speed information from all the multipaths. Intuitively, we analyze the acoustic channel from a \textit{statistical} perspective, where the multipath information is aggregated together for speed estimation. In other words, we leverage the diffusiveness of the room to infer the speed. To this end, it is practical for realistic room environments and robust to multipath effects. At the same time, we do not need to focus on the specific material properties within the room, as our method integrates information holistically, ensuring that no single material acts as a bottleneck.
}

\head{Remarks} 
\sysname is the first to employ the \textit{sound diffusion model} for speed estimation and \textit{comprehensively} integrate it with the acoustic channel.
The proposed model fundamentally advances acoustic speed estimation by exploring sound pressure theory and leveraging all multipath reflections.
As shown in \fig\ref{fig: dfs_acf}, \eqn\eqref{eq:acf_sound_2d} and \eqref{eq:acf_sound_3d}, \rev{we leverage all the multipath components and integrate over all directions.} 
This approach diverges completely from traditional DFS methods, marking a new foundation for acoustic speed estimation.

\section{\sysname Design}
\label{sec:design}

This section presents a pipeline of novel techniques that translate our theoretical model into a practical system for robust and accurate speed estimation. 

\subsection{Baseband Sequence Selection}
\label{sec:kasami}
\head{Transmitted Sequence}
CSI is by default not available in acoustic sensing, and we send certain signals to probe the channel. 
Normally, there are three main types of waveform in acoustic sensing \cite{caiUbiquitousAcousticSensing2019}: pure-tone, Pseudo-Noise (PN) code, and FMCW.  
While FMCW signals are widely used in acoustic sensing \cite{chengPushLimitDeviceFree2021, liLASensePushingLimits2022}, they are not standard impulse signals and thus introduce CIR estimation errors \imwut{\cite{maoDeepRangeAcousticRanging2020, 4686885}}. 
As for the sole impulse signal, the energy of the short-time signal would fade out quickly.
Therefore, we seek the PN sequence for CIR estimation in \sysname. 
PN sequence consists of equally spaced Dirac impulses, the signs of which alternate with specific rules. 
It is a noise-like signal with statistical randomness. 
We choose the Kasami sequence \cite{kasamiWEIGHTDISTRIBUTIONFORMULA1966} as our sensing waveform, mainly due to its superior orthogonality and tolerance to interference. 
Particularly, the mutual orthogonality of Kasami sequences is critical to our OTDM design, as detailed in \S\ref{sec:otdm}.

\head{Modulation to Inaudible Band}
\sysname uses only the inaudible sounds for sensing and modulates the sensing signals on the acoustic band of 17 kHz to 24 kHz, the pseudo-ultrasonic band supported by most commodity devices today. 
Since the PN code, including Kasami, has a spreading spectrum over the whole band, we should modulate the Kasami sequence to our desired band. 
The modulation process is shown in Figure \ref{fig:modulation_scheme}. 
The conversion of the full-band signal into a band signal is achieved by either temporal \cite{yunStrataFineGrainedAcousticbased2017} or spectral \cite{sunVSkinSensingTouch2018} interpolation. 
In \sysname, we use the frequency domain interpolation. 
We perform $N$-point 
FFT to obtain the frequency domain information, where $N$ is the original length of the sequence. 
Zero Padding is performed between positive and negative frequencies, creating a new sequence of length $N_s$. 
Then we perform IFFT to obtain the temporal signal. 
The band of the interpolated signal is thus transformed to $\frac{N}{N_s}f_s$. For a single-sequence signal, we can multiply it by the carrier signal to move the frequency band towards the central frequency $f_c$. 
In \sysname, we choose $f_c = 20.25$kHz. We use 63-point Kasami sequences and modulate them to $N_s = 512$, achieving a bandwidth of 5.9kHz. The transmission signal is coded via PCM and decoded at the receiver's end.
\begin{figure}[t]
    \centering
    \includegraphics[width=0.8\linewidth]{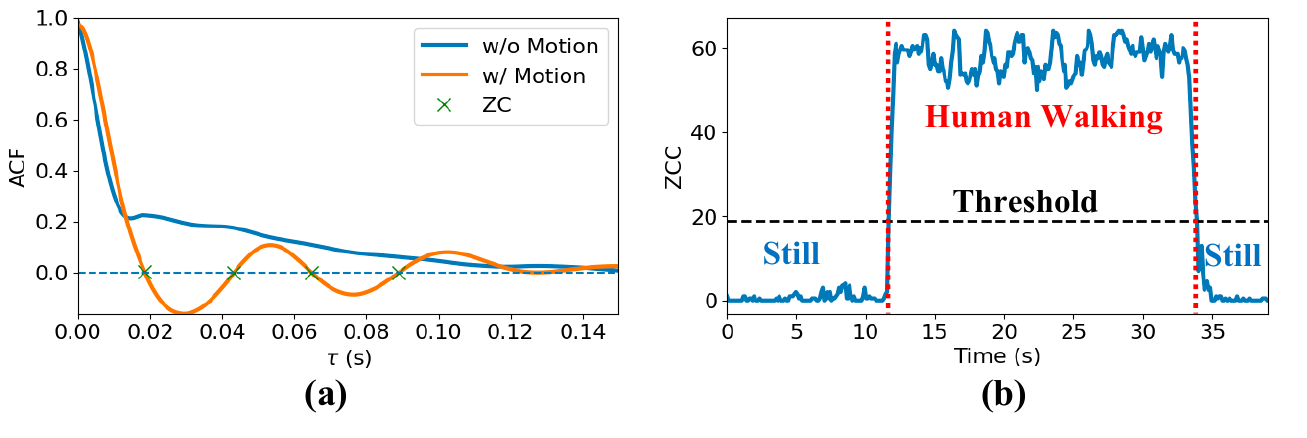}
    \caption{Illustration of Motion Detection. \rm{(a) ACF w/ and w/o movement. ACF without motion has considerably lower ZC counts. (b) Extracted the ZCC feature to detect motion.
    }}
    \label{fig:motion indicator}
\end{figure}

\subsection{Transmission Scheme}
\label{subsec: tx_design}
In this part, we will elaborate on our transmission scheme design. 
We will then describe the channel estimation principles.

\head{Single-Speaker OTDM}
We stack two orthogonal signals as a complex signal and pass it into the I/Q modulator for transmission.
Specifically, we can treat the orthogonal sequences $s_1(t)$ and $s_2(t)$ as the real and imaginary parts of a complex signal, \ie, $s(t) = s_1(t) + js_2(t)$. 
To delay $s_2(t)$, we shift the sequence by padding zeros to the front of $s_2(t)$. 
Practically, however, a speaker can only transmit a single sequence of real signals. 
To transmit the resultant complex signal over a speaker, we pass it through an I/Q modulator to convert it into a real one. 
As illustrated in \fig\ref{fig:modulation_scheme}, we additionally multiply $s_1(t)$ and $s_2(t)$ with a carrier signal of orthogonal phase and obtain the modulated real signal: $s(t) = s_1(t)\cos(2\pi f_c t) - s_2(t)\sin(2\pi f_c t)$. 
With such modulation, the two components remain orthogonal to each other, while the modulated signal can be transmitted through a single speaker. 
At Rx, the received signal is first demodulated by the carrier signal with a low-pass filter (LPF). 
We can then separate the overlapped signal and estimate the CSI for each sequence at the receiver's side, as shown in \fig\ref{fig:modulation_scheme}. 
By doing so, we can obtain two separated signals $r_1(t)$ and $r_2(t)$. Then, channel estimation will be performed as follows.

\begin{figure}[t]
    \centering
    \includegraphics[width=0.8\linewidth]{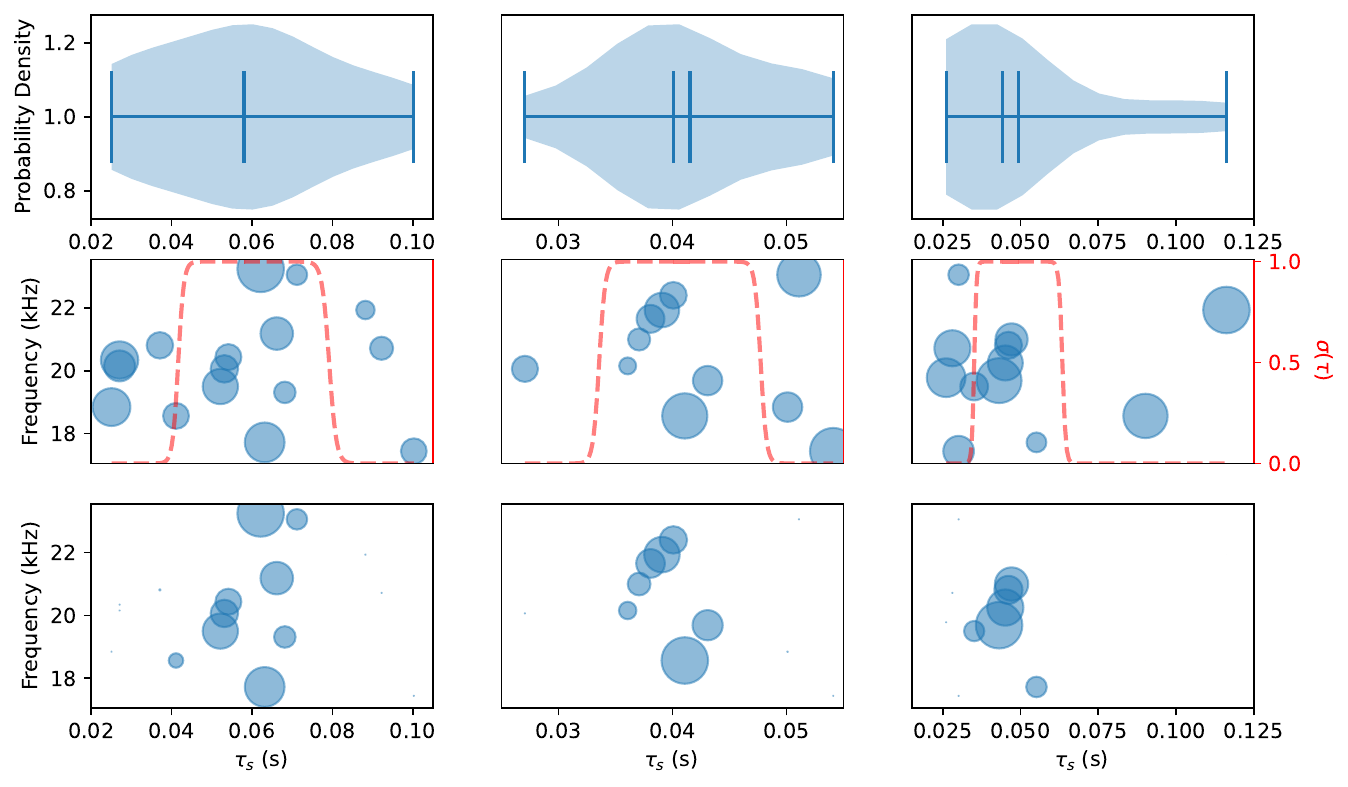}
    \caption{Peak prominence weighting examples. \rm{Top: Violin plots of the distribution of the delays of the first peak. Middle: Original weight distribution, where the circle size indicates the weights. The red dashed line represents the fitted $\sigma(\tau)$. Bottom: Adjusted prominence weights.}}
    \label{fig:MRC_violin}
\end{figure}

\head{Channel Estimation} 
To capture the channel properties, we generate the Kasami sequence $s(t)$ of length $N_s$ as the probe signal and transmit it over a speaker. 
The signal interacts with the environment before arriving at the microphone, where it undergoes scattering in the sound diffusion field. At the receiver end, we acquire $r(t)$ and separate it into $r_1(t)$ and $r_2(t)$, respectively. The CIR $h(t)$ is then estimated as the correlation between them, \ie, $h_i(t) = r_i(t) \ast s(t), i \in \{1,2\}$. Then CSI $H_i(f,t)$ is obtained by transforming $h_i(t)$ into the frequency domain via FFT. $H_1(f,t)$ and $H_2(f,t)$ will be then merged using the OTDM scheme to acquire the boosted channel estimation $H(f,t)$, as illustrated in \S\ref{sec:otdm}.

\imwut{
\head{Synchronization between Microphone and Speaker Pairs}
While the microphone and speaker are connected to the same controller and should theoretically be temporally aligned, hardware imperfections can introduce small time discrepancies between the Tx and Rx pair. Traditional FMCW tracking algorithms usually require precise synchronization to correctly determine the start time of IF signals. However, our approach minimizes the impact of synchronization issues.
At a high level, any discrepancies manifest as phase offsets in the asynchronized Channel State Information (CSI). This is not a significant problem for \sysname because we do not introduce blank intervals and instead focus on extracting channel power to determine speed.
Specifically, the asynchronous CSI would be 
\begin{equation}
    H_d(f) = H(f) \cdot \exp(-j \cdot 2 \pi f \cdot \tau)
\end{equation}
where $\tau$ is the temporal offset.
And the channel power, as denoted in \eqref{eq:CFR and sound pressure} can be written as 
\begin{equation}
\begin{aligned}
    G_d(f) &= |H_d(f)|^2 + \epsilon(f,t)  \\
            &= |H(f)|^2 + \epsilon(f,t) = G(f),
\end{aligned}    
\end{equation}
which is equivalent to the synchronized version. Therefore, the temporal offsets do not affect the magnitude of the channel power, ensuring reliable speed acquisition even without precise synchronization.
}

\subsection{Speed Estimation}
\label{subsec:speed_est}
Now we present how to estimate speed from the CSI time series, given the model described in \S\ref{sec:model}. 
We devise a robust motion indicator to detect movements. We perform speed estimation when motion is detected. 
We further enhance frequency diversity to achieve robust speed estimation.

\head{Motion Indicator} 
A robust motion indicator can help ensure only valid measurements are used for speed estimation, reducing the system overhead while improving estimation accuracy. 
Therefore, we propose a novel motion indicator called Zero Crossing Count (ZCC) in \sysname. As described in \sec\ref{sec:model}, ACF of channel power $\psi_G(f,\tau)$ is the function of moving speed $v$. When there is no motion, the ACF curve would be overall flat, resulting in no obvious peaks or zero crossings. In contrast, in the case of motion, the ACF will exhibit peaks and valleys, resulting in significantly more zero crossings, as shown in \fig\ref{fig:motion indicator}. 
To identify motion, we can perform Zero Crossing (ZC) analysis on the ACF and count dominant ZCs. 
\fig\ref{fig:motion indicator} shows an example of the ACF and ZCC values, indicating a clear difference between motive and still states. 
Hence, we can find a threshold to reliably detect motion from the ZCC values. 

\begin{figure*}[t]
\centering
   \begin{subfigure}{\linewidth}
        \centering
        \begin{subfigure}{.58\linewidth}
            \includegraphics[width=\linewidth]{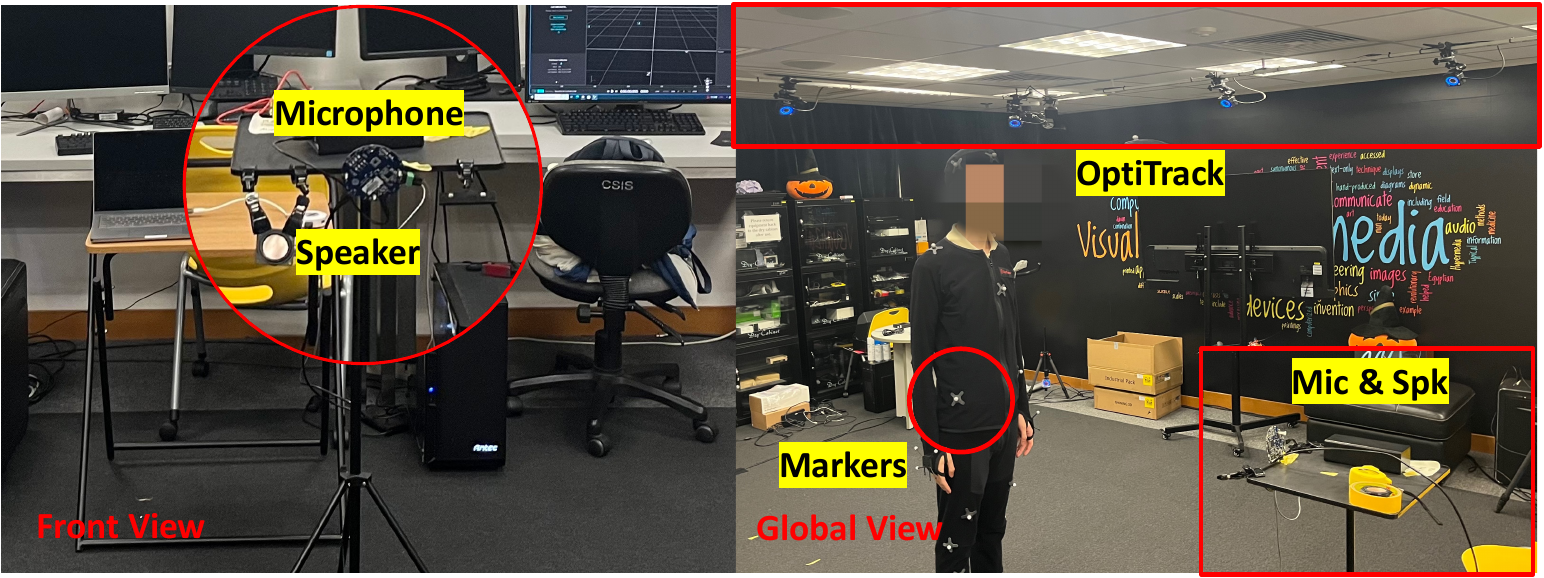}
            \caption{Overall Setting.}
            \label{subfig:exp}
        \end{subfigure}\hfill
        \begin{subfigure}{.41\linewidth}
            \includegraphics[width=\linewidth]{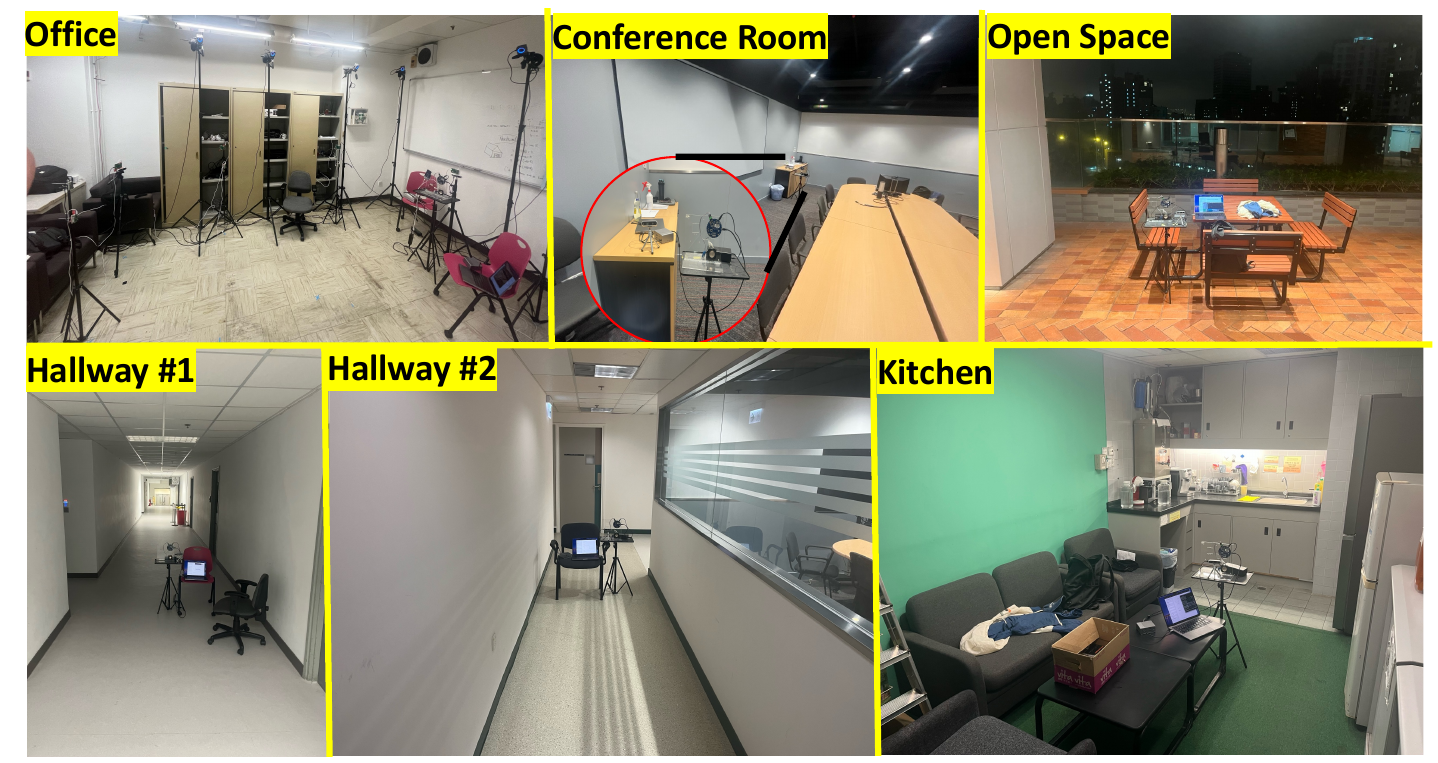}
            \caption{\imwut{Different Environments.}}
            \label{subfig:env_b}
        \end{subfigure}\hfill
        \label{fig:env}
    \end{subfigure}\hfill
    \caption{\imwut{Experiment Setting.}}
    \label{fig:exp_setting}
\end{figure*}

\head{Weighted Subcarrier Combining} 
Proposed model accounts for all multipath components for speed estimation. 
Recall \eqn\eqref{eq:acf_csi_velocity}, however, the theoretical model estimates speed from a single subcarrier. 
Given multiple subcarriers available, we can further embrace frequency diversity to boost speed estimation. 
Instead of performing speed estimation on each subcarrier and fusing the estimates, we propose to properly combine $\psi_G(f,t)$ calculated on each subcarrier to enhance the SNR of the speed signal. 
Formally, we aim to obtain the boosted ACF as $\hat{\psi}_G(\tau)=\sum w(f) \psi_G(f,\tau)$, where
\begin{equation}
\label{eq:mrc}
\begin{gathered}
    \max_{w} \quad \hat{\psi}_G(\tau_s) =  \sum_{i} w(f_i) \psi_G(f_i, \tau_{s_i}),\\
   \begin{aligned}
        & \text{s.t.} \quad \sum_{i} w(f) = 1, w(f) \geq 0, \forall f \in \mathbb{F},\\
    \end{aligned} 
\end{gathered}
\end{equation}
where $\mathbb{F}$ denotes the subcarrier set. $\tau_s$ is the first peak of $\hat{\psi}_G(\tau_s)$. The key is to find optimal weights $w(f)$ for effective combining. 
The desired goal is to obtain prominent peaks in the ACF for accurate and robust speed estimation. 
Therefore, we propose a novel weighting method based on the original peak prominence. 
The idea is that larger weights should be given to the ACF that features more prominent first peaks. 
Denote 
$\kappa(t)$ as the prominence of the detected first peak in $\psi(f,\tau)$. 
Intuitively, we can simply set $w(f)=\kappa(t)$ and combine all subcarriers. 
However, as shown in \fig\ref{fig:MRC_violin}, we observe that some ACF may have prominent peaks largely deviated from the centered delay of the majority of peaks. 
This would enlarge the noise and lower the combined first peak. 
To this end, we design a compensatory weight decaying algorithm. 
Specifically, we calculate the mean and median lags of all the identified peaks, and their lower or upper quartiles. 
We keep the prominence for peaks whose delays between the mean and median untouched, reduce the weights for those between the mean/median and the quartiles, and discard those falling outside the quartiles. 
Therefore, we use a sigmoid function as the decaying curve: $\sigma(\tau) = 1 / (1 + \exp(-a (\tau-b)))$, where $\tau$ is the peak value and $a$ and $b$ are two parameters that can be fitted. 
The adjusted prominence values are used as the weights for calculation in \eqn\eqref{eq:mrc}.

\head{Frequency Alignment} 
The weighted combining will enhance the speed signals and reduce the noise, only if the signals are synchronized. 
However, significant subcarrier frequency variations misalign the ACFs across subcarriers, particularly the first peak, for the same speed. 
Take two subcarriers on 20 kHz and 24 kHz as an example. Considering a speed $v=0.5 m/s$, the peaks will appear at a delay $\tau=x_0\lambda(f)/2\pi v$.
The wavelength $\lambda(f)$ depends on the carrier frequency, and will be 1.715 cm for 20 kHz and 1.429 cm for 24 kHz, leading to two considerably different delays 
for the first peaks in the respective ACFs. 
Therefore, we need to eliminate the subcarrier frequency differences and align all the ACFs. 
We scale the ACF in the lag domain and align them with respect to the first peaks. 
Particularly, we choose a virtual frequency as a reference, denoted as $f_{ref}$, and scale the ACF on all subcarriers to the same $f_{ref}$, \ie, $\psi(f,\tau) \to \psi(f_{ref},\tau')$. 
Recall \eqn\eqref{eq:speed_tau}, we obtain the aligned first peak as, 
\begin{equation}
    v(f,\tau_s) = \frac{x_0 c}{2 \pi  f \tau_s} = \frac{x_0 c}{2 \pi  f_{ref}\tau_s^\prime} = v(\tau_s^\prime),
\end{equation}
where $\tau_s^\prime = \frac{f}{ f_{ref}} \tau_s$ represents the aligned first peak. 
Practically, it is implemented by interpolation.
Then the weighted combining can be done on the aligned ACF.

\begin{figure*}[t]
    \centering
    \begin{minipage}[b]{0.32\textwidth}
    \centering
    \includegraphics[width=0.9\linewidth]{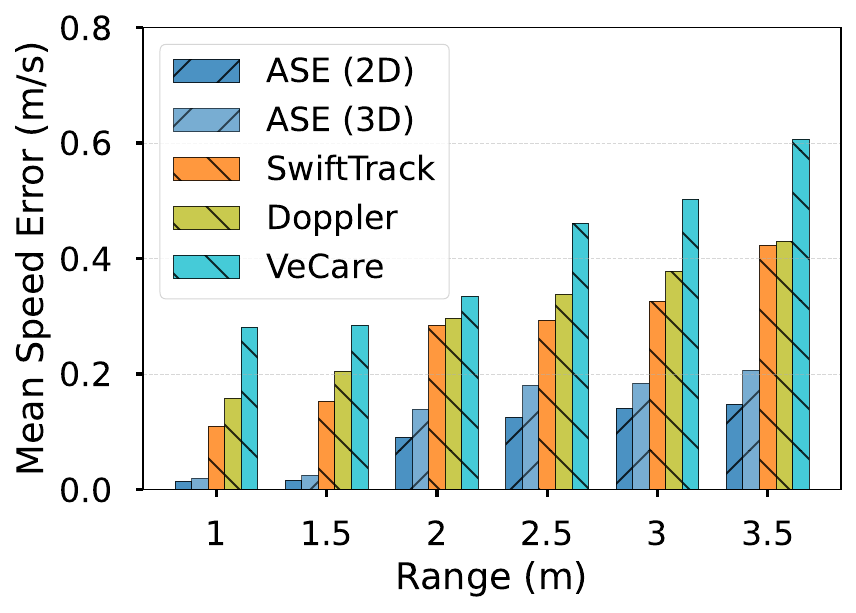}
    \caption{\rev{MSE-DW.}}
    \label{fig:dw_v_mean_error}
  \end{minipage}
  \hfill
  \begin{minipage}[b]{0.32\textwidth}
    \centering
    \includegraphics[width=0.9\linewidth]{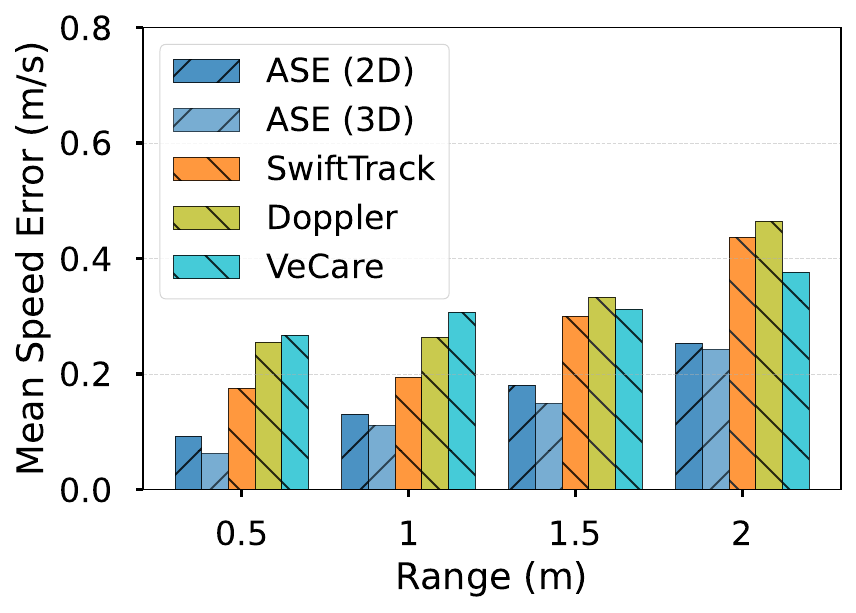}
    \caption{\rev{MSE-CW.}}
    \label{fig:cw_v_mean_error}
  \end{minipage}
  \hfill
  \begin{minipage}[b]{0.32\textwidth}
    \centering
    \includegraphics[width=0.9\linewidth]{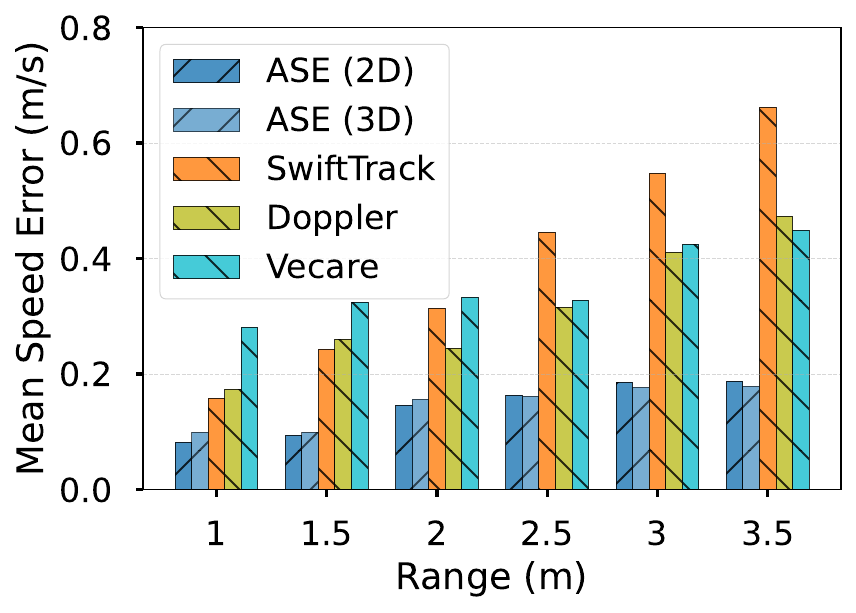}
    \caption{\rev{MSE-RW.}
    }
    \label{fig:rw_v_mean_error}
  \end{minipage}
\end{figure*}

\section{Evaluations}
\label{sec:exp}

\subsection{Implementation}
\label{sec:impl}

\head{Hardware} We implement \sysname with programmable smart speaker prototype, \ie, a UMA-8-SP USB microphone \cite{USBAudioStreaming}, and AS05308AS-R speaker\cite{SpeakersReceiversAS05308ASR}. 
Notably, we only \textbf{use one microphone channel} in our experiments. 
The microphone and the speaker are co-located. The hardware is connected to a power hub and MacBook Pro 2021.

\head{Software} We implement the algorithms of \sysname in Python and Matlab. Specifically, we develop a sound playing and recording program with Python and implement the pipeline of processing algorithms with MATLAB. 
We apply a 1-s window to compute ACF with a step of 0.1s. 
We perform motion detection, based on ZCC, to determine whether a user is moving and perform speed estimation when motion is detected. 
Before weighted subcarrier combining, we also perform outlier detection to sift out some abnormal ACFs, which occasionally appear as either a near-linear trend or a zig-zag spike. 

\head{Data Collection} As shown in \fig\ref{fig:exp_setting}, we mainly conduct our experiment in a 4 $\times$ 4m room equipped with the OptiTrack \cite{MotionCaptureSystems}, which serves as the ground truth.
It is a camera-based precise motion capture system. 
We derive speed from the solved whole-body skeleton coordinates from OptiTrack. 
To calibrate and fuse the data across cameras, participants wear specialized clothes with visual markers.
 The experiment is conducted in a regular office environment and is subject to various background noises, including footsteps in the corridor and the sound of the room air conditioner. 
 During the experiment, only the test participant is walking while others in the room are stationary. 
We have gained IRB approval from our university for data collection.
In total, we have collected over 700 minutes of moving data traces. 
We evaluate the performance of \sysname under various settings. 

\head{Metrics}
We use Mean Speed Error (MSE) between the estimated speed and the ground truth
and detection rate (\ie, how often \sysname can reliably detect a speed) as the main evaluation metrics. 
Our results demonstrate the effectiveness of our approach in accurately estimating speed and detecting movement in a daily living environment. We present detailed results below.

\begin{figure*}[t]
    \centering
      \begin{minipage}[b]{0.24\textwidth}
        \includegraphics[width=\linewidth]{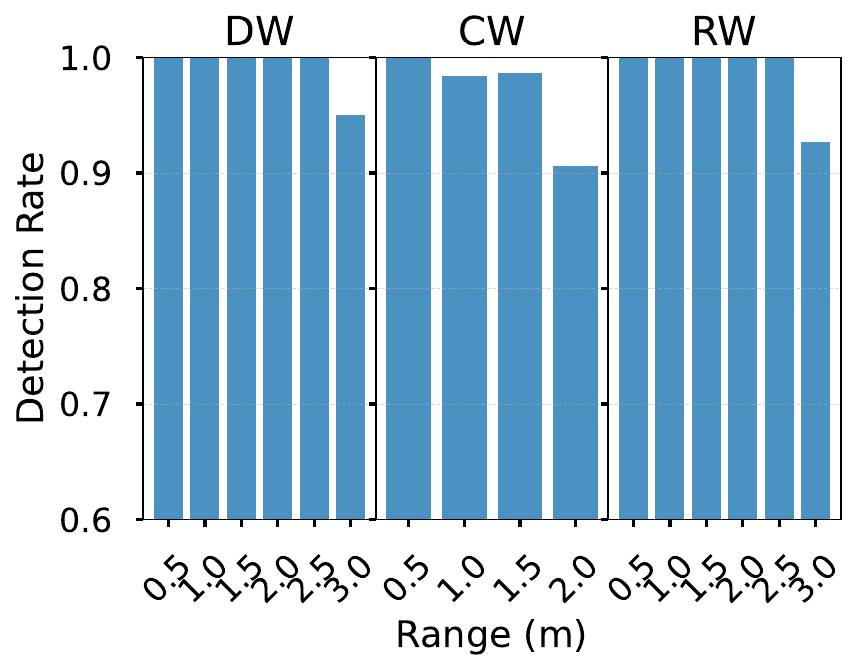}
        \caption{\rev{Detection Rate.}}
        \label{fig:dw_dr}
      \end{minipage}
      \hfill
      \begin{minipage}[b]{0.24\linewidth}
            \centering
          \includegraphics[width=\linewidth]{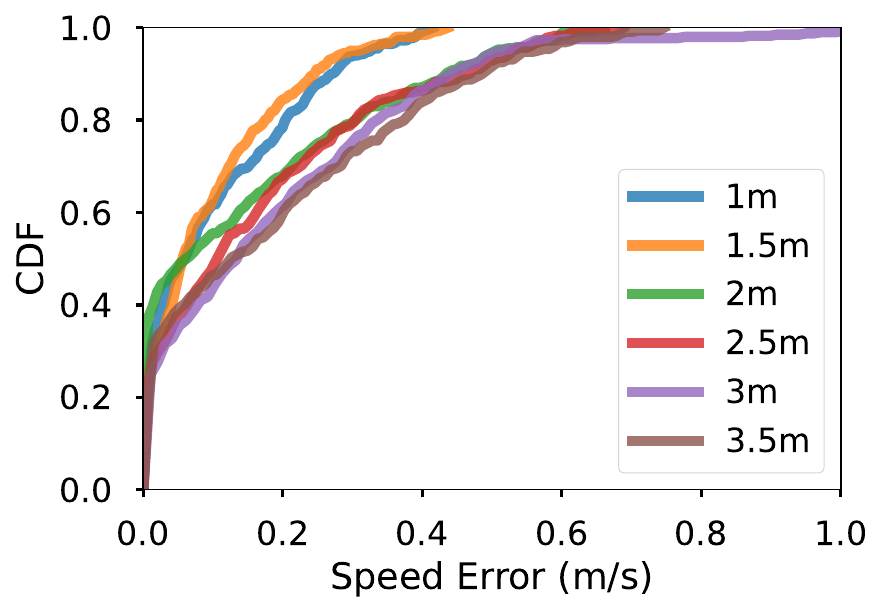}
          \caption{CDF vs. Distances.}
            \label{fig:rw_ase_cdf}
        \end{minipage}
      \hfill
      \begin{minipage}[b]{0.24\textwidth}
        \includegraphics[width=\linewidth]{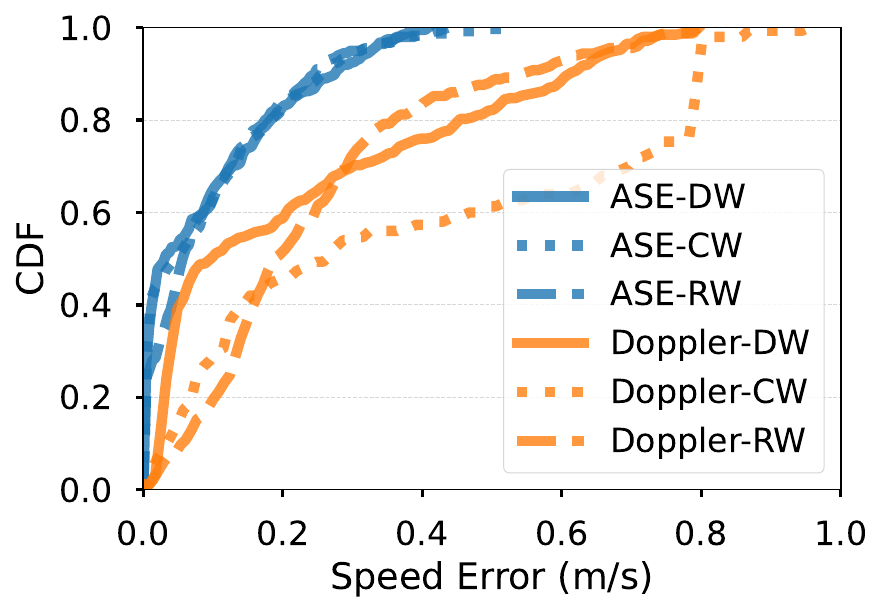}
        \caption{CDF vs. Walk Means.}
        \label{fig:combined_cdf}
      \end{minipage}
      \hfill
      \begin{minipage}[b]{0.24\textwidth}
        \includegraphics[width=\linewidth]{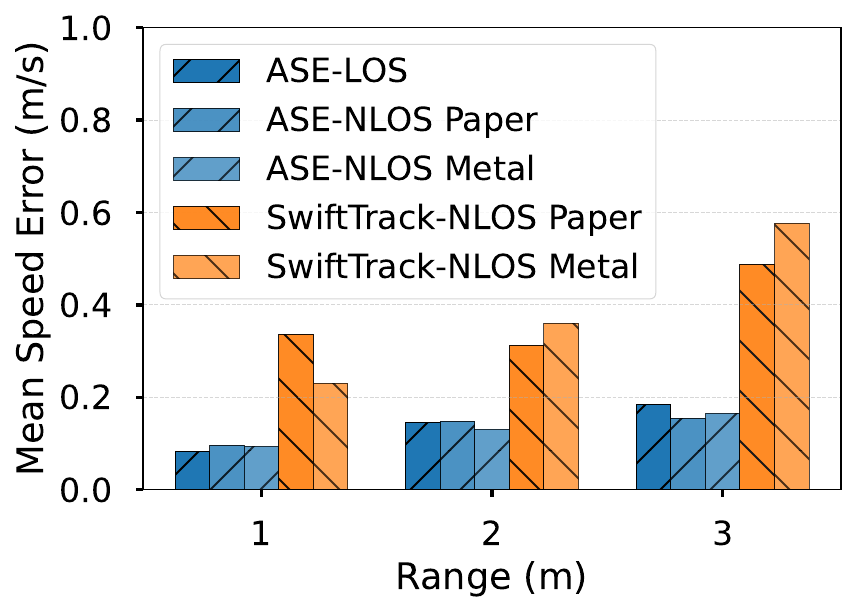}
        \caption{MSE of NLOS Case.}
        \label{fig:nlos_v_mean_error}
      \end{minipage}
\end{figure*}

\subsection{Overall Performance}
\label{subsec:overall_performance}
Our evaluation of \sysname considers the coverage and orientation of the audio devices by involving three different scenarios: direct walk (\ie, a straight line towards the devices), circle walk (\ie, around the devices with varying radii) and random walk (\ie, zig-zag walk, back-and-forth walk and run, \etc).

\head{Direct Walk (DW)}
We place the device at the center of one wall in the room.  
Participants are asked to walk towards the audio device 
from 1m to 3.5m.
We then calculate the mean speed error in  \fig\ref{fig:dw_v_mean_error} and detection rate in \fig\ref{fig:dw_dr}. 
Our results demonstrate that \sysname can estimate speeds reliably up to a distance of 3.5 m, with an average speed error of 0.08 m/s. 
When the object is close, 
the mean speed error is 0.016 m/s, which increases to 0.14m/s at 3.5m. 
Our system achieved an average detection rate of 99.0\% in the case of direct walking, with a slight degradation to 95.0\% at a distance of 3.5m.

\head{Circle Walk (CW)}
We conduct experiments in a circle-walk scenario. Participants were instructed to walk around the device at varying radii between 0.5m and 2m. The mean error and detection rate
are depicted in \fig\ref{fig:cw_v_mean_error} and \fig\ref{fig:dw_dr} respectively. While DFS often struggles to estimate speed on a circular path, \sysname achieves a mean speed error of 0.13m/s within 1 meter, 2 times lower than that of DFS.

\head{Random Walk (RW)}
We comprehensively evaluate the performance of \sysname in random walk scenarios. 
To do so, we let participants move freely within the experimental room,  and summarize the results in \fig\ref{fig:rw_v_mean_error} and \fig\ref{fig:dw_dr}. 
As seen, our approach achieved an average detection rate of 98.8\%, with a perfect detection rate of 100\% within 3m.
The total average mean error is 0.13m/s, which lowers to 0.08m/s when participants are walking within a shorter range. 
We also portray the CDF in \fig\ref{fig:rw_ase_cdf}, which shows a 90\%-tile error of less than 0.2m/s. 
The results demonstrate \sysname's robust performance in realistic environments for practical applications. 

\head{2D vs 3D model} We evaluate both 2D and 3D models in different scenarios, as illustrated in \fig\ref{fig:dw_v_mean_error}, \fig\ref{fig:cw_v_mean_error} and \fig\ref{fig:rw_v_mean_error}. The 3D model shows a lower mean speed error of 2.2 cm/s in circle walk scenarios compared to the 2D model. In random walk scenarios, the 3D model is more effective at longer distances. This evaluation underscores the feasibility of both models and their applicability to different scenarios.

\subsection{Comparative Study}

We focus on acoustic-based speed estimation and 
compare \sysname with \rev{three} baselines: 1) SwiftTrack \cite{zhang2023addressing}: An acoustic speed algorithm for practical fast motion tracking, 2) DFS: A widely used method, implemented based on CAT \cite{maoCATHighprecisionAcoustic2016}, and
3) VeCare \cite{zhang2023vecare}: A recent work of a statistical acoustic sensing framework for child presence detection. The former two are DFS-based speed estimation approaches.
For a fair comparison, we transmit modulated Zadoff-Chu (ZC) signals 
as in \cite{zhang2023addressing} with the same frame length, \ie, 10ms. We use unmixed CSI and extracted the frequency shift peak in the DFS spectrum for Doppler. We compare baselines under scenarios in \S\ref{subsec:overall_performance}, and also test the Non-Line-of-Sight (NLoS) condition. We apply the same settings as VeCare. 
We will first compare SwiftTrack and Doppler and leave VeCare as the last part.

\head{Different Walk Means} 
We compare \sysname with SwiftTrack and Doppler in three means: direct walk, circle walk and random walk. Our results are shown in \fig\ref{fig:dw_v_mean_error}, \fig\ref{fig:cw_v_mean_error} and \fig\ref{fig:rw_v_mean_error} respectively. 
While \sysname achieves a median error of 0.08m/s in direct walking, SwiftTrack and Doppler yield errors of 0.26m/s and 0.30m/s, respectively. 
Moreover, \sysname significantly surpasses SwiftTrack and Doppler by 68.9\% and 101\% for circle walk and by 176.7\% and 146.2\% for random walk.
We further depict the CDF of \sysname against Doppler in \fig\ref{fig:combined_cdf}. It shows DFS exhibits poor performance in circle walk while \sysname retains consistent performance across various walking means. 
It is plausible for worse performance of SwiftTrack against Doppler in random walk, given its speed estimation algorithm is tailored for gesture tracking.
\begin{figure*}[t]
    \centering
    \begin{minipage}[b]{0.46\linewidth}
        \centering
          \includegraphics[width=0.8\linewidth]{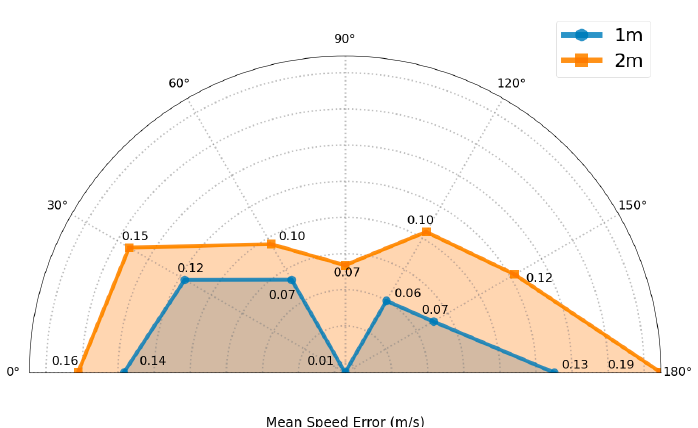}
          \caption{\rev{Impact of different directing angles.}}
        \label{fig:dw_angle_error}
    \end{minipage}
    \hfill
    \begin{minipage}[b]{0.24\linewidth}
        \centering
        \includegraphics[width=\linewidth]{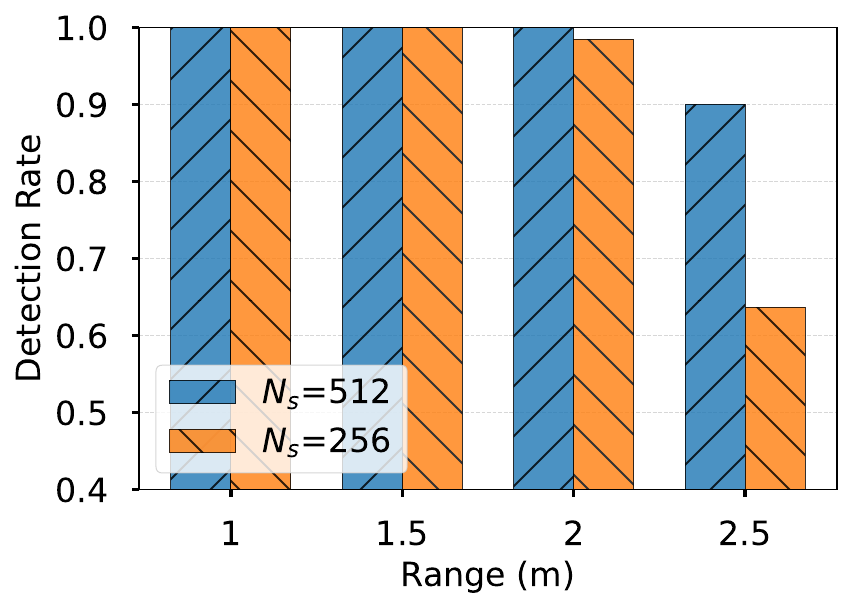}
        \caption{Impact of sequence lengths.}
        \label{fig:frame_len_detection_rate}
    \end{minipage}
    \hfill
    \begin{minipage}[b]{0.24\linewidth}
        \centering
        \includegraphics[width=\linewidth]{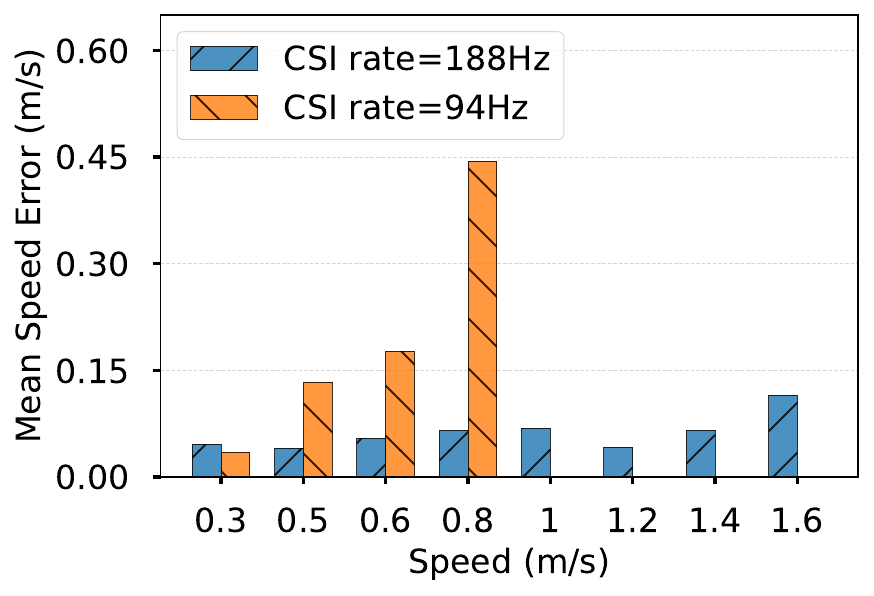}
        \caption{Impact of different CSI rates.}
        \label{fig:error_diff_csi}
    \end{minipage}
\end{figure*}

\head{Sensing Distance} 
A unique advantage of \sysname is the enlarged sensing coverage, and we verify this by evaluation over different distances, as illustrated in \fig\ref{fig:dw_v_mean_error}, \fig\ref{fig:cw_v_mean_error} and \fig\ref{fig:rw_v_mean_error}. 
While the error generally increases with respect to distances for all the methods, \sysname consistently outperforms the baseline methods at all distances, with more significant performance gains at larger distances. 
As shown in \fig\ref{fig:dw_v_mean_error}, \sysname achieves a mean speed error less than 0.15m/s at 3.5 m, while SwiftTrack and Doppler soar to 0.42m/s and 0.43m/s, respectively. 
Overall, both SwiftTrack and Doppler degrade significantly beyond the distance of 1.5m.
\sysname obtains superior performance because our model leverages all multipath signals, which get more observations across the room and expand the range to accommodate room-scale sensing. 

\head{LoS vs. NLoS} We assess our system in NLoS settings with obstructions like a metal panel or a paper bag in front of the speaker and microphone.
Despite sound's known vulnerability to absorption \cite{tang2017acoustic}, \sysname remains effective in NLoS scenarios, exhibiting an average error rate of 0.13m/s and a standard deviation of merely 0.003m/s, as illustrated in \fig\ref{fig:nlos_v_mean_error}. In contrast, the performance of SwiftTrack drops by 14.4\% and 11.2\% under metal and paper obstructions, respectively. We benefit from the effective use of multipath reflections, which is particularly helpful in NLoS scenarios.

\head{Comparison with VeCare} VeCare includes a statistical framework for presence detection, which mentions speed components. We implement the algorithm and evaluate in three scenarios. As shown in \fig\ref{fig:dw_v_mean_error}, \fig\ref{fig:cw_v_mean_error} and \fig\ref{fig:rw_v_mean_error}, the performance of VeCare is much worse than our \sysname. Notably, the performance of \sysname is 149.9\% better than VeCare in random walk scenarios. VeCare, which extrapolates Wi-Fi sensing methods without comprehensive modeling, is not targeted for speed estimation and does not include specific designs. 
These results highlight the novelty and superiority of our speed estimation techniques and theoretical model.

\subsection{Benchmark Study}
\label{subsec: benchmark}
In this section, we present a detailed analysis of various impact factors in \sysname. 
We begin by investigating the impact of different movement directions on the accuracy of our approach. Next, we examine the performance of our system extending to dual speakers to conduct the experiment. We also study the impact of various acoustic factors, including sound level, sound amplitude, and interference sources.

\begin{figure*}[t]
    \centering
    \begin{minipage}{0.33\linewidth}
        \centering
      \includegraphics[width=\linewidth]{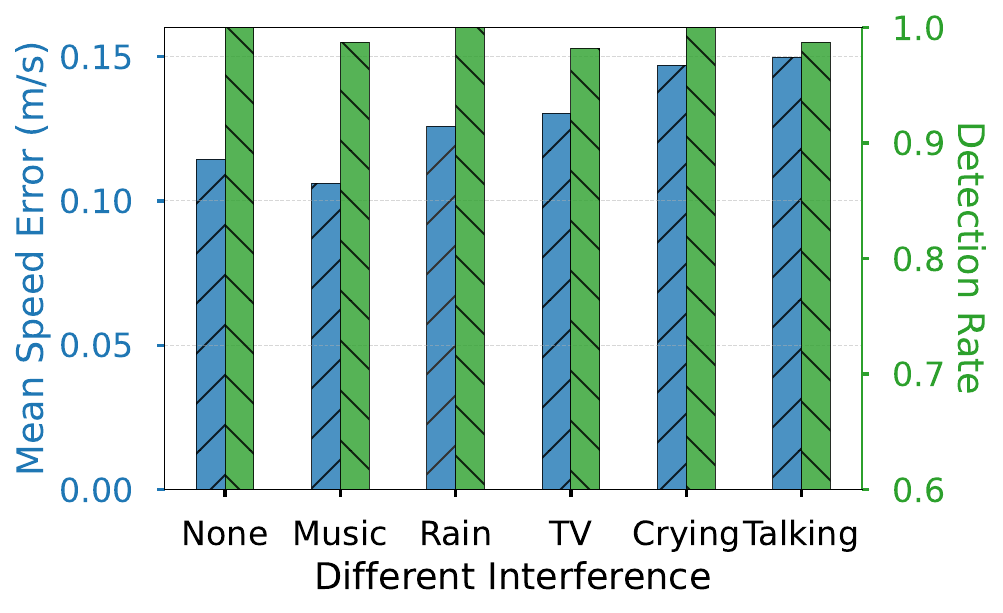}
      \caption{Impact of different interference sources.}
        \label{fig:rw_diff_interf_error_dr}
    \end{minipage}
    \hfill
    \begin{minipage}{0.33\linewidth}
        \centering
      \includegraphics[width=\linewidth]{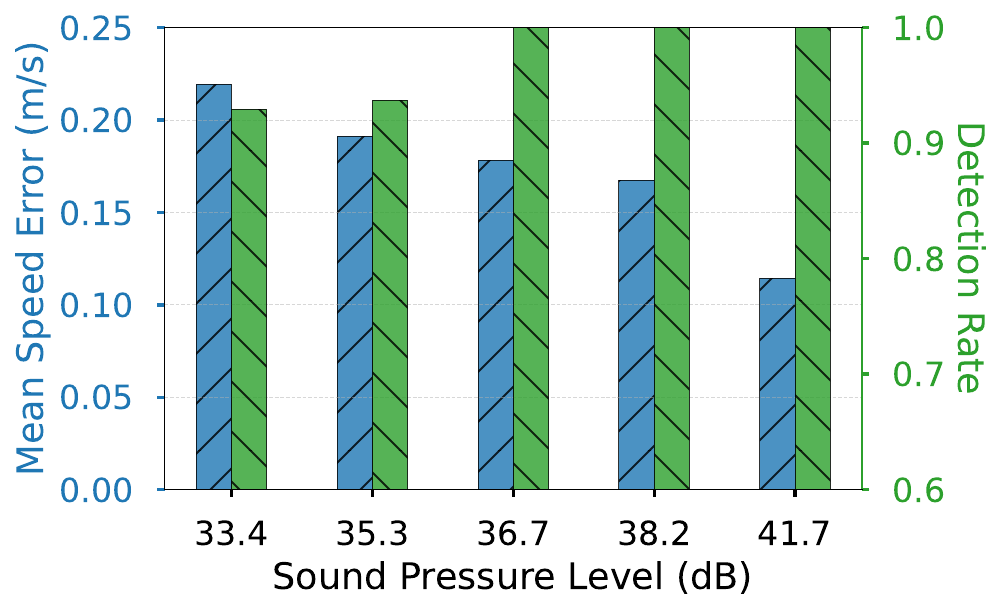}
      \caption{Impact of different sound volume levels.}
        \label{fig:rw_diff_volume_error_dr}
    \end{minipage}
    \hfill
    \begin{minipage}{0.33\linewidth}
        \centering
      \includegraphics[width=\linewidth]{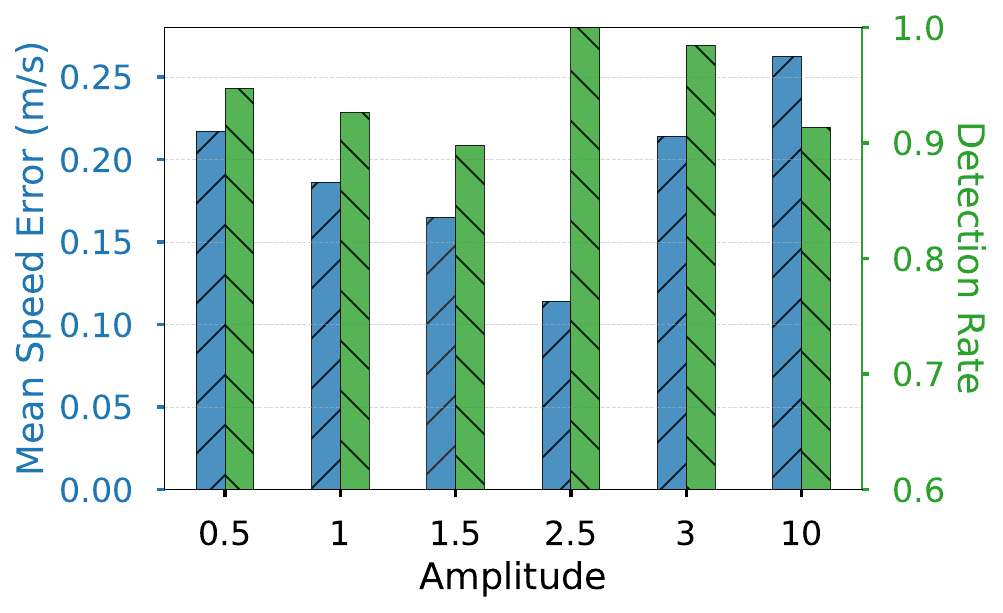}
      \caption{Impact of amplitudes of original sequence.}
        \label{fig:rw_diff_amp_error_dr}
    \end{minipage}
\end{figure*}

\head{Different Orientations}
We evaluate the impact of different walking orientations at various angles with respect to the audio devices. 
We evaluate different angles, from $0^\circ$ to $180^\circ$ with an increase of $30^\circ$, where $90^\circ$ is defined as facing directly towards the device. 
As illustrated in \fig\ref{fig:dw_angle_error}, the smallest speed error is observed at $90^\circ$. The errors increase as the orientation shifts away from $90^\circ$ towards both sides, yet are still acceptable even when at $0^\circ$ or $180^\circ$.

\head{Sound Level}
We modulate the transmission signal to the inaudible band to minimize interference with human hearing for daily use as detailed in \sec\ref{sec:impl}, rendering the sound nearly inaudible. 
 To evaluate the impact of different sound levels, we transmit signals at 
 sound pressure levels of 33.4dB, 35.3dB, 36.7dB, 38.2dB, and 41.7dB, respectively. 
 As shown in \fig\ref{fig:rw_diff_volume_error_dr}, the results indicate that the mean speed error decreases as the sound volume increases, as does the detection rate. Notably, our system's maximum sound pressure of 41.7dB is significantly lower than most previous acoustic sensing studies \cite{yunStrataFineGrainedAcousticbased2017,zhang2023vecare, wangContactlessInfantMonitoring2019a,maoAIMAcousticImaging2018}. 
 Despite the sound leakage problem on commodity devices \cite{li2022experience}, the sound noise of our system can be fairly neglected in commonly used settings.

\head{Amplitude of Kasami Sequence}
Besides the volume of the device, the sound power is also determined by the amplitude of the original sequence. We hereby experiment with different amplitudes, shown in \fig\ref{fig:rw_diff_amp_error_dr}. This is not the final amplitude of the transmitted signal, as it must go through band modulation and PCM before transmission.
As seen, the performance decreases first and then increases as amplitudes increase, and an amplitude of 2.5 strikes the best for both speed accuracy and detection rate. 
Amplitudes smaller than 2.5 will result in low sound power and thus large errors, while overlarge amplitudes will lead to distortions of the signal \cite{li2022experience} due to the hardware limitation. 
In our case, 
sounds with amplitudes larger than 2.5 would be clipped off. 
Hence, 
we choose 2.5 as the amplitude of the original PN sequence.

\head{Interferences} 
We also conduct experiments to evaluate the performance under different interference sources. We choose common sources of interference in indoor environments, including music, rain, TV broadcasting, baby crying, and human talking, illustrated in \fig\ref{fig:rw_diff_interf_error_dr}. 
This indicates that our system is robust to noise and interference, and can accurately estimate the speed even in the presence of interference.

\head{Sequence Lengths}
We then study the influence of different sequence lengths $N_s$. As discussed above, the sequence length impacts the sensing coverage and CSI rate. When the sequence length is reduced, the CSI rate would increase, but the sensing coverage would diminish simultaneously. To investigate it, we conduct experiments using sequence lengths of 512 and 256 and mainly examine the detection rate to study the coverage. 
As shown in \fig\ref{fig:frame_len_detection_rate}, using both lengths achieves great performance for small coverage, \eg, within 1.5 m. 
However, the detection rate for the sequence length of 256 drops significantly at larger distances of 2m and 2.5m. 
Particularly, at a distance of 2.5m, the detection rate drops to 63.6\% for $N_s = 256$ while the rate still remains at 99\% for $N_s = 512$. 
\sysname defaults to $N_s = 512$ to strike a balance between sensing coverage and CSI rate.

\head{CSI Rate} 
We now study how CSI rates will impact the performance for different speeds. 
To control the moving speed, we use a programmable rail track with a plate on it. 
We test with speeds of 0.3m/s to 1.6m/s, 
and compare the performance with a CSI rate of 94 Hz and 188 Hz, respectively. 
We apply the OTDM scheme to achieve the CSI rate of 188 Hz. 
As shown in \fig\ref{fig:error_diff_csi}, a low CSI rate cannot measure large speeds and will produce large errors. 
The mean errors for speeds of 0.6m/s and 0.8m/s with a CSI rate of 94Hz are 0.18m/s and 0.44m/s, respectively. Comparatively, the corresponding errors with that of 188Hz are 0.055m/s and 0.066m/s, respectively.  Notably, our system maintains a relatively low average error of 0.11m/s even at 1.6m/s. 
Overall, the results clearly demonstrate the need for a sufficient CSI rate for speed estimation and justify the effectiveness of the proposed OTDM scheme. In \sysname, we can hold speed up to 1.6m/s using single-speaker, adequate for capturing indoor walk speed.

\begin{figure*}[t]
    \centering
    \begin{minipage}{0.24\linewidth}
        \includegraphics[width=\linewidth]{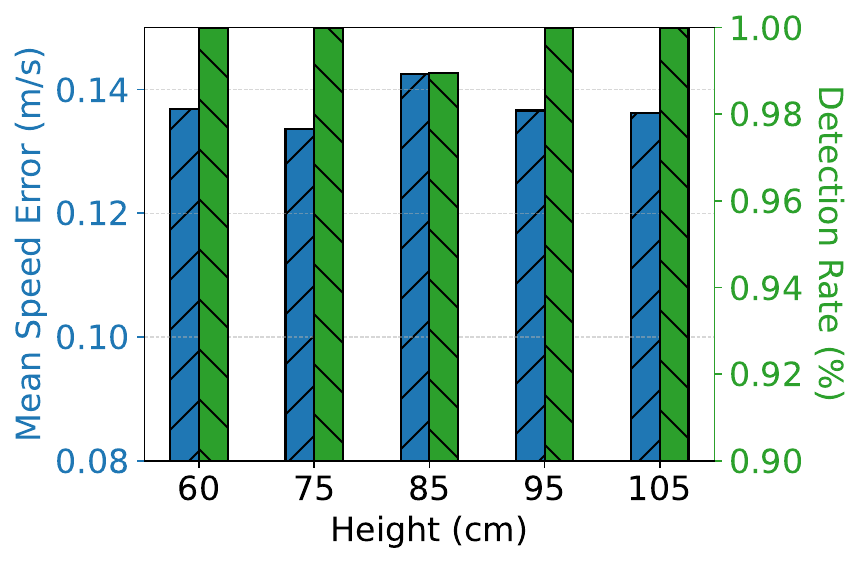}
        \caption{\imwut{Different Height.}}
        \label{fig:height_v_dr}
    \end{minipage}
    \hfill
    \begin{minipage}{0.24\linewidth}
        \includegraphics[width=\linewidth]{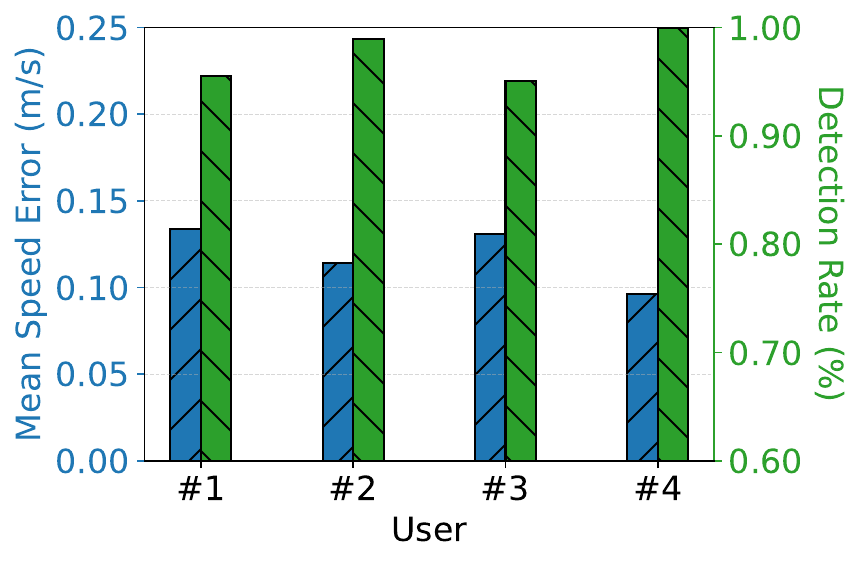}
        \caption{\imwut{Different Users.}}
        \label{fig:rw_diff_user}
    \end{minipage}
    \hfill
    \begin{minipage}{0.49\linewidth}
         \centering
    \includegraphics[width=\linewidth]{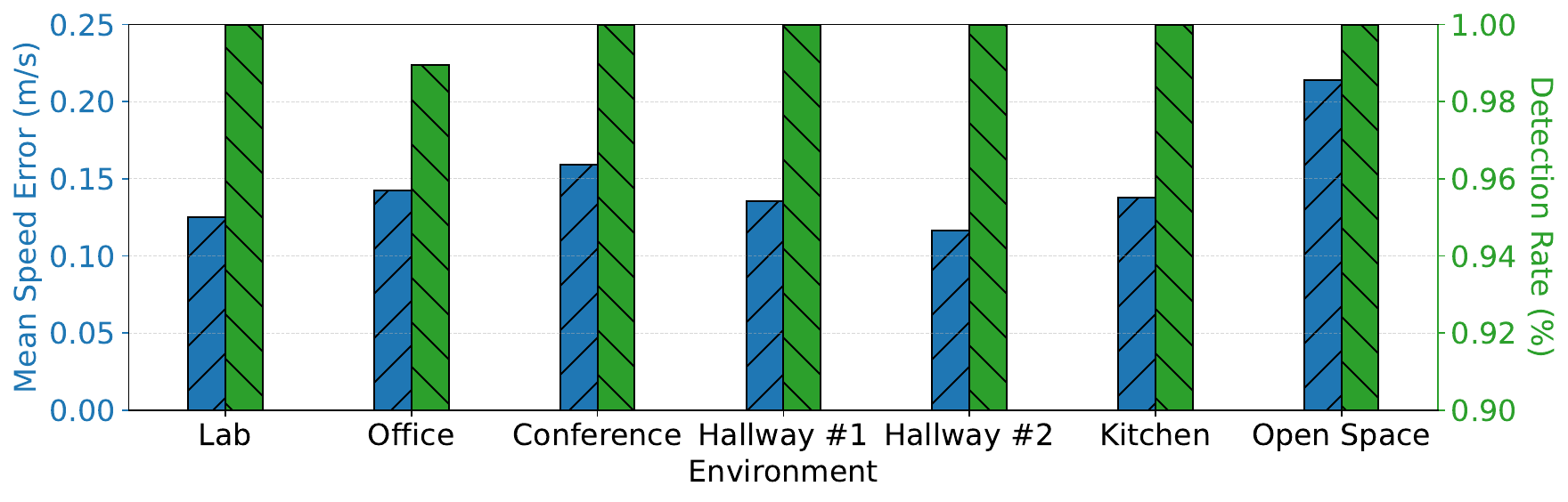}
      \caption{\imwut{Different Environment and Locations.}}
        \label{fig:diff_env}
    \end{minipage}
\end{figure*}

\head{Different Subcarrier Combining Weight} We test different subcarrier combining weight methods.  Our results show that using prominence as the weight leads to a mean speed error of 0.13m/s, whereas using motion statistics simply \cite{zhang2023vecare} results in an error of 0.19m/s. Besides, we compare different weight decaying algorithms. We observe a 15\% decrease in mean speed error when applying sigmoid weight decay. These results confirm the effectiveness of our subcarrier combination weight in \sysname.

\imwut{
\head{Different Device Height} To investigate the impact of varying the height of the transmitter and receiver pair on system performance, we conducted experiments using \sysname with different hardware heights, as shown in Figure \ref{fig:height_v_dr}. By utilizing a tripod, we controlled the height range from 80 to 115 cm.
The results indicate that \sysname retains consistent accuracy under changes in heights, with a mean of 0.13 m/s and a standard deviation of 0.002 m/s. Despite these variations, the average detection rate remains high at 99.8\%. This suggests that the height of the devices does not significantly affect the overall performance of \sysname.

}

\head{Different Users}
We also examine how \sysname performs for different users. 
We recruit four volunteers and let them walk freely with varying speeds, orientations, and trajectories. 
\imwut{
The results are shown in \fig\ref{fig:rw_diff_user}. Our approach achieves 80\% tile error of 0.2m/s. The detection rates of three users are 95.55\%, 98.95\% 95.03\% and 100\%, respectively. Accordingly, the mean error of four users is 0.12m/s with an average detection rate of 97.4\%. These findings demonstrate the effectiveness and accuracy of \sysname in estimating the speed of human movement in indoor environments. Our approach achieves high detection rates and low error rates across multiple users with varying movement patterns and speeds. These results have important implications for the use of \sysname in real-world applications, such as surveillance, security, and healthcare, where accurate estimation of human movement speed is critical.}

\imwut{\head{Different Environments and Locations} To evaluate \sysname's robustness across various settings, we conduct experiments in multiple environments, as illustrated in \fig\ref{subfig:env_b}. We deploy devices in a lab, office, conference room, hallways, kitchens, and an open space. These locations differ in geometry, building materials, floor heights, and ambient noise levels. Except for the lab and office, we leverage the depth camera to get the range information and then convert it to speed. During the tests, participants walk randomly within each environment.
We compute the mean speed error and detection rate for all scenarios, as summarized in \fig\ref{fig:diff_env}. In indoor environments, the average speed error is 0.136 m/s with a standard deviation of 0.01 m/s, which aligns with previous benchmark results. The corresponding detection rate averages 0.998, with a standard deviation of 0.003. These results demonstrate that \sysname consistently performs well across varied indoor settings, proving its robustness.
We also evaluate \sysname in an open space, where the mean speed error increases to 0.21 m/s, although the detection rate remains high. This outcome is expected given the relative scarcity of reflectors in open environments, which can impact speed estimation accuracy. Nevertheless, the consistently high detection rate highlights the robustness of our motion detection algorithm. Future work can explore the development of advanced speed estimation methods tailored for open areas.

}

\begin{figure*}[t]
    \centering
    \begin{minipage}{0.32\linewidth}
        \includegraphics[width=0.9\linewidth]{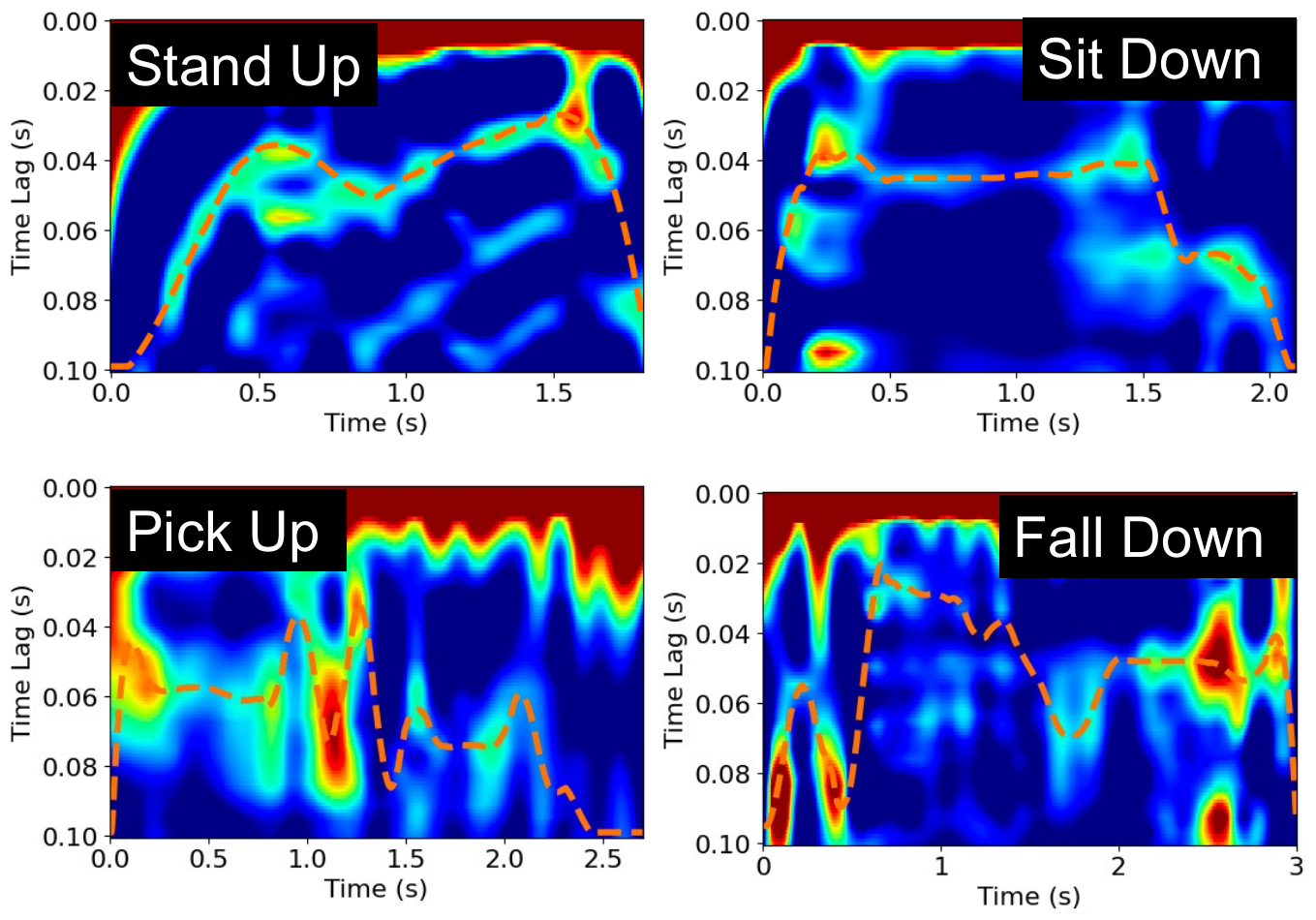}
        \caption{Speed profile of different human activities.}
        \label{fig:acf_har}
    \end{minipage}
    \hfill
    \begin{minipage}{0.33\linewidth}
         \centering
     \includegraphics[width=0.9\linewidth]{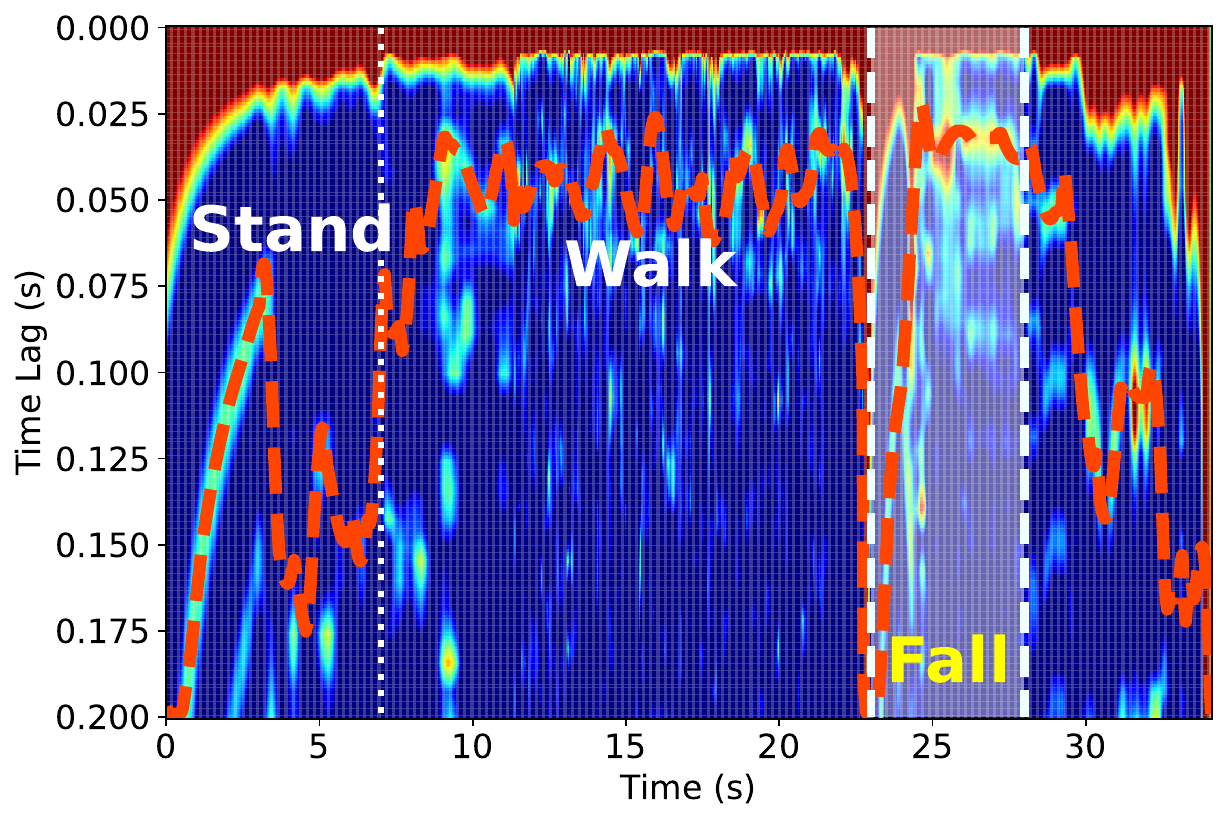}
      \caption{Speed profile of continuous activities.}
        \label{fig:har_acf_combined}
    \end{minipage}
    \hfill
    \begin{minipage}{0.32\linewidth}
        \centering
        \includegraphics[width=0.9\linewidth]{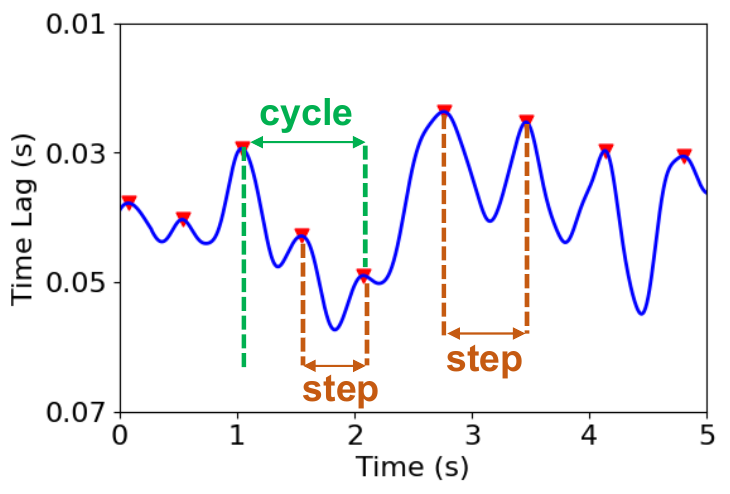}
        \caption{Gait step and Cycle.}
        \label{fig:gait}
    \end{minipage}
\end{figure*}

\imwut{
\head{Runtime Analysis}
In this section, we evaluate the runtime performance on various platforms. We first test the algorithm's performance on a MacBook 2021 (with Apple M1 Pro chip), analyzing a 10-second audio segment using a 1-second sliding window with a 0.1-second step. The algorithm demonstrates exceptional efficiency, achieving a total runtime of 2.299 seconds, well below the 10-second data duration, which confirms its capability for real-time processing. The decoding of OTDM takes 0.0999 seconds, and the ACF part finishes in 0.831 seconds. Peak prominence and weight computations are notably fast at 0.098 seconds and 0.067 seconds, respectively. Additionally, memory usage is a modest 387.56 MB, well within the capacity of modern systems, making this algorithm highly suitable for efficient, real-time processing. Additionally, we deploy our system on an embedded device, \ie, Raspberry Pi Compute Module 4, which is equipped with 4-core ARM-v8 processors. For the same 10-second audio segment with identical windowing parameters, the algorithm processes the respiration data with an overall runtime of 3.912 seconds, still significantly faster than the data duration, ensuring real-time performance even on resource-constrained hardware. The decoding takes 0.384 seconds, and ACF calculations total 0.440 seconds.  Memory usage on the Raspberry Pi is optimized at 318.57 MB, demonstrating the algorithm’s ability to operate efficiently within the limited resources of an embedded system. These results highlight the algorithm’s robustness and adaptability, enabling real-time speed analysis across both high-performance and embedded platforms.

}

\subsection{Case Study}

In this section, we present a case study on three applications of human walking speed: activity recognition, fall detection, and gait recognition,
to showcase various \sysname-enabled applications.

\head{Human Activity Recognition}
Speed profiles act as the core of human activity recognition \cite{raeis2021human,ouyang2022cosmo,shen2018virtual, ferlini2021eargate, xu2023practically}.
As shown in \fig\ref{fig:acf_har}, the speed profiles (the dashed lines) generated by \sysname can effectively distinguish different daily activities such as standing up, sitting down, and picking up objects. 
For instance, the speed profile of standing up shows a slower increase compared to sitting down. Picking up an object, however, forms a 'valley' in the profile, reflecting the bending and rising motion. 
\fig\ref{fig:har_acf_combined} provides an example of recognizing different activities through continuous monitoring. 
We keep it as a future direction to recognize different activities with \sysname deployed on commodity smart speakers. 

\head{Fall Detection} Fall detection is a vital yet challenging task. 
Passive speed estimation provides an effective way to detect falls \cite{hu2020wifi, zheng2019zero, ji2022sifall, lianFallDetectionInaudible2021}, mainly because falls induce a distinct speed pattern compared to normal daily activities. 
As shown in \fig\ref{fig:acf_har} and \fig\ref{fig:har_acf_combined}, a fall induces a unique pattern in the ACF matrix. The speed changes rapidly from a lower rate to a significantly higher rate within a brief period, yielding a large instantaneous acceleration. 
Following speed suggests further movement on the ground, before eventually decreasing to zero.

Based on these observations, we employ deep learning for proof-of-concept fall detection. As the ACF matrix contains speed information, we leverage pretrained MobileNet-v2 \cite{sandler2018mobilenetv2} as the ACF matrix encoder to classify the fall events. 
To increase data availability, we develop a recording platform that connects multiple off-the-shelf audio devices to a server via MQTT \cite{light2017mosquitto}. 
In total, we have 305 samples, with 175 representing fall events. We achieve the F-1 score of 0.92, demonstrating the effectiveness of \sysname for feature extractions.

\head{Gait Analysis} 
Gait analysis is important in healthcare, especially for patients with Parkinson's disease \cite{akhtaruzzaman2016gait, xuAcousticIDGaitbasedHuman2019,chiang20203d}. 
Intuitively, the walking speeds estimated by \sysname allow us to perform gait analysis. 
As a proof-of-concept, we recognize gait steps and cycles by detecting the peak in the speed profile, as shown in \fig\ref{fig:gait}. 
Our analysis of 10 traces reports an average gait cycle time of 1.00s with a variance of 1.2ms, demonstrating the potential of \sysname for gait applications.

\section{Discussions}
\label{sec:limitation}

We conclude some discussions and limitations for further exploration.

\head{Multiple Speeds} \sysname assumes the human body shares the same speed $v$ for walk speed estimation. The rationale behind our assumption is that we target room-scale sensing with relatively large coverage of several meters. In our proposed cases, the body speed dominates limb speeds. Moreover, a similar assumption is also 
 implicitly adopted by many other works \cite{zhang2023addressing, li2022room, niu2022rethinking}, \ie, hand-gesture works usually neglect body motions. As the next step, we are actively exploring multi-speed \sysname. Potentially, \eqn\eqref{eqn: multi_speed} suggests the potential to decompose the ACFs to account for varying speeds, which indicates an opportunity for future research to refine speed estimation by considering multiple speed scenarios.

\head{Multi-target Scenario} Current \sysname is not designed for multi-person scenarios. In most cases, indoor walking speed downstream tasks, such as fall detection and gait monitoring, involve a single person. Audio devices enable interaction in real-time and alert when abnormal conditions are detected. Therefore, \sysname can already foster many applications. Yet it is also potentially extend it to multi-person sensing in the future, \eg, by separating multipath signals for different ranges.  We also look forward to future work addressing the issue of multi-person movement.

\head{Multiple Speakers} In \sysname, we use OTDM with a single speaker and two channels, offering practical room-scale sensing. While this setup is common, OTDM can be extended to multiple speakers with advanced modulation scheme. Future work can explore OTDM+OFDM design for higher CSI rates.

\rev{
\head{Wi-Fi Speed Estimation} Numerous studies have utilized Wi-Fi for velocity capture \cite{wuGaitWayMonitoringRecognizing2021a, niu2022rethinking, li2024wifi}. However, the acoustic method offers several distinct advantages. 
Typically, commodity acoustic devices come with a co-located microphone and speaker, allowing sensing with a single device, \eg, Amazon Echo Dot, without any hardware/firmware changes. 
In contrast, Wi-Fi sensing generally necessitates two separate transceivers and relies on specific chipsets (e.g., Intel 5300, Qualcomm Atheros NICs) that require firmware adjustments.
In addition, acoustic devices bring side benefits by inherently making a voice interface available \cite{xu2022hearing,wang2019voicepop,zhou2021inferring, zhang2016dopenc}. For instance, besides running acoustic fall detection, a smart speaker can facilitate emergency calls directly. Therefore, it is imperative to enable accurate speed estimation for acoustic sensing. This work is orthogonal to existing Wi-Fi-based approaches.
}

\rev{
\head{Phase-based Estimation} Recent studies \cite{zhang2023addressing, li2024wifi} have employed phase difference to address the challenges posed by high speed scenarios. They implicitly acquire speed by calculating acceleration or displacement. In contrast, our proposed framework, OTDM plus sound diffusion model, directly estimates speed without the need for further transformations. While our current approach utilizes channel amplitude, it is promising to extend this design to incorporate phase-based speed estimation, which we intend to explore in future work.
}

\section{Related Works}
\label{sec:related_works}
\head{Acoustic Sensing}
Acoustic sensing has been extensively explored recently
and realize multiple applications, including indoor localization \cite{lianEchoSpotSpottingYour2021,gao2022mom,maoDeepRangeAcousticRanging2020,mao2017indoor}, fine-grained gesture tracking \cite{wangDevicefreeGestureTracking2016,nandakumarFingerIOUsingActive2016,maoCATHighprecisionAcoustic2016,yunTurningMobileDevice2015,sunVSkinSensingTouch2018,wangMilliSonicPushingLimits2019,zhang2023addressing,li2022room,yang2023biocase}, vital signs monitoring \cite{wangCFMCWBasedContactless2018,liLASensePushingLimits2022,qianAcousticcardiogramMonitoringHeartbeats2018,wanRespTrackerMultiuserRoomscale2021,wangContactlessInfantMonitoring2019a, wang2023df, kim2023low}, silent speech enhancement \cite{sunUltraSESinglechannelSpeech2021a,fu2022svoice, yu2025uspeech},  sound source localization \cite{shenVoiceLocalizationUsing2020, wang2021mavl, fan2023towards, veluri2024look, veluri2023real}, and communication \cite{zhang2023acoustic,chen2022underwater, tian2025aquascope}. 
The latest work VeCare \cite{zhang2023vecare} introduces statistical acoustic sensing for presence detection by extrapolating statistical WiFi sensing. 
\sysname significantly differs from VeCare in multiple important aspects: 
1) VeCare neither constructs the comprehensive model nor considers the fundamental differences between acoustic and electromagnetic waves. Conversely, \sysname integrates a novel sound diffusion model with the acoustic channel for speed estimation for the first time. 
2) VeCare mainly targets motion detection and breathing rate estimation, but does not address speed estimation (despite limited preliminary explorations). Differently, \sysname not only establishes the theoretical model but also builds practical techniques for location-independent and large-coverage speed estimation, achieving remarkably better performance. 
3) Additionally, we propose the first-of-its-kind OTDM design to increase the CSI rate, enabling previously challenging high speed estimation in a unique manner. 
Other than the key differences, we believe \sysname complements VeCare towards a comprehensive framework for motion detection, vital sign monitoring, and speed estimation, in a distinct paradigm differing from the prevalent literature in acoustic sensing.

\head{Speed Estimation}
Speed is vital information in human sensing, and has been exploited in various applications like fall detection \cite{huWiFiBasedPassiveFall2020,palipanaFallDeFiUbiquitousFall2018,wangWiFallDeviceFreeFall2017,lianFallDetectionInaudible2021}, interactive games \cite{qian2017inferring}, gait monitoring \cite{umairbinaltafAcousticGaitsGait2015,wuGaitWayMonitoringRecognizing2021a, xuAcousticIDGaitbasedHuman2019}, and localization \cite{qian2018widar2,mao2017indoor}. 
Speed estimation, however, is a challenging and enduring task. Camera-based approaches, such as VICON \cite{ViconAwardWinning}, require professional devices and specialized calibration, which cannot be used in ubiquitous settings. 
In wireless sensing, speed is generally derived using DFS, regardless of Wi-Fi \cite{qian2018widar2,xiong2013arraytrack,sun2015widraw}, acoustic \cite{yunTurningMobileDevice2015,maoCATHighprecisionAcoustic2016, sunVSkinSensingTouch2018,zhang2023addressing,maoCATHighprecisionAcoustic2016, shenVoiceLocalizationUsing2020,wang2021mavl}, or mmWave radar \cite{blanco2022augmenting,wu2020mmtrack} signals. \rev{Recently, some work \cite{zhang2023addressing, li2024wifi} also acknowledge the problems of low CSI rate and have proposed using phase difference to estimate acceleration or displacement, indirectly inferring speed. However, they do not jump out of the DFS framework, which inherently limits observations to radial speed components. Moreover, inferring speed from displacement or acceleration introduces additional transformations, thereby increasing cumulative error \cite{maoCATHighprecisionAcoustic2016}. In contrast, \sysname introduces a novel and explicit speed estimation framework that statistically aggregates all paths by modeling the sound diffusion field, thereby overcoming these limitations.}
Despite the advances in statistical wireless sensing \cite{zhangWiSpeedStatisticalElectromagnetic2018,wuGaitWayMonitoringRecognizing2021a}, sound waves and EM waves differ in nature, and acoustic speed estimation incurs different challenges.
 We build the acoustic diffusion speed estimation model plus OTDM design, a new way of acoustic speed estimation.

\section{Conclusions}
\label{sec:conclusion}

In this paper, we present \sysname, an end-to-end system for acoustic speed estimation. We identify two research problems in the acoustic speed estimation: insufficient CSI rate to capture the speed and the oversimplified signal model that fails to capture the entire speed. To this end, we propose a novel OTDM scheme to achieve a higher CSI rate for speed estimation. Inspired by the sound diffusion field, we establish a comprehensive theoretical model that enables full speed estimation from the spatial correlation of sounds. 
We extensively evaluate \sysname on commodity devices, which achieves an average accuracy of 13 cm/s for normal walking speed estimation. 
Overall, the proposed \sysname goes beyond the DFS-based paradigm and can open up new directions in acoustic sensing.

\begin{acks}
We sincerely thank all the anonymous reviewers for their valuable feedback throughout the submission process. We also wish to express our gratitude to the staff members at HKU who generously provided the initial space for our experiments. In addition, we are grateful to the lab members who assisted with proofreading the manuscript. 
This work is supported in part by NSFC under Grant No. 62222216, Hong Kong RGC GRF under Grant No. 17212224 and Healthy Longevity Catalyst Awards under Grant No. HLCA/E-712/22.
\end{acks}

\normalem
\bibliographystyle{ACM-Reference-Format}
\bibliography{refs}


\begin{thebibliography}{100}


\ifx \showCODEN    \undefined \def \showCODEN     #1{\unskip}     \fi
\ifx \showISBNx    \undefined \def \showISBNx     #1{\unskip}     \fi
\ifx \showISBNxiii \undefined \def \showISBNxiii  #1{\unskip}     \fi
\ifx \showISSN     \undefined \def \showISSN      #1{\unskip}     \fi
\ifx \showLCCN     \undefined \def \showLCCN      #1{\unskip}     \fi
\ifx \shownote     \undefined \def \shownote      #1{#1}          \fi
\ifx \showarticletitle \undefined \def \showarticletitle #1{#1}   \fi
\ifx \showURL      \undefined \def \showURL       {\relax}        \fi
\providecommand\bibfield[2]{#2}
\providecommand\bibinfo[2]{#2}
\providecommand\natexlab[1]{#1}
\providecommand\showeprint[2][]{arXiv:#2}

\bibitem[X4M(2021)]%
        {X4M03LaonuriCom2021}
 \bibinfo{year}{2021}\natexlab{}.
\newblock \bibinfo{title}{{X4M03} – laonuri.com}.
\newblock \bibinfo{howpublished}{\url{https://www.laonuri.com/product/x4m03/}}.
\newblock
\urldef\tempurl%
\url{https://www.laonuri.com/product/x4m03/}
\showURL{%
\tempurl}


\bibitem[Spe(2022)]%
        {SpeakersReceiversAS05308ASR}
 \bibinfo{year}{2022}\natexlab{}.
\newblock \bibinfo{title}{Speakers \& {Receivers} {\textbar} {AS05308AS}-{R}}.
\newblock \bibinfo{howpublished}{\url{https://puiaudio.com/product/speakers-and-receivers/AS05308AS-R}}.
\newblock
\urldef\tempurl%
\url{https://puiaudio.com/product/speakers-and-receivers/AS05308AS-R}
\showURL{%
\tempurl}


\bibitem[IWR(2023)]%
        {IWR1843DataSheet2023}
 \bibinfo{year}{2023}\natexlab{}.
\newblock \bibinfo{title}{{IWR1843} data sheet, product information and support {\textbar} {TI}.com}.
\newblock \bibinfo{howpublished}{\url{https://www.ti.com/product/IWR1843}}.
\newblock
\urldef\tempurl%
\url{https://www.ti.com/product/IWR1843}
\showURL{%
\tempurl}


\bibitem[Mot(2023)]%
        {MotionCaptureSystems}
 \bibinfo{year}{2023}\natexlab{}.
\newblock \bibinfo{title}{Motion {Capture} {Systems}}.
\newblock \bibinfo{howpublished}{\url{http://optitrack.com/index.html}}.
\newblock
\urldef\tempurl%
\url{http://optitrack.com/index.html}
\showURL{%
\tempurl}


\bibitem[USB(2023)]%
        {USBAudioStreaming}
 \bibinfo{year}{2023}\natexlab{}.
\newblock \bibinfo{title}{{USB} {Audio} {Streaming}: {UMA}-8-{SP} {USB} mic array}.
\newblock \bibinfo{howpublished}{\url{https://www.minidsp.com/products/usb-audio-interface/uma-8-sp-detail}}.
\newblock
\urldef\tempurl%
\url{https://www.minidsp.com/products/usb-audio-interface/uma-8-sp-detail}
\showURL{%
\tempurl}


\bibitem[Vic(2023)]%
        {ViconAwardWinning}
 \bibinfo{year}{2023}\natexlab{}.
\newblock \bibinfo{title}{Vicon {\textbar} {Award} {Winning} {Motion} {Capture} {Systems}}.
\newblock \bibinfo{howpublished}{\url{https://www.vicon.com/}}.
\newblock
\urldef\tempurl%
\url{https://www.vicon.com/}
\showURL{%
\tempurl}


\bibitem[Akhtaruzzaman et~al\mbox{.}(2016)]%
        {akhtaruzzaman2016gait}
\bibfield{author}{\bibinfo{person}{MD Akhtaruzzaman}, \bibinfo{person}{Amir~Akramin Shafie}, {and} \bibinfo{person}{Md~Raisuddin Khan}.} \bibinfo{year}{2016}\natexlab{}.
\newblock \showarticletitle{Gait analysis: Systems, technologies, and importance}.
\newblock \bibinfo{journal}{\emph{Journal of Mechanics in Medicine and Biology}} \bibinfo{volume}{16}, \bibinfo{number}{07} (\bibinfo{year}{2016}), \bibinfo{pages}{1630003}.
\newblock


\bibitem[Blanco et~al\mbox{.}(2022)]%
        {blanco2022augmenting}
\bibfield{author}{\bibinfo{person}{Alejandro Blanco}, \bibinfo{person}{Pablo~Jim{\'e}nez Mateo}, \bibinfo{person}{Francesco Gringoli}, {and} \bibinfo{person}{Joerg Widmer}.} \bibinfo{year}{2022}\natexlab{}.
\newblock \showarticletitle{Augmenting mmWave localization accuracy through sub-6 GHz on off-the-shelf devices}. In \bibinfo{booktitle}{\emph{Proceedings of the 20th Annual International Conference on Mobile Systems, Applications and Services}}. \bibinfo{pages}{477--490}.
\newblock


\bibitem[Cai et~al\mbox{.}(2022)]%
        {caiUbiquitousAcousticSensing2019}
\bibfield{author}{\bibinfo{person}{Chao Cai}, \bibinfo{person}{Rong Zheng}, {and} \bibinfo{person}{Jun Luo}.} \bibinfo{year}{2022}\natexlab{}.
\newblock \showarticletitle{Ubiquitous acoustic sensing on commodity iot devices: A survey}.
\newblock \bibinfo{journal}{\emph{IEEE Communications Surveys \& Tutorials}} \bibinfo{volume}{24}, \bibinfo{number}{1} (\bibinfo{year}{2022}), \bibinfo{pages}{432--454}.
\newblock


\bibitem[Chen et~al\mbox{.}(2017)]%
        {chenEchoTrackAcousticDevicefree2017}
\bibfield{author}{\bibinfo{person}{Huijie Chen}, \bibinfo{person}{Fan Li}, {and} \bibinfo{person}{Yu Wang}.} \bibinfo{year}{2017}\natexlab{}.
\newblock \showarticletitle{{EchoTrack}: {Acoustic} device-free hand tracking on smart phones}. In \bibinfo{booktitle}{\emph{{IEEE} {INFOCOM} 2017 - {IEEE} {Conference} on {Computer} {Communications}}}. \bibinfo{pages}{1--9}.
\newblock
\href{https://doi.org/10.1109/INFOCOM.2017.8057101}{doi:\nolinkurl{10.1109/INFOCOM.2017.8057101}}


\bibitem[Chen et~al\mbox{.}(2022)]%
        {chen2022underwater}
\bibfield{author}{\bibinfo{person}{Tuochao Chen}, \bibinfo{person}{Justin Chan}, {and} \bibinfo{person}{Shyamnath Gollakota}.} \bibinfo{year}{2022}\natexlab{}.
\newblock \showarticletitle{Underwater messaging using mobile devices}. In \bibinfo{booktitle}{\emph{Proceedings of the ACM SIGCOMM 2022 Conference}}. \bibinfo{pages}{545--559}.
\newblock


\bibitem[Cheng and Lou(2021)]%
        {chengPushLimitDeviceFree2021}
\bibfield{author}{\bibinfo{person}{Haiming Cheng} {and} \bibinfo{person}{Wei Lou}.} \bibinfo{year}{2021}\natexlab{}.
\newblock \showarticletitle{Push the {Limit} of {Device}-{Free} {Acoustic} {Sensing} on {Commercial} {Mobile} {Devices}}. In \bibinfo{booktitle}{\emph{{IEEE} {INFOCOM} 2021 - {IEEE} {Conference} on {Computer} {Communications}}}. \bibinfo{pages}{1--10}.
\newblock
\href{https://doi.org/10.1109/INFOCOM42981.2021.9488703}{doi:\nolinkurl{10.1109/INFOCOM42981.2021.9488703}}
\newblock
\shownote{ISSN: 2641-9874}.


\bibitem[Chiang et~al\mbox{.}(2020)]%
        {chiang20203d}
\bibfield{author}{\bibinfo{person}{Ting-Hui Chiang}, \bibinfo{person}{Yi-Juan Su}, \bibinfo{person}{Huan-Ruei Shiu}, {and} \bibinfo{person}{Yu-Chee Tseng}.} \bibinfo{year}{2020}\natexlab{}.
\newblock \showarticletitle{3D Gait Tracking by Acoustic Doppler Effects}. In \bibinfo{booktitle}{\emph{2020 42nd Annual International Conference of the IEEE Engineering in Medicine \& Biology Society (EMBC)}}. IEEE, \bibinfo{pages}{3146--3149}.
\newblock


\bibitem[Faiz et~al\mbox{.}(2012)]%
        {faiz2012measurement}
\bibfield{author}{\bibinfo{person}{Adil Faiz}, \bibinfo{person}{Jo{\"e}l Ducourneau}, \bibinfo{person}{Adel Khanfir}, {and} \bibinfo{person}{Jacques Chatillon}.} \bibinfo{year}{2012}\natexlab{}.
\newblock \showarticletitle{Measurement of sound diffusion coefficients of scattering furnishing volumes present in workplaces}. In \bibinfo{booktitle}{\emph{Acoustics 2012}}.
\newblock


\bibitem[Fan et~al\mbox{.}(2023)]%
        {fan2023towards}
\bibfield{author}{\bibinfo{person}{Tingchao Fan}, \bibinfo{person}{Huangwei Wu}, \bibinfo{person}{Meng Jin}, \bibinfo{person}{Tao Chen}, \bibinfo{person}{Longfei Shangguan}, \bibinfo{person}{Xinbing Wang}, {and} \bibinfo{person}{Chenghu Zhou}.} \bibinfo{year}{2023}\natexlab{}.
\newblock \showarticletitle{Towards Spatial Selection Transmission for Low-end IoT devices with SpotSound}. In \bibinfo{booktitle}{\emph{Proceedings of the 29th Annual International Conference on Mobile Computing and Networking}}.
\newblock


\bibitem[Ferlini et~al\mbox{.}(2021)]%
        {ferlini2021eargate}
\bibfield{author}{\bibinfo{person}{Andrea Ferlini}, \bibinfo{person}{Dong Ma}, \bibinfo{person}{Robert Harle}, {and} \bibinfo{person}{Cecilia Mascolo}.} \bibinfo{year}{2021}\natexlab{}.
\newblock \showarticletitle{EarGate: gait-based user identification with in-ear microphones}. In \bibinfo{booktitle}{\emph{Proceedings of the 27th Annual International Conference on Mobile Computing and Networking}}. \bibinfo{pages}{337--349}.
\newblock


\bibitem[Fritz and Lusardi(2009)]%
        {fritz2009white}
\bibfield{author}{\bibinfo{person}{Stacy Fritz} {and} \bibinfo{person}{Michelle Lusardi}.} \bibinfo{year}{2009}\natexlab{}.
\newblock \showarticletitle{White paper:“walking speed: the sixth vital sign”}.
\newblock \bibinfo{journal}{\emph{Journal of geriatric physical therapy}} \bibinfo{volume}{32}, \bibinfo{number}{2} (\bibinfo{year}{2009}), \bibinfo{pages}{2--5}.
\newblock


\bibitem[Fu et~al\mbox{.}(2022)]%
        {fu2022svoice}
\bibfield{author}{\bibinfo{person}{Yongjian Fu}, \bibinfo{person}{Shuning Wang}, \bibinfo{person}{Linghui Zhong}, \bibinfo{person}{Lili Chen}, \bibinfo{person}{Ju Ren}, {and} \bibinfo{person}{Yaoxue Zhang}.} \bibinfo{year}{2022}\natexlab{}.
\newblock \showarticletitle{SVoice: Enabling Voice Communication in Silence via Acoustic Sensing on Commodity Devices}. In \bibinfo{booktitle}{\emph{Proceedings of the 20th ACM Conference on Embedded Networked Sensor Systems}}. \bibinfo{pages}{622--636}.
\newblock


\bibitem[Gao et~al\mbox{.}(2022)]%
        {gao2022mom}
\bibfield{author}{\bibinfo{person}{Zhihui Gao}, \bibinfo{person}{Ang Li}, \bibinfo{person}{Dong Li}, \bibinfo{person}{Jialin Liu}, \bibinfo{person}{Jie Xiong}, \bibinfo{person}{Yu Wang}, \bibinfo{person}{Bing Li}, {and} \bibinfo{person}{Yiran Chen}.} \bibinfo{year}{2022}\natexlab{}.
\newblock \showarticletitle{Mom: Microphone based 3d orientation measurement}. In \bibinfo{booktitle}{\emph{2022 21st ACM/IEEE International Conference on Information Processing in Sensor Networks (IPSN)}}. IEEE, \bibinfo{pages}{132--144}.
\newblock


\bibitem[Ghosh et~al\mbox{.}(2019)]%
        {ghosh2019ultrasense}
\bibfield{author}{\bibinfo{person}{Arindam Ghosh}, \bibinfo{person}{Amartya Chakraborty}, \bibinfo{person}{Dhruv Chakraborty}, \bibinfo{person}{Mousumi Saha}, {and} \bibinfo{person}{Sujoy Saha}.} \bibinfo{year}{2019}\natexlab{}.
\newblock \showarticletitle{UltraSense: A non-intrusive approach for human activity identification using heterogeneous ultrasonic sensor grid for smart home environment}.
\newblock \bibinfo{journal}{\emph{Journal of Ambient Intelligence and Humanized Computing}} (\bibinfo{year}{2019}), \bibinfo{pages}{1--22}.
\newblock


\bibitem[Gong et~al\mbox{.}(2022)]%
        {gongBreathMentorAcousticbasedDiaphragmatic2022}
\bibfield{author}{\bibinfo{person}{Yanbin Gong}, \bibinfo{person}{Qian Zhang}, \bibinfo{person}{Bobby~H.P. NG}, {and} \bibinfo{person}{Wei Li}.} \bibinfo{year}{2022}\natexlab{}.
\newblock \showarticletitle{{BreathMentor}: {Acoustic}-based {Diaphragmatic} {Breathing} {Monitor} {System}}.
\newblock \bibinfo{journal}{\emph{Proceedings of the ACM on Interactive, Mobile, Wearable and Ubiquitous Technologies}} \bibinfo{volume}{6}, \bibinfo{number}{2} (\bibinfo{date}{July} \bibinfo{year}{2022}), \bibinfo{pages}{53:1--53:28}.
\newblock
\href{https://doi.org/10.1145/3534595}{doi:\nolinkurl{10.1145/3534595}}


\bibitem[Hu et~al\mbox{.}(2020a)]%
        {hu2020wifi}
\bibfield{author}{\bibinfo{person}{Yuqian Hu}, \bibinfo{person}{Feng Zhang}, \bibinfo{person}{Chenshu Wu}, \bibinfo{person}{Beibei Wang}, {and} \bibinfo{person}{KJ~Ray Liu}.} \bibinfo{year}{2020}\natexlab{a}.
\newblock \showarticletitle{A WiFi-based passive fall detection system}. In \bibinfo{booktitle}{\emph{ICASSP 2020-2020 IEEE International Conference on Acoustics, Speech and Signal Processing (ICASSP)}}. IEEE, \bibinfo{pages}{1723--1727}.
\newblock


\bibitem[Hu et~al\mbox{.}(2021)]%
        {hu2021defall}
\bibfield{author}{\bibinfo{person}{Yuqian Hu}, \bibinfo{person}{Feng Zhang}, \bibinfo{person}{Chenshu Wu}, \bibinfo{person}{Beibei Wang}, {and} \bibinfo{person}{KJ~Ray Liu}.} \bibinfo{year}{2021}\natexlab{}.
\newblock \showarticletitle{DeFall: Environment-independent passive fall detection using WiFi}.
\newblock \bibinfo{journal}{\emph{IEEE Internet of Things Journal}} \bibinfo{volume}{9}, \bibinfo{number}{11} (\bibinfo{year}{2021}), \bibinfo{pages}{8515--8530}.
\newblock


\bibitem[Hu et~al\mbox{.}(2020b)]%
        {huWiFiBasedPassiveFall2020}
\bibfield{author}{\bibinfo{person}{Yuqian Hu}, \bibinfo{person}{Feng Zhang}, \bibinfo{person}{Chenshu Wu}, \bibinfo{person}{Beibei Wang}, {and} \bibinfo{person}{K.~J. Ray~Liu}.} \bibinfo{year}{2020}\natexlab{b}.
\newblock \showarticletitle{A {WiFi}-{Based} {Passive} {Fall} {Detection} {System}}. In \bibinfo{booktitle}{\emph{{ICASSP} 2020 - 2020 {IEEE} {International} {Conference} on {Acoustics}, {Speech} and {Signal} {Processing} ({ICASSP})}}. \bibinfo{publisher}{IEEE}, \bibinfo{address}{Barcelona, Spain}, \bibinfo{pages}{1723--1727}.
\newblock
\showISBNx{978-1-5090-6631-5}
\href{https://doi.org/10.1109/ICASSP40776.2020.9054753}{doi:\nolinkurl{10.1109/ICASSP40776.2020.9054753}}


\bibitem[Ji et~al\mbox{.}(2022)]%
        {ji2022sifall}
\bibfield{author}{\bibinfo{person}{Sijie Ji}, \bibinfo{person}{Yaxiong Xie}, {and} \bibinfo{person}{Mo Li}.} \bibinfo{year}{2022}\natexlab{}.
\newblock \showarticletitle{SiFall: Practical Online Fall Detection with RF Sensing}. In \bibinfo{booktitle}{\emph{Proceedings of the 20th ACM Conference on Embedded Networked Sensor Systems}}. \bibinfo{pages}{563--577}.
\newblock


\bibitem[Kasami(1966)]%
        {kasamiWEIGHTDISTRIBUTIONFORMULA1966}
\bibfield{author}{\bibinfo{person}{Tadao Kasami}.} \bibinfo{year}{1966}\natexlab{}.
\newblock \bibinfo{booktitle}{\emph{{WEIGHT} {DISTRIBUTION} {FORMULA} {FOR} {SOME} {CLASS} {OF} {CYCLIC} {CODES}:}}.
\newblock \bibinfo{type}{{T}echnical {R}eport}. \bibinfo{institution}{Defense Technical Information Center}, \bibinfo{address}{Fort Belvoir, VA}.
\newblock
\href{https://doi.org/10.21236/AD0632574}{doi:\nolinkurl{10.21236/AD0632574}}


\bibitem[Kim et~al\mbox{.}(2023)]%
        {kim2023low}
\bibfield{author}{\bibinfo{person}{Maruchi Kim}, \bibinfo{person}{Anran Wang}, \bibinfo{person}{Srdjan Jelacic}, \bibinfo{person}{Andrew Bowdle}, \bibinfo{person}{Shyamnath Gollakota}, {and} \bibinfo{person}{Kelly Michaelsen}.} \bibinfo{year}{2023}\natexlab{}.
\newblock \showarticletitle{A Low-power wearable acoustic device for accurate invasive arterial pressure monitoring}.
\newblock \bibinfo{journal}{\emph{Communications Medicine}} \bibinfo{volume}{3}, \bibinfo{number}{1} (\bibinfo{year}{2023}), \bibinfo{pages}{70}.
\newblock


\bibitem[Kuttruff(2000)]%
        {kuttruffRoomAcoustics2000}
\bibfield{author}{\bibinfo{person}{Heinrich Kuttruff}.} \bibinfo{year}{2000}\natexlab{}.
\newblock \showarticletitle{Room acoustics}.
\newblock  (\bibinfo{year}{2000}).
\newblock
\newblock
\shownote{Publisher: Spon press UK}.


\bibitem[Leighton and Sands(1965)]%
        {leightonFeynmanLecturesPhysics1965}
\bibfield{author}{\bibinfo{person}{Robert~B Leighton} {and} \bibinfo{person}{Matthew Sands}.} \bibinfo{year}{1965}\natexlab{}.
\newblock \bibinfo{booktitle}{\emph{The {Feynman} lectures on physics}}.
\newblock \bibinfo{publisher}{Addison-Wesley Boston, MA, USA}.
\newblock
\showISBNx{0-201-02014-9}


\bibitem[Li et~al\mbox{.}(2022a)]%
        {li2022experience}
\bibfield{author}{\bibinfo{person}{Dong Li}, \bibinfo{person}{Shirui Cao}, \bibinfo{person}{Sunghoon~Ivan Lee}, {and} \bibinfo{person}{Jie Xiong}.} \bibinfo{year}{2022}\natexlab{a}.
\newblock \showarticletitle{Experience: practical problems for acoustic sensing}. In \bibinfo{booktitle}{\emph{Proceedings of the 28th Annual International Conference on Mobile Computing And Networking}}. \bibinfo{pages}{381--390}.
\newblock


\bibitem[Li et~al\mbox{.}(2022b)]%
        {liLASensePushingLimits2022}
\bibfield{author}{\bibinfo{person}{Dong Li}, \bibinfo{person}{Jialin Liu}, \bibinfo{person}{Sunghoon~Ivan Lee}, {and} \bibinfo{person}{Jie Xiong}.} \bibinfo{year}{2022}\natexlab{b}.
\newblock \showarticletitle{{LASense}: {Pushing} the {Limits} of {Fine}-grained {Activity} {Sensing} {Using} {Acoustic} {Signals}}.
\newblock \bibinfo{journal}{\emph{Proceedings of the ACM on Interactive, Mobile, Wearable and Ubiquitous Technologies}} \bibinfo{volume}{6}, \bibinfo{number}{1} (\bibinfo{date}{March} \bibinfo{year}{2022}), \bibinfo{pages}{21:1--21:27}.
\newblock


\bibitem[Li et~al\mbox{.}(2022c)]%
        {li2022room}
\bibfield{author}{\bibinfo{person}{Dong Li}, \bibinfo{person}{Jialin Liu}, \bibinfo{person}{Sunghoon~Ivan Lee}, {and} \bibinfo{person}{Jie Xiong}.} \bibinfo{year}{2022}\natexlab{c}.
\newblock \showarticletitle{Room-Scale Hand Gesture Recognition Using Smart Speakers}. In \bibinfo{booktitle}{\emph{Proceedings of the 20th ACM Conference on Embedded Networked Sensor Systems}}. \bibinfo{pages}{462--475}.
\newblock


\bibitem[Li et~al\mbox{.}(2024)]%
        {li2024wifi}
\bibfield{author}{\bibinfo{person}{Wenwei Li}, \bibinfo{person}{Ruiyang Gao}, \bibinfo{person}{Jie Xiong}, \bibinfo{person}{Jiarun Zhou}, \bibinfo{person}{Leye Wang}, \bibinfo{person}{Xingjian Mao}, \bibinfo{person}{Enze Yi}, {and} \bibinfo{person}{Daqing Zhang}.} \bibinfo{year}{2024}\natexlab{}.
\newblock \showarticletitle{WiFi-CSI Difference Paradigm: Achieving Efficient Doppler Speed Estimation for Passive Tracking}.
\newblock \bibinfo{journal}{\emph{Proceedings of the ACM on Interactive, Mobile, Wearable and Ubiquitous Technologies}} \bibinfo{volume}{8}, \bibinfo{number}{2} (\bibinfo{year}{2024}), \bibinfo{pages}{1--29}.
\newblock


\bibitem[Lian et~al\mbox{.}(1 01)]%
        {lianEchoSpotSpottingYour2021}
\bibfield{author}{\bibinfo{person}{Jie Lian}, \bibinfo{person}{Jiadong Lou}, \bibinfo{person}{Li Chen}, {and} \bibinfo{person}{Xu Yuan}.} \bibinfo{year}{2021-01-01}\natexlab{}.
\newblock \showarticletitle{{EchoSpot}: {Spotting} {Your} {Locations} via {Acoustic} {Sensing}}.
\newblock \bibinfo{journal}{\emph{Proceedings of the ACM on Interactive, Mobile, Wearable and Ubiquitous Technologies}} \bibinfo{volume}{5}, \bibinfo{number}{3} (\bibinfo{date}{Sept.} \bibinfo{year}{2021-01-01}), \bibinfo{pages}{113:1--113:21}.
\newblock
\href{https://doi.org/10.1145/3478095}{doi:\nolinkurl{10.1145/3478095}}


\bibitem[Lian et~al\mbox{.}(2021)]%
        {lianFallDetectionInaudible2021}
\bibfield{author}{\bibinfo{person}{Jie Lian}, \bibinfo{person}{Xu Yuan}, \bibinfo{person}{Ming Li}, {and} \bibinfo{person}{Nian-Feng Tzeng}.} \bibinfo{year}{2021}\natexlab{}.
\newblock \showarticletitle{Fall {Detection} via {Inaudible} {Acoustic} {Sensing}}.
\newblock \bibinfo{journal}{\emph{Proceedings of the ACM on Interactive, Mobile, Wearable and Ubiquitous Technologies}} \bibinfo{volume}{5}, \bibinfo{number}{3} (\bibinfo{date}{Sept.} \bibinfo{year}{2021}), \bibinfo{pages}{1--21}.
\newblock
\showISSN{2474-9567}
\href{https://doi.org/10.1145/3478094}{doi:\nolinkurl{10.1145/3478094}}


\bibitem[Light(2017)]%
        {light2017mosquitto}
\bibfield{author}{\bibinfo{person}{Roger~A Light}.} \bibinfo{year}{2017}\natexlab{}.
\newblock \showarticletitle{Mosquitto: server and client implementation of the MQTT protocol}.
\newblock \bibinfo{journal}{\emph{Journal of Open Source Software}} \bibinfo{volume}{2}, \bibinfo{number}{13} (\bibinfo{year}{2017}), \bibinfo{pages}{265}.
\newblock


\bibitem[Mao et~al\mbox{.}(2016)]%
        {maoCATHighprecisionAcoustic2016}
\bibfield{author}{\bibinfo{person}{Wenguang Mao}, \bibinfo{person}{Jian He}, {and} \bibinfo{person}{Lili Qiu}.} \bibinfo{year}{2016}\natexlab{}.
\newblock \showarticletitle{{CAT}: high-precision acoustic motion tracking}. In \bibinfo{booktitle}{\emph{Proceedings of the 22nd {Annual} {International} {Conference} on {Mobile} {Computing} and {Networking}}}. \bibinfo{publisher}{ACM}, \bibinfo{address}{New York City New York}, \bibinfo{pages}{69--81}.
\newblock
\showISBNx{978-1-4503-4226-1}
\href{https://doi.org/10.1145/2973750.2973755}{doi:\nolinkurl{10.1145/2973750.2973755}}


\bibitem[Mao et~al\mbox{.}(2020)]%
        {maoDeepRangeAcousticRanging2020}
\bibfield{author}{\bibinfo{person}{Wenguang Mao}, \bibinfo{person}{Wei Sun}, \bibinfo{person}{Mei Wang}, {and} \bibinfo{person}{Lili Qiu}.} \bibinfo{year}{2020}\natexlab{}.
\newblock \showarticletitle{{DeepRange}: {Acoustic} {Ranging} via {Deep} {Learning}}.
\newblock \bibinfo{journal}{\emph{Proceedings of the ACM on Interactive, Mobile, Wearable and Ubiquitous Technologies}} \bibinfo{volume}{4}, \bibinfo{number}{4} (\bibinfo{date}{Dec.} \bibinfo{year}{2020}), \bibinfo{pages}{1--23}.
\newblock
\showISSN{2474-9567}
\href{https://doi.org/10.1145/3432195}{doi:\nolinkurl{10.1145/3432195}}


\bibitem[Mao et~al\mbox{.}(2018)]%
        {maoAIMAcousticImaging2018}
\bibfield{author}{\bibinfo{person}{Wenguang Mao}, \bibinfo{person}{Mei Wang}, {and} \bibinfo{person}{Lili Qiu}.} \bibinfo{year}{2018}\natexlab{}.
\newblock \showarticletitle{{AIM}: {Acoustic} {Imaging} on a {Mobile}}. In \bibinfo{booktitle}{\emph{Proceedings of the 16th {Annual} {International} {Conference} on {Mobile} {Systems}, {Applications}, and {Services}}}. \bibinfo{publisher}{ACM}, \bibinfo{address}{Munich Germany}, \bibinfo{pages}{468--481}.
\newblock
\showISBNx{978-1-4503-5720-3}
\href{https://doi.org/10.1145/3210240.3210325}{doi:\nolinkurl{10.1145/3210240.3210325}}


\bibitem[Mao et~al\mbox{.}(2017)]%
        {mao2017indoor}
\bibfield{author}{\bibinfo{person}{Wenguang Mao}, \bibinfo{person}{Zaiwei Zhang}, \bibinfo{person}{Lili Qiu}, \bibinfo{person}{Jian He}, \bibinfo{person}{Yuchen Cui}, {and} \bibinfo{person}{Sangki Yun}.} \bibinfo{year}{2017}\natexlab{}.
\newblock \showarticletitle{Indoor follow me drone}. In \bibinfo{booktitle}{\emph{Proceedings of the 15th annual international conference on mobile systems, applications, and services}}. \bibinfo{pages}{345--358}.
\newblock


\bibitem[Middleton et~al\mbox{.}(2015)]%
        {middleton2015walking}
\bibfield{author}{\bibinfo{person}{Addie Middleton}, \bibinfo{person}{Stacy~L Fritz}, {and} \bibinfo{person}{Michelle Lusardi}.} \bibinfo{year}{2015}\natexlab{}.
\newblock \showarticletitle{Walking speed: the functional vital sign}.
\newblock \bibinfo{journal}{\emph{Journal of aging and physical activity}} \bibinfo{volume}{23}, \bibinfo{number}{2} (\bibinfo{year}{2015}), \bibinfo{pages}{314--322}.
\newblock


\bibitem[Nandakumar et~al\mbox{.}(2016)]%
        {nandakumarFingerIOUsingActive2016}
\bibfield{author}{\bibinfo{person}{Rajalakshmi Nandakumar}, \bibinfo{person}{Vikram Iyer}, \bibinfo{person}{Desney Tan}, {and} \bibinfo{person}{Shyamnath Gollakota}.} \bibinfo{year}{2016}\natexlab{}.
\newblock \showarticletitle{{FingerIO}: {Using} {Active} {Sonar} for {Fine}-{Grained} {Finger} {Tracking}}. In \bibinfo{booktitle}{\emph{Proceedings of the 2016 {CHI} {Conference} on {Human} {Factors} in {Computing} {Systems}}}. \bibinfo{publisher}{ACM}, \bibinfo{address}{San Jose California USA}, \bibinfo{pages}{1515--1525}.
\newblock
\showISBNx{978-1-4503-3362-7}
\href{https://doi.org/10.1145/2858036.2858580}{doi:\nolinkurl{10.1145/2858036.2858580}}


\bibitem[Niu et~al\mbox{.}(2022)]%
        {niu2022rethinking}
\bibfield{author}{\bibinfo{person}{Kai Niu}, \bibinfo{person}{Xuanzhi Wang}, \bibinfo{person}{Fusang Zhang}, \bibinfo{person}{Rong Zheng}, \bibinfo{person}{Zhiyun Yao}, {and} \bibinfo{person}{Daqing Zhang}.} \bibinfo{year}{2022}\natexlab{}.
\newblock \showarticletitle{Rethinking Doppler effect for accurate velocity estimation with commodity WiFi devices}.
\newblock \bibinfo{journal}{\emph{IEEE Journal on Selected Areas in Communications}} \bibinfo{volume}{40}, \bibinfo{number}{7} (\bibinfo{year}{2022}), \bibinfo{pages}{2164--2178}.
\newblock


\bibitem[Ouyang et~al\mbox{.}(2022)]%
        {ouyang2022cosmo}
\bibfield{author}{\bibinfo{person}{Xiaomin Ouyang}, \bibinfo{person}{Xian Shuai}, \bibinfo{person}{Jiayu Zhou}, \bibinfo{person}{Ivy~Wang Shi}, \bibinfo{person}{Zhiyuan Xie}, \bibinfo{person}{Guoliang Xing}, {and} \bibinfo{person}{Jianwei Huang}.} \bibinfo{year}{2022}\natexlab{}.
\newblock \showarticletitle{Cosmo: contrastive fusion learning with small data for multimodal human activity recognition}. In \bibinfo{booktitle}{\emph{Proceedings of the 28th Annual International Conference on Mobile Computing And Networking}}. \bibinfo{pages}{324--337}.
\newblock


\bibitem[Palipana et~al\mbox{.}(2018)]%
        {palipanaFallDeFiUbiquitousFall2018}
\bibfield{author}{\bibinfo{person}{Sameera Palipana}, \bibinfo{person}{David Rojas}, \bibinfo{person}{Piyush Agrawal}, {and} \bibinfo{person}{Dirk Pesch}.} \bibinfo{year}{2018}\natexlab{}.
\newblock \showarticletitle{{FallDeFi}: {Ubiquitous} {Fall} {Detection} using {Commodity} {Wi}-{Fi} {Devices}}.
\newblock \bibinfo{journal}{\emph{Proceedings of the ACM on Interactive, Mobile, Wearable and Ubiquitous Technologies}} \bibinfo{volume}{1}, \bibinfo{number}{4} (\bibinfo{date}{Jan.} \bibinfo{year}{2018}), \bibinfo{pages}{1--25}.
\newblock
\showISSN{2474-9567}
\urldef\tempurl%
\url{https://dl.acm.org/doi/10.1145/3161183}
\showURL{%
\tempurl}


\bibitem[Pierce(2019)]%
        {pierceAcousticsIntroductionIts2019}
\bibfield{author}{\bibinfo{person}{Allan~D Pierce}.} \bibinfo{year}{2019}\natexlab{}.
\newblock \bibinfo{booktitle}{\emph{Acoustics: an introduction to its physical principles and applications}}.
\newblock \bibinfo{publisher}{Springer}.
\newblock


\bibitem[Qian et~al\mbox{.}(2018a)]%
        {qianAcousticcardiogramMonitoringHeartbeats2018}
\bibfield{author}{\bibinfo{person}{Kun Qian}, \bibinfo{person}{Chenshu Wu}, \bibinfo{person}{Fu Xiao}, \bibinfo{person}{Yue Zheng}, \bibinfo{person}{Yi Zhang}, \bibinfo{person}{Zheng Yang}, {and} \bibinfo{person}{Yunhao Liu}.} \bibinfo{year}{2018}\natexlab{a}.
\newblock \showarticletitle{Acousticcardiogram: {Monitoring} {Heartbeats} using {Acoustic} {Signals} on {Smart} {Devices}}. In \bibinfo{booktitle}{\emph{{IEEE} {INFOCOM} 2018 - {IEEE} {Conference} on {Computer} {Communications}}}. \bibinfo{publisher}{IEEE}, \bibinfo{address}{Honolulu, HI}, \bibinfo{pages}{1574--1582}.
\newblock
\showISBNx{978-1-5386-4128-6}
\href{https://doi.org/10.1109/INFOCOM.2018.8485978}{doi:\nolinkurl{10.1109/INFOCOM.2018.8485978}}


\bibitem[Qian et~al\mbox{.}(2018b)]%
        {qian2018widar2}
\bibfield{author}{\bibinfo{person}{Kun Qian}, \bibinfo{person}{Chenshu Wu}, \bibinfo{person}{Yi Zhang}, \bibinfo{person}{Guidong Zhang}, \bibinfo{person}{Zheng Yang}, {and} \bibinfo{person}{Yunhao Liu}.} \bibinfo{year}{2018}\natexlab{b}.
\newblock \showarticletitle{Widar2. 0: Passive human tracking with a single Wi-Fi link}. In \bibinfo{booktitle}{\emph{Proceedings of the 16th Annual International Conference on Mobile Systems, Applications, and Services}}. \bibinfo{pages}{350--361}.
\newblock


\bibitem[Qian et~al\mbox{.}(2017)]%
        {qian2017inferring}
\bibfield{author}{\bibinfo{person}{Kun Qian}, \bibinfo{person}{Chenshu Wu}, \bibinfo{person}{Zimu Zhou}, \bibinfo{person}{Yue Zheng}, \bibinfo{person}{Zheng Yang}, {and} \bibinfo{person}{Yunhao Liu}.} \bibinfo{year}{2017}\natexlab{}.
\newblock \showarticletitle{Inferring motion direction using commodity wi-fi for interactive exergames}. In \bibinfo{booktitle}{\emph{Proceedings of the 2017 CHI conference on human factors in computing systems}}. \bibinfo{pages}{1961--1972}.
\newblock


\bibitem[Raeis et~al\mbox{.}(2021)]%
        {raeis2021human}
\bibfield{author}{\bibinfo{person}{Hossein Raeis}, \bibinfo{person}{Mohammad Kazemi}, {and} \bibinfo{person}{Shervin Shirmohammadi}.} \bibinfo{year}{2021}\natexlab{}.
\newblock \showarticletitle{Human activity recognition with device-free sensors for well-being assessment in smart homes}.
\newblock \bibinfo{journal}{\emph{IEEE Instrumentation \& Measurement Magazine}} \bibinfo{volume}{24}, \bibinfo{number}{6} (\bibinfo{year}{2021}), \bibinfo{pages}{46--57}.
\newblock


\bibitem[Rasmussen et~al\mbox{.}(2019)]%
        {rasmussen2019association}
\bibfield{author}{\bibinfo{person}{Line Jee~Hartmann Rasmussen}, \bibinfo{person}{Avshalom Caspi}, \bibinfo{person}{Antony Ambler}, \bibinfo{person}{Jonathan~M Broadbent}, \bibinfo{person}{Harvey~J Cohen}, \bibinfo{person}{Tracy d’Arbeloff}, \bibinfo{person}{Maxwell Elliott}, \bibinfo{person}{Robert~J Hancox}, \bibinfo{person}{HonaLee Harrington}, \bibinfo{person}{Sean Hogan}, {et~al\mbox{.}}} \bibinfo{year}{2019}\natexlab{}.
\newblock \showarticletitle{Association of neurocognitive and physical function with gait speed in midlife}.
\newblock \bibinfo{journal}{\emph{JAMA network open}} \bibinfo{volume}{2}, \bibinfo{number}{10} (\bibinfo{year}{2019}), \bibinfo{pages}{e1913123--e1913123}.
\newblock


\bibitem[Rosso et~al\mbox{.}(2017)]%
        {rosso2017slowing}
\bibfield{author}{\bibinfo{person}{Andrea~L Rosso}, \bibinfo{person}{Joe Verghese}, \bibinfo{person}{Andrea~L Metti}, \bibinfo{person}{Robert~M Boudreau}, \bibinfo{person}{Howard~J Aizenstein}, \bibinfo{person}{Stephen Kritchevsky}, \bibinfo{person}{Tamara Harris}, \bibinfo{person}{Kristine Yaffe}, \bibinfo{person}{Suzanne Satterfield}, \bibinfo{person}{Stephanie Studenski}, {et~al\mbox{.}}} \bibinfo{year}{2017}\natexlab{}.
\newblock \showarticletitle{Slowing gait and risk for cognitive impairment: the hippocampus as a shared neural substrate}.
\newblock \bibinfo{journal}{\emph{Neurology}} \bibinfo{volume}{89}, \bibinfo{number}{4} (\bibinfo{year}{2017}), \bibinfo{pages}{336--342}.
\newblock


\bibitem[Sandler et~al\mbox{.}(2018)]%
        {sandler2018mobilenetv2}
\bibfield{author}{\bibinfo{person}{Mark Sandler}, \bibinfo{person}{Andrew Howard}, \bibinfo{person}{Menglong Zhu}, \bibinfo{person}{Andrey Zhmoginov}, {and} \bibinfo{person}{Liang-Chieh Chen}.} \bibinfo{year}{2018}\natexlab{}.
\newblock \showarticletitle{Mobilenetv2: Inverted residuals and linear bottlenecks}. In \bibinfo{booktitle}{\emph{Proceedings of the IEEE conference on computer vision and pattern recognition}}. \bibinfo{pages}{4510--4520}.
\newblock


\bibitem[Shen et~al\mbox{.}(2018)]%
        {shen2018virtual}
\bibfield{author}{\bibinfo{person}{Dakun Shen}, \bibinfo{person}{Ian Markwood}, \bibinfo{person}{Dan Shen}, {and} \bibinfo{person}{Yao Liu}.} \bibinfo{year}{2018}\natexlab{}.
\newblock \showarticletitle{Virtual safe: Unauthorized walking behavior detection for mobile devices}.
\newblock \bibinfo{journal}{\emph{IEEE Transactions on Mobile Computing}} \bibinfo{volume}{18}, \bibinfo{number}{3} (\bibinfo{year}{2018}), \bibinfo{pages}{688--701}.
\newblock


\bibitem[Shen et~al\mbox{.}(2020)]%
        {shenVoiceLocalizationUsing2020}
\bibfield{author}{\bibinfo{person}{Sheng Shen}, \bibinfo{person}{Daguan Chen}, \bibinfo{person}{Yu-Lin Wei}, \bibinfo{person}{Zhijian Yang}, {and} \bibinfo{person}{Romit~Roy Choudhury}.} \bibinfo{year}{2020}\natexlab{}.
\newblock \showarticletitle{Voice localization using nearby wall reflections}. In \bibinfo{booktitle}{\emph{Proceedings of the 26th {Annual} {International} {Conference} on {Mobile} {Computing} and {Networking}}}. \bibinfo{publisher}{ACM}, \bibinfo{address}{London United Kingdom}, \bibinfo{pages}{1--14}.
\newblock
\showISBNx{978-1-4503-7085-1}
\href{https://doi.org/10.1145/3372224.3380884}{doi:\nolinkurl{10.1145/3372224.3380884}}


\bibitem[Sun and Zhang(2021)]%
        {sunUltraSESinglechannelSpeech2021a}
\bibfield{author}{\bibinfo{person}{Ke Sun} {and} \bibinfo{person}{Xinyu Zhang}.} \bibinfo{year}{2021}\natexlab{}.
\newblock \showarticletitle{{UltraSE}: single-channel speech enhancement using ultrasound}. In \bibinfo{booktitle}{\emph{Proceedings of the 27th {Annual} {International} {Conference} on {Mobile} {Computing} and {Networking}}}. \bibinfo{publisher}{ACM}, \bibinfo{address}{New Orleans Louisiana}, \bibinfo{pages}{160--173}.
\newblock
\showISBNx{978-1-4503-8342-4}


\bibitem[Sun et~al\mbox{.}(2018)]%
        {sunVSkinSensingTouch2018}
\bibfield{author}{\bibinfo{person}{Ke Sun}, \bibinfo{person}{Ting Zhao}, \bibinfo{person}{Wei Wang}, {and} \bibinfo{person}{Lei Xie}.} \bibinfo{year}{2018}\natexlab{}.
\newblock \showarticletitle{{VSkin}: {Sensing} {Touch} {Gestures} on {Surfaces} of {Mobile} {Devices} {Using} {Acoustic} {Signals}}. In \bibinfo{booktitle}{\emph{Proceedings of the 24th {Annual} {International} {Conference} on {Mobile} {Computing} and {Networking}}}. \bibinfo{publisher}{ACM}, \bibinfo{address}{New Delhi India}, \bibinfo{pages}{591--605}.
\newblock
\showISBNx{978-1-4503-5903-0}
\href{https://doi.org/10.1145/3241539.3241568}{doi:\nolinkurl{10.1145/3241539.3241568}}


\bibitem[Sun et~al\mbox{.}(2015)]%
        {sun2015widraw}
\bibfield{author}{\bibinfo{person}{Li Sun}, \bibinfo{person}{Souvik Sen}, \bibinfo{person}{Dimitrios Koutsonikolas}, {and} \bibinfo{person}{Kyu-Han Kim}.} \bibinfo{year}{2015}\natexlab{}.
\newblock \showarticletitle{Widraw: Enabling hands-free drawing in the air on commodity wifi devices}. In \bibinfo{booktitle}{\emph{Proceedings of the 21st Annual International Conference on Mobile Computing and Networking}}. \bibinfo{pages}{77--89}.
\newblock


\bibitem[Tang and Yan(2017)]%
        {tang2017acoustic}
\bibfield{author}{\bibinfo{person}{Xiaoning Tang} {and} \bibinfo{person}{Xiong Yan}.} \bibinfo{year}{2017}\natexlab{}.
\newblock \showarticletitle{Acoustic energy absorption properties of fibrous materials: A review}.
\newblock \bibinfo{journal}{\emph{Composites Part A: Applied Science and Manufacturing}}  \bibinfo{volume}{101} (\bibinfo{year}{2017}), \bibinfo{pages}{360--380}.
\newblock


\bibitem[Tian et~al\mbox{.}(2025)]%
        {tian2025aquascope}
\bibfield{author}{\bibinfo{person}{Beitong Tian}, \bibinfo{person}{Lingzhi Zhao}, \bibinfo{person}{Bo Chen}, \bibinfo{person}{Mingyuan Wu}, \bibinfo{person}{Haozhen Zheng}, \bibinfo{person}{Deepak Vasisht}, \bibinfo{person}{Francis~Y Yan}, {and} \bibinfo{person}{Klara Nahrstedt}.} \bibinfo{year}{2025}\natexlab{}.
\newblock \showarticletitle{AquaScope: Reliable Underwater Image Transmission on Mobile Devices}.
\newblock \bibinfo{journal}{\emph{arXiv preprint arXiv:2502.10891}} (\bibinfo{year}{2025}).
\newblock


\bibitem[Training(2015)]%
        {ProSoundTraining2015}
\bibfield{author}{\bibinfo{person}{Pro~Sound Training}.} \bibinfo{year}{2015}\natexlab{}.
\newblock \bibinfo{title}{Acoustical Scattering}.
\newblock \bibinfo{howpublished}{\url{https://www.prosoundtraining.com/2015/04/03/acoustical-scattering/}}.
\newblock


\bibitem[Umair Bin~Altaf et~al\mbox{.}(2015)]%
        {umairbinaltafAcousticGaitsGait2015}
\bibfield{author}{\bibinfo{person}{M. Umair Bin~Altaf}, \bibinfo{person}{Taras Butko}, {and} \bibinfo{person}{Biing-Hwang Juang}.} \bibinfo{year}{2015}\natexlab{}.
\newblock \showarticletitle{Acoustic {Gaits}: {Gait} {Analysis} {With} {Footstep} {Sounds}}.
\newblock \bibinfo{journal}{\emph{IEEE Transactions on Biomedical Engineering}} \bibinfo{volume}{62}, \bibinfo{number}{8} (\bibinfo{date}{Aug.} \bibinfo{year}{2015}), \bibinfo{pages}{2001--2011}.
\newblock
\showISSN{1558-2531}
\href{https://doi.org/10.1109/TBME.2015.2410142}{doi:\nolinkurl{10.1109/TBME.2015.2410142}}


\bibitem[Veluri et~al\mbox{.}(2023)]%
        {veluri2023real}
\bibfield{author}{\bibinfo{person}{Bandhav Veluri}, \bibinfo{person}{Justin Chan}, \bibinfo{person}{Malek Itani}, \bibinfo{person}{Tuochao Chen}, \bibinfo{person}{Takuya Yoshioka}, {and} \bibinfo{person}{Shyamnath Gollakota}.} \bibinfo{year}{2023}\natexlab{}.
\newblock \showarticletitle{Real-time target sound extraction}. In \bibinfo{booktitle}{\emph{ICASSP 2023-2023 IEEE International Conference on Acoustics, Speech and Signal Processing (ICASSP)}}. IEEE, \bibinfo{pages}{1--5}.
\newblock


\bibitem[Veluri et~al\mbox{.}(2024)]%
        {veluri2024look}
\bibfield{author}{\bibinfo{person}{Bandhav Veluri}, \bibinfo{person}{Malek Itani}, \bibinfo{person}{Tuochao Chen}, \bibinfo{person}{Takuya Yoshioka}, {and} \bibinfo{person}{Shyamnath Gollakota}.} \bibinfo{year}{2024}\natexlab{}.
\newblock \showarticletitle{Look once to hear: Target speech hearing with noisy examples}. In \bibinfo{booktitle}{\emph{Proceedings of the 2024 CHI Conference on Human Factors in Computing Systems}}. \bibinfo{pages}{1--16}.
\newblock


\bibitem[Viikari et~al\mbox{.}(2008)]%
        {4686885}
\bibfield{author}{\bibinfo{person}{Ville Viikari}, \bibinfo{person}{Kimmo Kokkonen}, {and} \bibinfo{person}{Johanna Meltaus}.} \bibinfo{year}{2008}\natexlab{}.
\newblock \showarticletitle{Optimized signal processing for FMCW interrogated reflective delay line-type SAW sensors}.
\newblock \bibinfo{journal}{\emph{IEEE Transactions on Ultrasonics, Ferroelectrics, and Frequency Control}} \bibinfo{volume}{55}, \bibinfo{number}{11} (\bibinfo{year}{2008}), \bibinfo{pages}{2522--2526}.
\newblock
\href{https://doi.org/10.1109/TUFFC.961}{doi:\nolinkurl{10.1109/TUFFC.961}}


\bibitem[Wan et~al\mbox{.}(2021)]%
        {wanRespTrackerMultiuserRoomscale2021}
\bibfield{author}{\bibinfo{person}{Haoran Wan}, \bibinfo{person}{Shuyu Shi}, \bibinfo{person}{Wenyu Cao}, \bibinfo{person}{Wei Wang}, {and} \bibinfo{person}{Guihai Chen}.} \bibinfo{year}{2021}\natexlab{}.
\newblock \showarticletitle{{RespTracker}: {Multi}-user {Room}-scale {Respiration} {Tracking} with {Commercial} {Acoustic} {Devices}}. In \bibinfo{booktitle}{\emph{{IEEE} {INFOCOM} 2021 - {IEEE} {Conference} on {Computer} {Communications}}}. \bibinfo{publisher}{IEEE}, \bibinfo{address}{Vancouver, BC, Canada}, \bibinfo{pages}{1--10}.
\newblock
\showISBNx{978-1-66540-325-2}
\href{https://doi.org/10.1109/INFOCOM42981.2021.9488881}{doi:\nolinkurl{10.1109/INFOCOM42981.2021.9488881}}


\bibitem[Wang and Gollakota(2019)]%
        {wangMilliSonicPushingLimits2019}
\bibfield{author}{\bibinfo{person}{Anran Wang} {and} \bibinfo{person}{Shyamnath Gollakota}.} \bibinfo{year}{2019}\natexlab{}.
\newblock \showarticletitle{{MilliSonic}: {Pushing} the {Limits} of {Acoustic} {Motion} {Tracking}}. In \bibinfo{booktitle}{\emph{Proceedings of the 2019 {CHI} {Conference} on {Human} {Factors} in {Computing} {Systems}}}. \bibinfo{publisher}{ACM}, \bibinfo{address}{Glasgow Scotland Uk}, \bibinfo{pages}{1--11}.
\newblock
\showISBNx{978-1-4503-5970-2}
\href{https://doi.org/10.1145/3290605.3300248}{doi:\nolinkurl{10.1145/3290605.3300248}}


\bibitem[Wang et~al\mbox{.}(2019b)]%
        {wangContactlessInfantMonitoring2019a}
\bibfield{author}{\bibinfo{person}{Anran Wang}, \bibinfo{person}{Jacob~E. Sunshine}, {and} \bibinfo{person}{Shyamnath Gollakota}.} \bibinfo{year}{2019}\natexlab{b}.
\newblock \showarticletitle{Contactless {Infant} {Monitoring} using {White} {Noise}}. In \bibinfo{booktitle}{\emph{The 25th {Annual} {International} {Conference} on {Mobile} {Computing} and {Networking}}}. \bibinfo{publisher}{ACM}, \bibinfo{address}{Los Cabos Mexico}, \bibinfo{pages}{1--16}.
\newblock
\showISBNx{978-1-4503-6169-9}
\href{https://doi.org/10.1145/3300061.3345453}{doi:\nolinkurl{10.1145/3300061.3345453}}


\bibitem[Wang et~al\mbox{.}(2023)]%
        {wang2023df}
\bibfield{author}{\bibinfo{person}{Lei Wang}, \bibinfo{person}{Tao Gu}, \bibinfo{person}{Wei Li}, \bibinfo{person}{Haipeng Dai}, \bibinfo{person}{Yong Zhang}, \bibinfo{person}{Dongxiao Yu}, \bibinfo{person}{Chenren Xu}, {and} \bibinfo{person}{Daqing Zhang}.} \bibinfo{year}{2023}\natexlab{}.
\newblock \showarticletitle{DF-Sense: Multi-user Acoustic Sensing for Heartbeat Monitoring with Dualforming}. In \bibinfo{booktitle}{\emph{Proceedings of the 21st Annual International Conference on Mobile Systems, Applications and Services}}. \bibinfo{pages}{1--13}.
\newblock


\bibitem[Wang et~al\mbox{.}(2022)]%
        {wang2022loear}
\bibfield{author}{\bibinfo{person}{Lei Wang}, \bibinfo{person}{Wei Li}, \bibinfo{person}{Ke Sun}, \bibinfo{person}{Fusang Zhang}, \bibinfo{person}{Tao Gu}, \bibinfo{person}{Chenren Xu}, {and} \bibinfo{person}{Daqing Zhang}.} \bibinfo{year}{2022}\natexlab{}.
\newblock \showarticletitle{Loear: Push the range limit of acoustic sensing for vital sign monitoring}.
\newblock \bibinfo{journal}{\emph{Proceedings of the ACM on Interactive, Mobile, Wearable and Ubiquitous Technologies}} \bibinfo{volume}{6}, \bibinfo{number}{3} (\bibinfo{year}{2022}), \bibinfo{pages}{1--24}.
\newblock


\bibitem[Wang and Sun(2021)]%
        {wang2021mavl}
\bibfield{author}{\bibinfo{person}{Mei Wang} {and} \bibinfo{person}{Wei Sun}.} \bibinfo{year}{2021}\natexlab{}.
\newblock \showarticletitle{MAVL: Multiresolution Analysis of Voice Localization.}
\newblock \bibinfo{journal}{\emph{In Proc. of NSDI}} (\bibinfo{year}{2021}).
\newblock


\bibitem[Wang et~al\mbox{.}(2019a)]%
        {wang2019voicepop}
\bibfield{author}{\bibinfo{person}{Qian Wang}, \bibinfo{person}{Xiu Lin}, \bibinfo{person}{Man Zhou}, \bibinfo{person}{Yanjiao Chen}, \bibinfo{person}{Cong Wang}, \bibinfo{person}{Qi Li}, {and} \bibinfo{person}{Xiangyang Luo}.} \bibinfo{year}{2019}\natexlab{a}.
\newblock \showarticletitle{Voicepop: A pop noise based anti-spoofing system for voice authentication on smartphones}. In \bibinfo{booktitle}{\emph{IEEE INFOCOM 2019-IEEE Conference on Computer Communications}}. IEEE, \bibinfo{pages}{2062--2070}.
\newblock


\bibitem[Wang et~al\mbox{.}(2018)]%
        {wangCFMCWBasedContactless2018}
\bibfield{author}{\bibinfo{person}{Tianben Wang}, \bibinfo{person}{Daqing Zhang}, \bibinfo{person}{Yuanqing Zheng}, \bibinfo{person}{Tao Gu}, \bibinfo{person}{Xingshe Zhou}, {and} \bibinfo{person}{Bernadette Dorizzi}.} \bibinfo{year}{2018}\natexlab{}.
\newblock \showarticletitle{C-{FMCW} {Based} {Contactless} {Respiration} {Detection} {Using} {Acoustic} {Signal}}.
\newblock \bibinfo{journal}{\emph{Proceedings of the ACM on Interactive, Mobile, Wearable and Ubiquitous Technologies}} \bibinfo{volume}{1}, \bibinfo{number}{4} (\bibinfo{date}{Jan.} \bibinfo{year}{2018}), \bibinfo{pages}{1--20}.
\newblock
\showISSN{2474-9567}
\href{https://doi.org/10.1145/3161188}{doi:\nolinkurl{10.1145/3161188}}


\bibitem[Wang et~al\mbox{.}(2015)]%
        {wang2015understanding}
\bibfield{author}{\bibinfo{person}{Wei Wang}, \bibinfo{person}{Alex~X Liu}, \bibinfo{person}{Muhammad Shahzad}, \bibinfo{person}{Kang Ling}, {and} \bibinfo{person}{Sanglu Lu}.} \bibinfo{year}{2015}\natexlab{}.
\newblock \showarticletitle{Understanding and modeling of wifi signal based human activity recognition}. In \bibinfo{booktitle}{\emph{Proceedings of the 21st annual international conference on mobile computing and networking}}. \bibinfo{pages}{65--76}.
\newblock


\bibitem[Wang et~al\mbox{.}(2016)]%
        {wangDevicefreeGestureTracking2016}
\bibfield{author}{\bibinfo{person}{Wei Wang}, \bibinfo{person}{Alex~X. Liu}, {and} \bibinfo{person}{Ke Sun}.} \bibinfo{year}{2016}\natexlab{}.
\newblock \showarticletitle{Device-free gesture tracking using acoustic signals}. In \bibinfo{booktitle}{\emph{Proceedings of the 22nd {Annual} {International} {Conference} on {Mobile} {Computing} and {Networking}}} \emph{(\bibinfo{series}{{MobiCom} '16})}. \bibinfo{publisher}{Association for Computing Machinery}, \bibinfo{address}{New York, NY, USA}, \bibinfo{pages}{82--94}.
\newblock
\showISBNx{978-1-4503-4226-1}
\href{https://doi.org/10.1145/2973750.2973764}{doi:\nolinkurl{10.1145/2973750.2973764}}


\bibitem[Wang et~al\mbox{.}(2017)]%
        {wangWiFallDeviceFreeFall2017}
\bibfield{author}{\bibinfo{person}{Yuxi Wang}, \bibinfo{person}{Kaishun Wu}, {and} \bibinfo{person}{Lionel~M. Ni}.} \bibinfo{year}{2017}\natexlab{}.
\newblock \showarticletitle{{WiFall}: {Device}-{Free} {Fall} {Detection} by {Wireless} {Networks}}.
\newblock \bibinfo{journal}{\emph{IEEE Transactions on Mobile Computing}} \bibinfo{volume}{16}, \bibinfo{number}{2} (\bibinfo{date}{Feb.} \bibinfo{year}{2017}), \bibinfo{pages}{581--594}.
\newblock
\showISSN{1558-0660}


\bibitem[Wang et~al\mbox{.}(2020)]%
        {wang2020uwhear}
\bibfield{author}{\bibinfo{person}{Ziqi Wang}, \bibinfo{person}{Zhe Chen}, \bibinfo{person}{Akash~Deep Singh}, \bibinfo{person}{Luis Garcia}, \bibinfo{person}{Jun Luo}, {and} \bibinfo{person}{Mani~B Srivastava}.} \bibinfo{year}{2020}\natexlab{}.
\newblock \showarticletitle{Uwhear: through-wall extraction and separation of audio vibrations using wireless signals}. In \bibinfo{booktitle}{\emph{Proceedings of the 18th Conference on Embedded Networked Sensor Systems}}. \bibinfo{pages}{1--14}.
\newblock


\bibitem[Wu et~al\mbox{.}(2022)]%
        {wu2022WifiCanDoMore}
\bibfield{author}{\bibinfo{person}{Chenshu Wu}, \bibinfo{person}{Beibei Wang}, \bibinfo{person}{Oscar~C Au}, {and} \bibinfo{person}{KJ~Ray Liu}.} \bibinfo{year}{2022}\natexlab{}.
\newblock \showarticletitle{Wi-Fi Can Do More: Toward Ubiquitous Wireless Sensing}.
\newblock \bibinfo{journal}{\emph{IEEE Communications Standards Magazine}} \bibinfo{volume}{6}, \bibinfo{number}{2} (\bibinfo{year}{2022}), \bibinfo{pages}{42--49}.
\newblock


\bibitem[Wu et~al\mbox{.}(2021)]%
        {wuGaitWayMonitoringRecognizing2021a}
\bibfield{author}{\bibinfo{person}{Chenshu Wu}, \bibinfo{person}{Feng Zhang}, \bibinfo{person}{Yuqian Hu}, {and} \bibinfo{person}{K.~J.~Ray Liu}.} \bibinfo{year}{2021}\natexlab{}.
\newblock \showarticletitle{{GaitWay}: {Monitoring} and {Recognizing} {Gait} {Speed} {Through} the {Walls}}.
\newblock \bibinfo{journal}{\emph{IEEE Transactions on Mobile Computing}} \bibinfo{volume}{20}, \bibinfo{number}{6} (\bibinfo{date}{June} \bibinfo{year}{2021}), \bibinfo{pages}{2186--2199}.
\newblock
\showISSN{1558-0660}


\bibitem[Wu et~al\mbox{.}(2020a)]%
        {wu2020gaitway}
\bibfield{author}{\bibinfo{person}{Chenshu Wu}, \bibinfo{person}{Feng Zhang}, \bibinfo{person}{Yuqian Hu}, {and} \bibinfo{person}{KJ~Ray Liu}.} \bibinfo{year}{2020}\natexlab{a}.
\newblock \showarticletitle{GaitWay: Monitoring and recognizing gait speed through the walls}.
\newblock \bibinfo{journal}{\emph{IEEE Transactions on Mobile Computing}} \bibinfo{volume}{20}, \bibinfo{number}{6} (\bibinfo{year}{2020}), \bibinfo{pages}{2186--2199}.
\newblock


\bibitem[Wu et~al\mbox{.}(2020b)]%
        {wu2020mmtrack}
\bibfield{author}{\bibinfo{person}{Chenshu Wu}, \bibinfo{person}{Feng Zhang}, \bibinfo{person}{Beibei Wang}, {and} \bibinfo{person}{KJ~Ray Liu}.} \bibinfo{year}{2020}\natexlab{b}.
\newblock \showarticletitle{mmTrack: Passive multi-person localization using commodity millimeter wave radio}. In \bibinfo{booktitle}{\emph{IEEE INFOCOM 2020-IEEE Conference on Computer Communications}}. IEEE, \bibinfo{pages}{2400--2409}.
\newblock


\bibitem[Xie et~al\mbox{.}(2023)]%
        {xie2023boosting}
\bibfield{author}{\bibinfo{person}{Binbin Xie}, \bibinfo{person}{Minhao Cui}, \bibinfo{person}{Deepak Ganesan}, \bibinfo{person}{Xiangru Chen}, {and} \bibinfo{person}{Jie Xiong}.} \bibinfo{year}{2023}\natexlab{}.
\newblock \showarticletitle{Boosting the Long Range Sensing Potential of LoRa}. In \bibinfo{booktitle}{\emph{Proceedings of the 21st Annual International Conference on Mobile Systems, Applications and Services}}. \bibinfo{pages}{177--190}.
\newblock


\bibitem[Xiong and Jamieson(2013)]%
        {xiong2013arraytrack}
\bibfield{author}{\bibinfo{person}{Jie Xiong} {and} \bibinfo{person}{Kyle Jamieson}.} \bibinfo{year}{2013}\natexlab{}.
\newblock \showarticletitle{$\{$ArrayTrack$\}$: A $\{$Fine-Grained$\}$ indoor location system}. In \bibinfo{booktitle}{\emph{10th USENIX Symposium on Networked Systems Design and Implementation (NSDI 13)}}. \bibinfo{pages}{71--84}.
\newblock


\bibitem[Xu et~al\mbox{.}(2022)]%
        {xu2022hearing}
\bibfield{author}{\bibinfo{person}{Chenhan Xu}, \bibinfo{person}{Tianyu Chen}, \bibinfo{person}{Huining Li}, \bibinfo{person}{Alexander Gherardi}, \bibinfo{person}{Michelle Weng}, \bibinfo{person}{Zhengxiong Li}, {and} \bibinfo{person}{Wenyao Xu}.} \bibinfo{year}{2022}\natexlab{}.
\newblock \showarticletitle{Hearing Heartbeat from Voice: Towards Next Generation Voice-User Interfaces with Cardiac Sensing Functions}. In \bibinfo{booktitle}{\emph{Proceedings of the 20th ACM Conference on Embedded Networked Sensor Systems}}. \bibinfo{pages}{149--163}.
\newblock


\bibitem[Xu et~al\mbox{.}(2023)]%
        {xu2023practically}
\bibfield{author}{\bibinfo{person}{Huatao Xu}, \bibinfo{person}{Pengfei Zhou}, \bibinfo{person}{Rui Tan}, {and} \bibinfo{person}{Mo Li}.} \bibinfo{year}{2023}\natexlab{}.
\newblock \showarticletitle{Practically Adopting Human Activity Recognition}. In \bibinfo{booktitle}{\emph{Proceedings of the 29th Annual International Conference on Mobile Computing and Networking}}.
\newblock


\bibitem[Xu et~al\mbox{.}(2019a)]%
        {xu2019acousticid}
\bibfield{author}{\bibinfo{person}{Wei Xu}, \bibinfo{person}{ZhiWen Yu}, \bibinfo{person}{Zhu Wang}, \bibinfo{person}{Bin Guo}, {and} \bibinfo{person}{Qi Han}.} \bibinfo{year}{2019}\natexlab{a}.
\newblock \showarticletitle{Acousticid: gait-based human identification using acoustic signal}.
\newblock \bibinfo{journal}{\emph{Proceedings of the ACM on Interactive, Mobile, Wearable and Ubiquitous Technologies}} \bibinfo{volume}{3}, \bibinfo{number}{3} (\bibinfo{year}{2019}), \bibinfo{pages}{1--25}.
\newblock


\bibitem[Xu et~al\mbox{.}(2019b)]%
        {xuAcousticIDGaitbasedHuman2019}
\bibfield{author}{\bibinfo{person}{Wei Xu}, \bibinfo{person}{ZhiWen Yu}, \bibinfo{person}{Zhu Wang}, \bibinfo{person}{Bin Guo}, {and} \bibinfo{person}{Qi Han}.} \bibinfo{year}{2019}\natexlab{b}.
\newblock \showarticletitle{{AcousticID}: {Gait}-based {Human} {Identification} {Using} {Acoustic} {Signal}}.
\newblock \bibinfo{journal}{\emph{Proceedings of the ACM on Interactive, Mobile, Wearable and Ubiquitous Technologies}} \bibinfo{volume}{3}, \bibinfo{number}{3} (\bibinfo{date}{Sept.} \bibinfo{year}{2019}), \bibinfo{pages}{1--25}.
\newblock
\showISSN{2474-9567}


\bibitem[Yang et~al\mbox{.}(2023)]%
        {yang2023biocase}
\bibfield{author}{\bibinfo{person}{Yilin Yang}, \bibinfo{person}{Xin Li}, \bibinfo{person}{Zhengkun Ye}, \bibinfo{person}{Yan Wang}, {and} \bibinfo{person}{Yingying Chen}.} \bibinfo{year}{2023}\natexlab{}.
\newblock \showarticletitle{BioCase: Privacy Protection via Acoustic Sensing of Finger Touches on Smartphone Case Mini-Structures}. In \bibinfo{booktitle}{\emph{Proceedings of the 21st Annual International Conference on Mobile Systems, Applications and Services}}. \bibinfo{pages}{397--409}.
\newblock


\bibitem[Yu et~al\mbox{.}(2025)]%
        {yu2025uspeech}
\bibfield{author}{\bibinfo{person}{Luca Jiang-Tao Yu}, \bibinfo{person}{Running Zhao}, \bibinfo{person}{Sijie Ji}, \bibinfo{person}{Edith~CH Ngai}, {and} \bibinfo{person}{Chenshu Wu}.} \bibinfo{year}{2025}\natexlab{}.
\newblock \showarticletitle{USpeech: Ultrasound-Enhanced Speech with Minimal Human Effort via Cross-Modal Synthesis}.
\newblock \bibinfo{journal}{\emph{Proceedings of the ACM on Interactive, Mobile, Wearable and Ubiquitous Technologies}} \bibinfo{volume}{9}, \bibinfo{number}{2} (\bibinfo{year}{2025}), \bibinfo{pages}{1--31}.
\newblock


\bibitem[Yuan et~al\mbox{.}(2024)]%
        {kuang2024brush}
\bibfield{author}{\bibinfo{person}{Kuang Yuan}, \bibinfo{person}{Mohamed Ibrahim}, \bibinfo{person}{Yiwen Song}, \bibinfo{person}{Guoxiang Deng}, \bibinfo{person}{Robert~A. Nerone}, \bibinfo{person}{Suvendra Vijayan}, \bibinfo{person}{Akshay Gadre}, {and} \bibinfo{person}{Swarun Kumar}.} \bibinfo{year}{2024}\natexlab{}.
\newblock \showarticletitle{ToMoBrush: Exploring Dental Health Sensing Using a Sonic Toothbrush}.
\newblock \bibinfo{journal}{\emph{Proc. ACM Interact. Mob. Wearable Ubiquitous Technol.}} \bibinfo{volume}{8}, \bibinfo{number}{3}, Article \bibinfo{articleno}{139} (\bibinfo{date}{Sept.} \bibinfo{year}{2024}), \bibinfo{numpages}{27}~pages.
\newblock
\href{https://doi.org/10.1145/3678505}{doi:\nolinkurl{10.1145/3678505}}


\bibitem[Yuan et~al\mbox{.}(2025)]%
        {kuang2025air}
\bibfield{author}{\bibinfo{person}{Kuang Yuan}, \bibinfo{person}{Dong Li}, \bibinfo{person}{Hao Zhou}, \bibinfo{person}{Zhehao Li}, \bibinfo{person}{Lili Qiu}, \bibinfo{person}{Swarun Kumar}, {and} \bibinfo{person}{Jie Xiong}.} \bibinfo{year}{2025}\natexlab{}.
\newblock \showarticletitle{WindDancer: Understanding Acoustic Sensing under Ambient Airflow}.
\newblock \bibinfo{journal}{\emph{Proc. ACM Interact. Mob. Wearable Ubiquitous Technol.}} \bibinfo{volume}{9}, \bibinfo{number}{2}, Article \bibinfo{articleno}{61} (\bibinfo{date}{June} \bibinfo{year}{2025}), \bibinfo{numpages}{25}~pages.
\newblock
\href{https://doi.org/10.1145/3729469}{doi:\nolinkurl{10.1145/3729469}}


\bibitem[Yun et~al\mbox{.}(2015)]%
        {yunTurningMobileDevice2015}
\bibfield{author}{\bibinfo{person}{Sangki Yun}, \bibinfo{person}{Yi-Chao Chen}, {and} \bibinfo{person}{Lili Qiu}.} \bibinfo{year}{2015}\natexlab{}.
\newblock \showarticletitle{Turning a {Mobile} {Device} into a {Mouse} in the {Air}}. In \bibinfo{booktitle}{\emph{Proceedings of the 13th {Annual} {International} {Conference} on {Mobile} {Systems}, {Applications}, and {Services}}}. \bibinfo{publisher}{ACM}, \bibinfo{address}{Florence Italy}, \bibinfo{pages}{15--29}.
\newblock
\showISBNx{978-1-4503-3494-5}
\urldef\tempurl%
\url{https://dl.acm.org/doi/10.1145/2742647.2742662}
\showURL{%
\tempurl}


\bibitem[Yun et~al\mbox{.}(2017)]%
        {yunStrataFineGrainedAcousticbased2017}
\bibfield{author}{\bibinfo{person}{Sangki Yun}, \bibinfo{person}{Yi-Chao Chen}, \bibinfo{person}{Huihuang Zheng}, \bibinfo{person}{Lili Qiu}, {and} \bibinfo{person}{Wenguang Mao}.} \bibinfo{year}{2017}\natexlab{}.
\newblock \showarticletitle{Strata: {Fine}-{Grained} {Acoustic}-based {Device}-{Free} {Tracking}}. In \bibinfo{booktitle}{\emph{Proceedings of the 15th {Annual} {International} {Conference} on {Mobile} {Systems}, {Applications}, and {Services}}}. \bibinfo{publisher}{ACM}, \bibinfo{address}{Niagara Falls New York USA}, \bibinfo{pages}{15--28}.
\newblock
\showISBNx{978-1-4503-4928-4}
\urldef\tempurl%
\url{https://dl.acm.org/doi/10.1145/3081333.3081356}
\showURL{%
\tempurl}


\bibitem[Zhang et~al\mbox{.}(2018)]%
        {zhangWiSpeedStatisticalElectromagnetic2018}
\bibfield{author}{\bibinfo{person}{Feng Zhang}, \bibinfo{person}{Chen Chen}, \bibinfo{person}{Beibei Wang}, {and} \bibinfo{person}{K.~J.~Ray Liu}.} \bibinfo{year}{2018}\natexlab{}.
\newblock \showarticletitle{{WiSpeed}: {A} {Statistical} {Electromagnetic} {Approach} for {Device}-{Free} {Indoor} {Speed} {Estimation}}.
\newblock \bibinfo{journal}{\emph{IEEE Internet of Things Journal}} \bibinfo{volume}{5}, \bibinfo{number}{3} (\bibinfo{date}{June} \bibinfo{year}{2018}), \bibinfo{pages}{2163--2177}.
\newblock
\showISSN{2327-4662}
\href{https://doi.org/10.1109/JIOT.2018.2826227}{doi:\nolinkurl{10.1109/JIOT.2018.2826227}}


\bibitem[Zhang et~al\mbox{.}(2016)]%
        {zhang2016dopenc}
\bibfield{author}{\bibinfo{person}{Huanle Zhang}, \bibinfo{person}{Wan Du}, \bibinfo{person}{Pengfei Zhou}, \bibinfo{person}{Mo Li}, {and} \bibinfo{person}{Prasant Mohapatra}.} \bibinfo{year}{2016}\natexlab{}.
\newblock \showarticletitle{DopEnc: Acoustic-based encounter profiling using smartphones}. In \bibinfo{booktitle}{\emph{Proceedings of the 22nd Annual International Conference on Mobile Computing and Networking}}. \bibinfo{pages}{294--307}.
\newblock


\bibitem[Zhang et~al\mbox{.}(2023a)]%
        {zhang2023vecare}
\bibfield{author}{\bibinfo{person}{Yi Zhang}, \bibinfo{person}{Weiying Hou}, \bibinfo{person}{Zheng Yang}, {and} \bibinfo{person}{Chenshu Wu}.} \bibinfo{year}{2023}\natexlab{a}.
\newblock \showarticletitle{VECARE: Statistical Acoustic Sensing for Automotive In-Cabin Monitoring}. In \bibinfo{booktitle}{\emph{USENIX NSDI}}.
\newblock


\bibitem[Zhang et~al\mbox{.}(2023b)]%
        {zhang2023addressing}
\bibfield{author}{\bibinfo{person}{Yongzhao Zhang}, \bibinfo{person}{Hao Pan}, \bibinfo{person}{Yi-Chao Chen}, \bibinfo{person}{Lili Qiu}, \bibinfo{person}{Yu Lu}, \bibinfo{person}{Guangtao Xue}, \bibinfo{person}{Jiadi Yu}, \bibinfo{person}{Feng Lyu}, {and} \bibinfo{person}{Haonan Wang}.} \bibinfo{year}{2023}\natexlab{b}.
\newblock \showarticletitle{Addressing Practical Challenges in Acoustic Sensing To Enable Fast Motion Tracking}. In \bibinfo{booktitle}{\emph{Proceedings of the 22nd International Conference on Information Processing in Sensor Networks}}. \bibinfo{pages}{82--95}.
\newblock


\bibitem[Zhang et~al\mbox{.}(2023c)]%
        {zhang2023acoustic}
\bibfield{author}{\bibinfo{person}{Yongzhao Zhang}, \bibinfo{person}{Yezhou Wang}, \bibinfo{person}{Lanqing Yang}, \bibinfo{person}{Mei Wang}, \bibinfo{person}{Yi-Chao Chen}, \bibinfo{person}{Lili Qiu}, \bibinfo{person}{Yihong Liu}, \bibinfo{person}{Guangtao Xue}, {and} \bibinfo{person}{Jiadi Yu}.} \bibinfo{year}{2023}\natexlab{c}.
\newblock \showarticletitle{Acoustic Sensing and Communication Using Metasurface}. In \bibinfo{booktitle}{\emph{20th USENIX Symposium on Networked Systems Design and Implementation (NSDI 23)}}. \bibinfo{pages}{1359--1374}.
\newblock


\bibitem[Zheng et~al\mbox{.}(2019)]%
        {zheng2019zero}
\bibfield{author}{\bibinfo{person}{Yue Zheng}, \bibinfo{person}{Yi Zhang}, \bibinfo{person}{Kun Qian}, \bibinfo{person}{Guidong Zhang}, \bibinfo{person}{Yunhao Liu}, \bibinfo{person}{Chenshu Wu}, {and} \bibinfo{person}{Zheng Yang}.} \bibinfo{year}{2019}\natexlab{}.
\newblock \showarticletitle{Zero-effort cross-domain gesture recognition with Wi-Fi}. In \bibinfo{booktitle}{\emph{Proceedings of the 17th annual international conference on mobile systems, applications, and services}}. \bibinfo{pages}{313--325}.
\newblock


\bibitem[Zhou et~al\mbox{.}(2021)]%
        {zhou2021inferring}
\bibfield{author}{\bibinfo{person}{Suping Zhou}, \bibinfo{person}{Jia Jia}, \bibinfo{person}{Zhiyong Wu}, \bibinfo{person}{Zhihan Yang}, \bibinfo{person}{Yanfeng Wang}, \bibinfo{person}{Wei Chen}, \bibinfo{person}{Fanbo Meng}, \bibinfo{person}{Shuo Huang}, \bibinfo{person}{Jialie Shen}, {and} \bibinfo{person}{Xiaochuan Wang}.} \bibinfo{year}{2021}\natexlab{}.
\newblock \showarticletitle{Inferring emotion from large-scale internet voice data: A semi-supervised curriculum augmentation based deep learning approach}. In \bibinfo{booktitle}{\emph{Proceedings of the AAAI conference on artificial intelligence}}, Vol.~\bibinfo{volume}{35}. \bibinfo{pages}{6039--6047}.
\newblock


\end{thebibliography}

\end{document}